\documentclass{article}
\usepackage[usenames,dvipsnames]{color}
\usepackage[utf8]{inputenc}
\usepackage[noadjust]{cite}
\usepackage{amsmath,amssymb,amsthm}
\usepackage{easybmat}
\usepackage{appendix}
\usepackage{hyperref}
\usepackage{cleveref}
\usepackage{graphicx}
\graphicspath{ {./images/} }
\numberwithin{equation}{section}
\usepackage{blindtext}
\usepackage[a4paper, total={6in, 8in}]{geometry}
\usepackage{graphicx}

\begin{document}

\title{Characterization of Quantumness of non-Gaussian states under the influence of Gaussian channel}
    \author{
Ramniwas Meena$^{a}$\thanks{E-mail: meena.53@iitj.ac.in}, 
Subhashish Banerjee$^{b}$\thanks{E-mail: subhashish@iitj.ac.in}
\\
$^{a,b}$Department of Physics, Indian Institute of Technology, Jodhpur, India-342030\\
}
\date{December 2022}


\maketitle
\begin{abstract}
    The impact of a noisy Gaussian channel on a wide range of non-Gaussian input states is studied in this work. The nonclassical nature of the states, both input and output, is developed by studying the corresponding photon statistics and quasi-probability distributions. It is found that photon addition has more robust quantum mechanical properties as compared to the photon subtraction case. The threshold value of the noise parameter corresponding to the transition from partial negative ($W$ and $P$) and zero ($Q$) to completely positive definite, at the center of phase space, depends not only on the average number of thermal photons in the state, but also on the squeezing parameter. In addition it is observed that the nonclassicality of the $k^{th}$ number filtrated thermal state could be further enhanced by adding photon(s).
\end{abstract}
\providecommand{\keywords}[1]
{
  \small	
  \textbf{\textit{Keywords---}} #1
}
\section{Introduction}

Quantum information is an application of quantum mechanics to information theory, arising from its fundamental properties \cite{nielsen_chuang_2010, wilde_2013}. It exploits concepts such as state superposition, entanglement, and wave function collapse to establish new paradigms in the field of information processing, for example, for computing applications \cite{gruska1999quantum}, cryptography \cite{bennett1992quantum}, and simulation of quantum phenomena \cite{buluta2009quantum}. To harness the quantum system behaviour it is necessary to focus on quantum state engineering \cite{dell2006multiphoton, makhlin2001quantum}. Quantum engineering advances these concepts by implementing them in a realistic scenario, as well as developing algorithms, protocols, devices and systems.

The existence of quantum systems in complete isolation is a very rare occurrence. Generally, the dynamics is that of an open system where the system of interest evolves under the influence of interactions with its surroundings \cite{subhashish2019open,louisell1973quantum}. It is essential to understand how the environment affects quantum systems, in order to investigate their properties. The idea of open quantum system plays an important role in quantum state engineering. States engineered, both Gaussian and non-Gaussian, are impacted by their environment which could be modelled by (non-)Gaussian channels \cite{paris2003purity, lvovsky2020production}. These play an important role in the physics of continuous-variable quantum systems.
Here, Gaussian channels have attracted considerable attention \cite{RevModPhys.84.621}. A study of Gaussian channels has been conducted to study their impact on the evolution of Gaussian states  in the physics of continuous-variable quantum systems \cite{RevModPhys.84.621}. Over the past decade, a number of studies have been carried out on Gaussian states, both from theoretical as well as experimental perspectives. Nevertheless, many quantum technologies \cite{o2009photonic} beyond the realm of Gaussian states, demand non-Gaussian elements to be introduced.\\

It should be noted that while Gaussian states have been a great success both experimentally and theoretically \cite{adesso2014continuous, ourjoumtsev2007increasing, laurat2005entanglement}, they have a significant disadvantage when it comes to quantum technology: all Gaussian measurements can be efficiently simulated \cite{PhysRevLett.88.097904}. A non-Gaussian operation was argued to be necessary for the implementation of a universal quantum computer in pioneering work on CV quantum computation \cite{PhysRevLett.82.1784}. It is particularly difficult to implement common schemes based on the cubic phase gate in realistic setups \cite{PhysRevLett.97.110501,PhysRevA.79.062318,PhysRevLett.112.120504,PhysRevA.95.052352}. In addition, the information encoded by these protocols is highly non-Gaussian in nature, such as the Gottesman-Kitaev-Preskill state \cite{PhysRevA.64.012310}. However, these states are quite difficult to create even though they could also be used to implement non-Gaussian gates \cite{PhysRevLett.123.200502}. Although non-Gaussian states present practical difficulties, in the CV regime one must venture into non-Gaussian territory to achieve a quantum computational advantage \cite{Kim_2008}. In this sense, it is important to have a general understanding of the properties of non-Gaussian states and the way they behave.
\\
In recent years, many proposals have been put forward to generate highly nonclassical quantum states by appropriate operators acting on light fields \cite{Xu_2014, Zhou, Yuan_Hong_Chun_2010, PhysRevA.82.053812, PhysRevA.84.012302, PhysRevA.82.043828, Zhou_2012, Lu_2014, WANG2015108}. With the advent of quantum state engineering \cite{PhysRevLett.71.1816, Sperling_2015, miranowicz2004dissipation, marchiolli2004engineering} and quantum information processing (\cite{pathak2013elements} and references therein), the study of non-classical properties of engineered quantum states have become a prominent ﬁeld \cite{agarwal2012quantum}. This is so because only the presence of nonclassical features in a quantum state can provide quantum supremacy. In the recent past, various techniques for quantum state engineering have been developed \cite{PhysRevLett.71.1816, Sperling_2015, miranowicz2004dissipation, marchiolli2004engineering , agarwal1991nonclassical, lee2010quantum}.
There are many methods of quantum state construction. Prominent among these are the use of the state superposition principle of quantum mechanics and operators acting on the light field state.
An important class of non-Gaussian states, the photon-added thermal state (PATS), which does not exhibit squeezing, was introduced in \cite{PhysRevA.43.492}. Multi-photon schemes can be realized in a laboratory, as the initial thermal (even coherent) fields contain a very small number of photons. The non-classicality of a single-photon-subtracted Gaussian state as well as a photon-added-then-subtracted thermal state was investigated in \cite{Yang:09}.
Photon subtraction and addition represented by bosonic annihilation  and creation operators $\hat{a}$ and $\hat{a}^{\dag}$, respectively, have been employed to transform a field state to a desired one. Focusing only on optics, these methods rely mainly on the use of beam splitters and detectors, as well as postselection measurements. Such techniques are useful in creating holes in the photon number distribution \cite{escher2004controlled} and in generating finite-dimensional quantum states \cite{miranowicz2004dissipation}, both of which are non-classical. They are also useful in realizing non-Gaussianity inducing operations, like photon addition and subtraction \cite{zavatta2004quantum, podoshvedov2014extraction}. Recently, nonclassical properties of photon-added and subtracted displaced Fock states were studied using various witnesses of lower- and higher-order nonclassicality \cite{Priya,MALPANI2020124964}. Hole burning operations are also very relevant, since the states investigated are extremely nonclassical when quantified through a nonclassicality measure \cite{Malpani2020,PhysRevA.100.023813}. The decoherence of  photon-added thermal, photon-added squeezed thermal states and photon-subtraction squeezed thermal states have been  studied in a photon-loss channel in terms of negativity of  Wigner function \cite{xu2019dynamical, hu2012nonclassicality, hu2010photon}.  However, the construction of photons filtered from the thermal state and subsequent photon addition on it and the investigation of its quantum properties have not yet been reported.
\\

In this work, our aim is to study the impact of a Gaussian channel on various non-Gaussian states {\it viz.} photon  added thermal state (PATS), photon-subtracted thermal state (PSTS), photon-added $k^{th}$-number filtered thermal state (PAKFTS), photon-added squeezed thermal state (PASTS) and photon-subtracted squeezed thermal state (PSSTS). 
By a comparison of input and output states for various parameters of nonclassicality witness, for example, sub-Poissonian photon statistics, Mandel $Q_{M}$ parameter, second ordered correlation function $g^{2}$, zeros of $Q$ function, negative region of Wigner $W$ and Sudarshan-Glauber $P$ functions, the effect of  noise on non-classicality or decoherence is brought out.\\

Making use of the technique of integration within an ordered product (IWOP) of operators \cite{Hong-IWOP}, quantum operators in optical fields are arranged into products in a unified manner  (normal ordering, antinormal ordering, Weyl (symmetric) ordering). This technique was proposed to utilize the power of Dirac’s symbolic method \cite{dirac1958section} and representation theory.
An ordered product symbol is constructed by arranging noncommutative operators in a way that enables them to commute by the IWOP technique. However, the nature of the operators remains unchanged, they are still $q$-numbers, instead of $c$-numbers. After the integration over $c$-numbers within an ordered product is performed, one can get rid of the normal ordering symbol.

The paper is organized as follows. The input non-Gaussian states are introduced in Sec. II, followed by an introduction of bosonic Gaussian channel in Sec. III. The resulting output states, after passing through the Gaussian channel, are then discussed. In Sec. V, we describe the phase space distribution for the input and output states. The photon statistics of the input and output states are studied in Section VI. We then make our conclusions.
 
\section{Input State}
 Here we introduce the input Gaussian states to be subsequently passed through the Gaussian quantum channel, discussed in Sec. III.  As examples of Gaussian states we consider the thermal and squeezed thermal states and apply
generalized $m$-photon addition/subtraction to the thermal and squeezed thermal states, to generate the corresponding non-Gaussian states. We also discuss hole burning in the thermal state.
 
 \subsection{Thermal State}
The Hamiltonian of a single mode harmonic oscillator in thermal equilibrium is $\hat{H}={\hat{a}^\dag\hat{a}\hbar\omega}$ and the state is described by the density operator $\hat{\rho}_{th}$ \cite{scully_zubairy_1997, puri2001mathematical},
\begin{align}
    \hat{\rho}_{th}={\left(1-e^{-\beta\hbar\omega}\right)}e^{-\beta\hbar\omega\hat{n}},\: \beta=\frac{1}{k_{B} T} \label{th},
\end{align}
where, $\omega$ represents the frequency of a single mode in thermal equilibrium at temperature $T$, $k_{B}$ is the Boltzmann constant, and  $\hat{n}=\hat{a}^\dag\hat{a}$, the number operator.\\ 
The Eq. \eqref{th} can be written as follows in normal order:
\begin{align}
    \hat{\rho}_{th}=A:e^{-A{\hat{a}^{\dag}\hat{a}}}:,\: A = \left( 1 - e^{- \beta\hbar\omega} \right) = \frac{1}{1 + n_{\text{th}}},
    \label{nth}
\end{align}
where, $n_{th}=\left[\exp(\beta\hbar\omega)-1\right]^{-1}$ is the average photon number in the thermal state and $:\hat{O} :$ is indicative of normal ordering of operator $\hat{O}$.

\paragraph{A. Photon Added Thermal State\\ }
The $m$-photon-added scheme, denoted by the mapping $\hat{\rho}\rightarrow\hat{a}^{\dag m}\hat{\rho}\hat{a}^{m}$, was first proposed by Agarwal and Tara \cite{PhysRevA.43.492}.
Theoretically, the photon added thermal state (PATS) can be obtained by repeatedly operating the photon creation operator $\hat{a}^\dag$ on the thermal state.
Thus, the density operator for photon added thermal state $\hat{\rho}_{PATS}$ is given by,
\begin{align}
    \hat{\rho}_{PATS}=\frac{\hat{a}^{\dag m} \hat{\rho}_{th}\hat{a}^{m}}{Tr\left[\hat{a}^{\dag m} \rho_{t h}\left(\hat{a}\right)^{m}\right]}=N_{m}^{-1}:\hat{a}^{\dag m} e^{-A\hat{a}^\dag\hat{a}}\hat{a}^{m}:
    \label{PATS}.
\end{align}

Here, $m$ is positive integer number and $N_{m} $ is the normalization constant which is determined in Appendix [A].

\paragraph{B. Photon Subtracted Thermal State\\}
The photon subtracted thermal State (PSTS) can be obtained by repeatedly operating the photon annihilator operator $\hat{a}$ on the thermal state. Hence the density operator for the photon subtracted thermal state $\hat{\rho}_{PSTS}$ is,
\begin{align*}
    \hat{\rho}_{PSTS}=\frac{\hat{a}^{m}\hat{\rho}_{th}\hat{a}^{\dag m }}{Tr [\hat{a}^{m}\hat{\rho}_{th}\hat{a}^{\dag m }]}.
\end{align*}
Any density operator $\hat{\rho}$ in terms of the coherent state, can be written as,
  \[ \hat{\rho}=\int{P(\alpha)|\alpha\rangle \langle\alpha| \frac{d^{2}\alpha}{\pi}},\]

where $P(\alpha)$ is a weight function sometimes known as the Glauber-Sudarshan $P$-function \cite{Gerry3chep}. The $P$-function is a prominent quasi-probability distribution \cite{THAPLIYAL2015261}.
The Eq. \eqref{th} can be written in terms of coherent state, as 
\begin{align}
    \hat{\rho}_{th}= \frac{1}{ n_{\text{th}}}\int{\frac{d^2 \alpha}{\pi}e^{-\frac{|\alpha|^{2}}{ n_{\text{th}}}}|\alpha\rangle\langle\alpha |}
    \label{cth}.
\end{align}
Using Eq. \eqref{cth}, we get following density operator  $ \hat{\rho}_{PSTS}$ for the photon subtraction thermal state,
\begin{equation}
    \hat{\rho}_{PSTS}=N_{m^-}^{-1}\hat{a}^{ m}\int{\frac{d^2 \alpha}{\pi}e^{-\frac{|\alpha|^{2}}{ n_{\text{th}}}}|\alpha\rangle\langle\alpha |}\hat{a}^{\dag m}= N_{m^-}^{-1}\int{\frac{d^2 \alpha}{\pi}{(\alpha^{\ast}\alpha)^m}e^{-\frac{|\alpha|^{2}}{ n_{\text{th}}}}|\alpha\rangle\langle\alpha |}
     \label{PSTS}   ,
\end{equation}
where, $m$ is positive integer number represent subtracted photons and  $N_{m^{-}}$ is the normalization constant which is determined in Appendix [A].

\paragraph{C. Photon Added Hole Burning Thermal State\\}
In terms of Fock basis, the thermal state can be described as:
\begin{align}
    \hat{\rho}_{th} = \left( 1 - e^{- \beta\hbar\omega} \right)\sum_{n = 0}^{}{e^{- \beta\hbar\omega\hat{n}}\left| n \right\rangle\left\langle n \right|} = \left( 1 - e^{- \beta\hbar\omega} \right)\sum_{n = 0}^{}{e^{- \beta\hbar\omega n}\left| n \right\rangle\left\langle n \right|} \label{eq8}.
\end{align}
There are a number of ways in which hole burning can be done for given states. It is possible to determine the hole burning state at a particular position (for example, the $k^{th}$ number state in the photon number distribution) by filtering the $k^{th}$-number state from the thermal state. This hole burning state, is also known as $k^{th}$-number filtered thermal state (KFTS).\\
In the KFTS, the density operator $\hat{\rho}_{KFTS}$ can be expressed as:
\begin{align*}
    \hat{\rho}_{\text{KFTS}} = \left( 1 - e^{- \beta\hbar\omega} \right)\sum_{n = 0,n \neq \ k}^{}{e^{- \beta\hbar\omega n}\left| n \right\rangle\left\langle n \right|} = \left( 1 - e^{- \beta\hbar\omega} \right)\bigg[\sum_{n = 0}^{}{e^{- \beta\hbar\omega n}\left| n \right\rangle\left\langle n \right|} - e^{- \beta\hbar\omega k}\left| k \right\rangle\left\langle k \right|\bigg] .
\end{align*}

With the use of IWOP techniques, the above equation can be written as follows:
\begin{align}
    \hat{\rho}_{KFTS}&=A\left[:e^{-A \hat{a}^{\dag}\hat{a}}: -\: \frac{e^{- \beta\hbar\omega k}}{k!}:{\hat{a}^{\dag k}} e^{- \hat{a}^{\dag}\hat{a}}{\hat{a}}^{k}:\right]  \label{9},
\end{align}
where, $A = \left( 1 - e^{- \beta\hbar\omega} \right) = \frac{1}{1 + n_{\text{th}}}.\\$

Theoretically, photon added KFTS  can be obtained by operating the creation operator on the  
$k^{th}$-number state filtration of thermal state. The density operator for photon added $k^{th}$-number filtered thermal state (PAKFTS) can be expressed as follows:
\begin{align}
    \hat{\rho}_{\text{PAKFTS}}=N_{\text{km}}^{- 1}\left[ :{{\hat{a}}^{\dag m}e^{- A{\hat{a}}^{\dag}\hat{a}}\hat{a}}^{m}:-\:{\frac{e^{- \beta \hbar\omega k}}{k!}:{{\hat{a}}^{\dag (m + k)}e}^{- {\hat{a}}^{\dag}\hat{a}}{\hat{a}}^{(m + k)}: }\right] \label{PAKFTS}.
\end{align}
Here, $N_{km}$ is the normalization constant which is determined in Appendix [A].

\subsection{Squeezed Thermal State}
The squeezed thermal state (STS) can be obtained from the thermal state by applying the single mode squeezing operator,
\begin{align*}
    \hat{\rho}_{s}={S_1}^{\dag}\hat{\rho}_{t h}S_1,
\end{align*}
where $S_1=\frac{1}{\sqrt\mu}\int dq\left|\left.\frac{q}{\mu}\right\rangle\right.\left\langle\left.q\right|\right.$; $\mu=e^{\lambda} $ is the single-mode squeezing operator with ${\lambda}$ being the squeezing parameter \cite{PhysRevD.35.1831,scully_zubairy_1997,Dodonov_2002}.\\
The above method enables one to obtain the normally ordered form of STS directly \cite{Hu:12},

\begin{align}
   \hat{\rho}_{s}=\frac{1}{\sqrt{A}}:\exp{\left[\frac{C}{2}(\hat{a}^{\dag 2}+\hat{a}^{2})+(B-1)\hat{a}^{\dag}\hat{a}\right]}: \label{sts},
\end{align}
where we have set,

$A=n_{th}^2+(2n_{th}+1){\cosh}^2\lambda$, $B=\frac{n_{th}}{A}(n_{th}+1)$, $C=\frac{2n_{th}+1}{2A}\sinh2\lambda$.

\paragraph{A. Photon Added Squeezed Thermal State\\}

The photon added squeezed thermal state (PASTS) can be obtained by repeatedly applying creator operators to squeezed thermal states.
From Eq. \eqref{sts} a photon added squeezed thermal state  $\hat{\rho}_{PASTS}$ can be obtained as:
\begin{align}
     \hat{\rho}_{PASTS}=\frac{N_{a,m}^{-1}}{\sqrt A}:\hat{a}^{\dag m}\exp{\left[\frac{C}{2}(\hat{a}^{\dag 2}+{\hat{a}}^2)+(B-1){\hat{a}}^\dag\hat{a}\right]} \hat{a}^{m}: 
     \label{PASTS}.
 \end{align}
Here, $m$ is a non-negative number and  $N_{a,m}$ is the normalization that is calculated in Appendix [A].

\paragraph{B. Photon Subtracted Squeezed Thermal State (PSSTS)\\}

Additionally, we introduce the photon subtracted thermal state (PSSTS), which can be generated by repeatedly applying the annihilator operator to squeezed thermal states.\\
From Eq. \eqref{sts} the photon added squeezed thermal state  $\hat{\rho}_{PASTS}$ can be obtained as:
\begin{align}
    \hat{\rho}_{PSSTS}=\frac{\hat{a}^{m}\hat{\rho}_{s}\hat{a}^{\dag m }}{Tr[\hat{a}^{m}\hat{\rho}_{s}\hat{a}^{\dag m 
     }]}=\frac{N_{a,m^{-}}^{-1}}{\sqrt A}\hat{a}^{ m}:\exp{\left[\frac{C}{2}(\hat{a}^{\dag 2}+{\hat{a}}^2)+(B-1){\hat{a}}^\dag\hat{a}\right]}: \hat{a}^{\dag m} \label{prePSSTS}.
\end{align}
Using the over-completeness relation for the coherent states, expressed as an integral over
the complex $\alpha$-plane, as \[\int{\frac{d^2 \alpha}{\pi}|\alpha\rangle\langle\alpha|}=1\] in 
Eq. \eqref{prePSSTS}, we have:
 \begin{align}
    \hat{\rho}_{PSSTS} &= \frac{N_{a,m^{-}}^{-1}}{\sqrt A}\int\int\frac{d^2\alpha}{\pi}\frac{d^2\beta}{\pi}\hat{a}^{ 
     m}|\alpha\rangle\langle\alpha|:\exp{\left[\frac{C}{2}(\hat{a}^{\dag 
     2}+{\hat{a}}^2)+(B-1){\hat{a}}^\dag\hat{a}\right]}:|\beta\rangle\langle\beta| \hat{a}^{\dag m} \nonumber \\
     & =\frac{N_{a,m^{-}}^{-1}}{\sqrt 
     A}\int\int\frac{d^2\alpha}{\pi}\frac{d^2\beta}{\pi}(\beta^\ast\alpha)^m\exp{\left[\frac{C}{2}(\alpha^{\ast 
     2}+{\beta}^2)+B\alpha^\ast\beta-\frac{|\alpha|^2+|\beta|^2}{2}\right]}|\alpha\rangle\langle\beta| .
     \label{PSSTS}
 \end{align}
$N_{a,m^{-}}$ is the normalization calculated in Appendix [A].

\section{Bosonic Gaussian Channels}

Bosonic Gaussian channels are the results of zero-mean random Gaussian shifts in the phase space of bosonic modes, which provide a completely positive map (CPM) on the phase space \cite{PhysRevA.50.3295}.
By applying this CPM to an input state $/hat[/rho]$, the output state $/mathrm{/Phi}_s/left(/hat{/rho}/right)$ may be obtained:
\begin{align}
    \mathrm{\Phi}_s\left(\hat{\rho}\right)=\int{\hat{D}_z\hat{\rho}\ {\hat{D}_z}^\dag} G\left(z\right)d^2z \label{Noise}.
\end{align}
Here $\hat{D}_z=e^{z\hat{a}^\dag-z^{\ast}\hat{a}}$ is the displacement operator in the phase space of the bosonic mode, where the bosonic creation and annihilation operators obey the commutation relation $[\hat{a},\hat{a}^\dag]=1$, and the classical-noise Gaussian distribution,
$G\left(z\right)=\frac{e^{-|z|^2/s}}{\pi s}$,
has zero mean and phase-space variance $\sigma^{2}=s$. The complex number $\alpha=(p+iq)/\sqrt{2}$ corresponds to the phase-space points of a single-mode harmonic oscillator described by position $q$ and momentum $p$. 
\begin{figure}
    \centering
    \includegraphics[width=.8\textwidth]{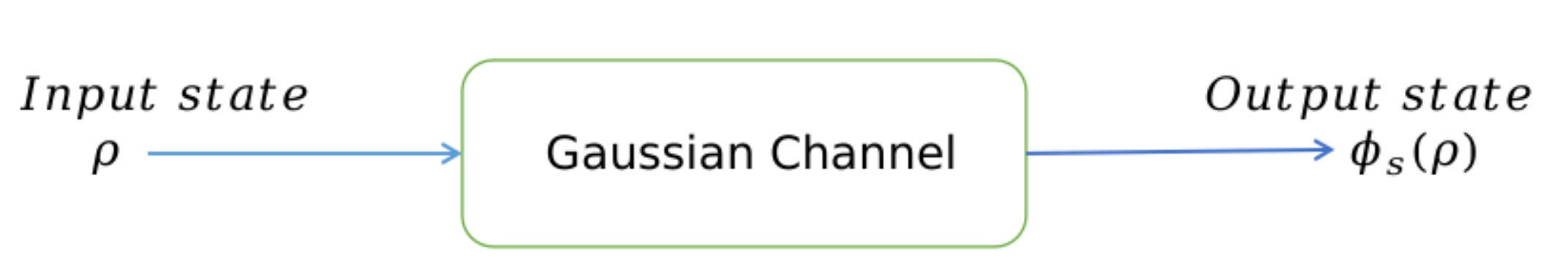}
    \caption{Noisy Gaussian Channel}
    \label{channel}
\end{figure}
It is also possible to write the Eq. \eqref{Noise} in terms of Kraus operators as:
\begin{align}
    \mathrm{\Phi}_s\left(\hat{\rho}\right)=\sum_{i} \hat{K}_i \rho {\hat{K}_i}^\dag, 
\end{align}
where, the Kraus operators $\hat{K}_i = \sqrt{G\left(z\right)\ }\hat{D}(z)$.
Here, the Gaussian channel $\phi_s(\rho)$ satisfies the semigroup property with respect to the parameter $s$ (with $s \geq 0$). Hence $s$ can be used to parameterize the evolution time \cite{zhang2021quantifying}. 
\section{Output State}

As the input state passes through the noisy Gaussian channel, it gets influenced by the channel resulting in the corresponding output state at the other end, see Fig.~(\ref{channel}). We now provide analytic expressions for various output states corresponding to the input states. This is made possible by the use of IWOP techniques.
\subsection{Thermal State}

\paragraph{A. Photon Added Thermal State \\}
From Eq. \eqref{PATS}, we have,
\begin{align}
     \hat{D}_{z}\hat{\rho}_{PATS} {\hat{D}_z}^\dag=N_{m}^{-1}:{{({\hat{a}}^\dag-z^\ast)}^m \exp{\bigg[-A({\hat{a}}^\dag-z^\ast){(\hat{a}-z)}\bigg]}{(\hat{a}-z)}^m}: \label{DzPATS}.
\end{align}
Here, we use the following properties of displacement operator for normal ordered function $F^n(\hat{a},\hat{a}^{\dag})$-
\begin{align}
    \hat{D_z}F^{n}(\hat{a},\hat{a}^{\dag})\hat{{D_z}}^\dag=F^{n}(\hat{a}-z,\hat{a}^{\dag}-z^{\ast}) \label{Dzf}.
\end{align}
From Eqs. \eqref{Noise} and \eqref{DzPATS}, shifting, $\left( \hat{a} - z \right) \rightarrow - z$,
\begin{align*}
    \mathrm
    {\Phi}_s\left(\hat{\rho}_{PATS}\right)=N_{m}^{-1}\int{\frac{d^2z}{\pi s}:\left(z^\ast z\right)^m \exp\bigg[-\left(A+1/s\right)|z|^2 +\frac{{\hat{a}}^\dag z}{s}+\frac{\hat{a}z^\ast}{s}-\frac{{\hat{a}}^\dag\hat{a}}{s}\bigg]}:.
\end{align*}

\fbox{%
\parbox{1.0\linewidth}{%
\begin{align}
    \int \frac{d^2 z}{\pi}z^{n}z^{\ast n} \exp[A|z|^2+Bz+Cz^{\ast}]=e^{-BC/A}\sum_{l=0}{\frac{n!m!}{l!(n-l)!(m-l)!(-A)^{n+m-l+1}}B^{m-l}C^{n-l}}\label{MIFS},
\end{align}
$Re(A)<0$.
\begin{align}
    \int{\frac{d^{2}z}{\pi}\exp{\big[\zeta|z|^{2}+\xi z+\eta z^{\ast}+f z^{2}+g z^{\ast 2}}\big]}=\frac{1}{\sqrt{\zeta^{2}-4 f g}}\exp{\left[\frac{-\zeta\xi\eta+\xi^{2}g+\eta^{2}f}{\zeta^{2}-4 f g}\right]} \label{MIF}  
\end{align}	
   whose convergent condition is $Re(\zeta\mp f\mp g)<0$ and $ Re\left(\frac{\zeta^{2}-4fg}{\zeta\mp f\mp g}\right)<0 $.%
	}%
	}
 \\

Using Eq. \eqref{MIFS}, along with the IWOP technique, we get the following output state for PATS
\begin{align}
    \mathrm
{\Phi}_s\left(\hat{\rho}_{PATS}\right)=N_{m}^{-1}:\exp^{-\frac{A{\hat{a}}^\dag\hat{a}}{As+1}}\sum_{l=0}^{m}\frac{{m!}^2\left({\hat{a}}^\dag\right)^{m-l}\left(\hat{a}\right)^{m-l}s^l}{l!{\left(m-l\right)!)}^2\left(As+1\right)^{2m-l+1}}: \label{NPATS}.
\end{align}

\paragraph{B. PSTS\\}
For photon subtracted thermal state from Eq. \eqref{PSTS},
\begin{align*}
   \hat{D}_{z}\hat{\rho}_{PSTS} {\hat{D}_z}^\dag=N_{m^-}^{-1}\int{\frac{d^2 
   \alpha}{\pi}{(\alpha^{\ast}\alpha)^m} e^{-\frac{|\alpha|^2}{n_{th}}}\hat{D}_{z}|\alpha\rangle\langle\alpha| {\hat{D}_z}^
   \dag}.
\end{align*}
Using the following property for the displacement operator,
\begin{align}
    \hat{D}_{z}|\alpha \rangle=\exp{\bigg[\frac{1}{2}(z\alpha^\ast-z^\ast \alpha)\bigg]} |{\alpha+z}\rangle
    \label{Dz0},
\end{align}
the output state for PSTS can be written as (using Eqs. \eqref{PSTS}, \eqref{Dz0} and \eqref{Noise}),
\begin{align}
    {\Phi}_s\left(\hat{\rho}\right)=N_{m^-}^{-1}\int\int{\frac{d^2 \alpha}{\pi}\frac{d^2 
    z}{\pi s}{(\alpha^{\ast}\alpha)^m}\exp{\bigg[-\frac{|\alpha|^2}{n_{th}}-\frac{|z|^2}{s}\bigg]}|{\alpha+z}\rangle\langle{\alpha+z}|} \label{NPSTS}.
\end{align}

\paragraph{\textbf{C. PAKFTS\\}}
From Eq. \eqref{PAKFTS}, we have,
\begin{equation}
    {\hat{D}}_{z}^{\dag}\hat{\rho}_{\text{KFTS}}{\hat{D}}_{z}^{\dag} = N_{\text{km}}^{- 1}\left( {\hat{D}}_{z}:{{\hat{a}}^{\dag m}e^{- A{\hat{a}}^{\dag}\hat{a}}\hat{a}}^{m}:{\hat{D}}_{z}^{\dag} - \frac{e^{- \beta\hbar\omega k}}{k!}{\hat{D}}_{z}:{{\hat{a}}^{\dag (m + k)}e}^{- {\hat{a}}^{\dag}\hat{a}}{\hat{a}}^{(m + k)}:{\hat{D}}_{z}^{\dag} \right) \label{DzPAKFTS}.
\end{equation}
From Eqs. \eqref{Dzf}, \eqref{PAKFTS} and \eqref{DzPAKFTS},
shifting, $\left( \hat{a} - z \right) \rightarrow - z$,
\begin{align*}
    \phi_{s}{\left(\hat{\rho}_{\text{PAKFTS}}\right)}=N_{\text{km}}^{- 1} \int{\left(:(z^\ast z)^{m} e^{-A |z|^{2}+{\frac{(\hat{a}^\dag-z\ast)(\hat{a}-z)}{s}}}:-\:\frac{e^{-\beta\hbar\omega k}}{k!}:(z^\ast z)^{m+k} e^{- |z|^{2}+{\frac{(\hat{a}^\dag-z\ast)(\hat{a}-z)}{s}}}:\right)\frac{d^{2}z}{\pi s}}.
\end{align*}
Using Eq. \eqref{MIF}, the output state for PAKFTS can be written as
\begin{multline}
      \phi_{s}{\left(\hat{\rho}_{\text{PAKFTS}}\right)}=N_{\text{km}}^{- 1} \bigg[:e^{-\frac{A\hat{a}^{\dag}\hat{a}}{As+1}}\sum_{l=0}^{m}\frac{m!^{2}\hat{a}^{\dag m-l}\hat{a}^{m-l}s^{l}}{l! (m-l)!^{2}(As+1)^{2m-l+1}}:\\
      -\:\frac{e^{-\beta\hbar\omega k}}{k!}:e^{-\frac{\hat{a}^{\dag}\hat{a}}{s+1}}\sum_{l=0}^{m+k}\frac{(m+k)!^{2}\hat{a}^{\dag m+k-l}\hat{a}^{m+k-l}s^{l}}{l! (m+k-l)!^{2}(s+1)^{2(m+k)-l+1}}:\bigg] \label{NPAKFTS}.
\end{multline}

\subsection{For Squeezed Thermal State}

\paragraph{A. Noisy PASTS\\}

For PASTS, we have (using Eqs. \eqref{Dzf} and \eqref{PASTS}),
\begin{align}
    \hat{D_z}\hat{\rho}_{PASTS}\hat{{D_z}}^\dag
    =\frac{N_{a,m}^{-1}}{\sqrt A}:\left({\hat{a}}^\dag-z^\ast\right)^m \exp\left[\frac{C}{2}(({{\hat{a}}^\dag-z^\ast)}^2+\left(\hat{a}-z\right)^2)+(B-1)({\hat{a}}^\dag-z^\ast)(\hat{a}-z)\right]{(\hat{a}-z)}^{m}: \label{DzPASTS}.
\end{align}
 
From Eqs. \eqref{Noise} and \eqref{DzPASTS},
\begin{align*}
    {\Phi}_s\left(\hat{\rho}_{PASTS}\right)
      &=\frac{N_{a,m}^{-1}}{\sqrt A}\int{\frac{d^2z}{\pi\ s}:{({\hat{a}}^\dag-z^\ast)}^{m}\exp{\left[\frac{C}{2}(({{\hat{a}}^\dag-z^\ast)}^2+{(\hat{a}-z)}^2)+(B-1)({\hat{a}}^\dag-z^\ast)(\hat{a}-z)-\frac{\left|z\right|^2}{s}\right]}{(\hat{a}-z)}^m:}.
\end{align*}

  The advantage of using the IWOP technique is that the operators $\hat{a}$ and $\hat{a}^{\dag}$ can be regraded as parameters in the integrations and can be permuted, allowing for the shifting
  $( z-\hat{a}\longrightarrow z) $,
\begin{align*}
    {\Phi}_s\left(\hat{\rho}_{PASTS}\right) & =\frac{N_{a,m}^{-1}}{\sqrt A}\int{\frac{d^2z}{\pi s}:{(z^\ast z)}^m \exp\left[\frac{C}{2}({z^\ast}^2+z^2) + \left(B-1-1/s\right)|z|^{2} +\frac{{\hat{a}}^\dag z}{s}+\frac{\hat{a}z^\ast}{s}-\frac{{\hat{a}}^\dag\hat{a}}{s}\right]}:\\
    &=\frac{N_{a,m}^{-1}}{\sqrt A}\partial_{Y}^{m}\int{\frac{d^2z}{\pi s}:{ \exp\left[\frac{C}{2}({z^\ast}^2+z^2) - Y |z|^{2} +\frac{{\hat{a}}^\dag z}{s}+\frac{\hat{a}z^\ast}{s}-\frac{{\hat{a}}^\dag\hat{a}}{s}\right]}:},
\end{align*}
 
 where $Y=(-B+1+1/s)$.
 
Using the Eq. \eqref{MIF}, we get the following output state for PASTS
\begin{align}
    \mathrm{\Phi}_s\left(\hat{\rho}_{PASTS}\right)=\frac{N_{a,m}^{-1}}{s\sqrt A}:\exp\left(-\frac{{\hat{a}}^\dag\hat{a}}{s}\right)\partial_Y^{m}\left[{(Y^2- C^2)}^{-1/2} \exp\left[\frac{\frac{C}{2}({{\hat{a}}^{\dag 2}}+{\hat{a}}^2) - Y {\hat{a}}^\dag\hat{a}}{s^2(Y^2-C^2)}\right]\right]: \label{NPASTS}.
\end{align}

\paragraph{B. PSSTS\\}

From Eq. \eqref{PSSTS} we have,
\begin{align*}
    \hat{D}_{z}\hat{\rho}_{\text{PSSTS}}0{\hat{D}_{z}}^{\dag}&=\frac{N_{a,m^{-}}^{-1}}{\sqrt{A}}\int{\frac{d^{2}\alpha}{\pi}{\frac{d^2\beta}{\pi}}(\beta^\ast\alpha)^{m}\exp{\left[\frac{C}{2}(\alpha^{\ast 2}+\beta^{2})+B\alpha^{\ast} \beta-\frac{|\alpha|^2+|\beta|^2}{2}\right]}{\hat{D}_{z}}{|{\alpha}\rangle\langle{\beta}|}{{\hat{D}_{z}}^\dag}}.
\end{align*}
From the above and Eq. \eqref{Noise}, the output state for PSSTS is seen to be
\begin{align*}
    \phi_{s}\left(\hat{\rho}_{PSSTS}\right)=\frac{N_{a,m^{-}}^{-1}}{\sqrt A}\int\frac{d^2 z}{s\pi}\frac{d^{2}\alpha}{\pi}\frac{d^{2}\beta}{\pi}(\beta^{\ast}\alpha)^{m} \exp{\left[\frac{C}{2}(\alpha^{\ast 2}+\beta^{2})+B \alpha^{\ast} \beta-\frac{|\alpha|^{2}+|\beta|^{2}}{2}-\frac{|z|^2}{s}\right]}{\hat{D}_{z}}|{\alpha}\rangle\langle{\beta}|{\hat{D_z}^{\dag}}.
\end{align*}
Using the Eq. \eqref{Dz0}, the above equation can be expressed as
\begin{align}
    \phi_s(\hat{\rho}_{PSSTS})
    & =\frac{N_{a,m^{-}}^{-1}}{\sqrt A}\int\frac{d^2z}{s\pi}\frac{d^2\alpha}{\pi}\frac{d^2\beta}{\pi}
    (\beta^\ast\alpha)^m \nonumber \\
    &\times\exp{\left[\frac{C}{2}(\alpha^{\ast 2}+{\beta}^2)+B\alpha^\ast\beta+\frac{1}{2}\{z(\alpha^\ast-\beta^\ast)-z^\ast(\alpha-\beta)\}-\frac{|\alpha|^2+|\beta|^2}{2}-\frac{|z|^2}{s}\right]}|{\alpha+z}\rangle\langle{\beta+z}|.
\end{align}

\section{Quasi-Probability Distribution}
An important aspect in the quest for the understanding of the quantumness of states, discussed above, would be the study of the corresponding quasi-probability distributions, for example, the $W$, $P$ and $Q$ distributions. These can be studied on a common platform by the use of  Cahill and Glauber's $\kappa $-parameterized function \cite{cahill1969ordered},
\begin{align}
     \chi\left(\gamma,\kappa\right)={Tr\left[\hat{\rho} \exp{\left(\gamma \hat{a}^\dag -\gamma^* \hat{a}+\left(\kappa/2\right) {|\gamma|}^2 \right)} \right]} \label{CF},
\end{align}
where $\kappa=-1,\:0,\:1$ corresponds to $Q$, $W$ and $P$ quasi-probability distributions, respectively.
 $P_\kappa\left(\alpha\right)$ are connected to the family of characteristics functions in CV systems through the complex Fourier transform,
\begin{align}
 P_\kappa\left(\alpha\right)=\frac{1}{\pi} \int \exp\left(\gamma^\ast\alpha-\gamma\alpha^\ast\right)\chi\left(\gamma,\kappa \right) \frac{d^2\gamma}{\pi} \label{QP}.
\end{align}

These distributions are called quasi-probability distributions, because even though they sum up to unity, their behavior is not entirely consistent with that expected of probability distributions. In particular, there are (infinitely many) quantum states $\rho$ for which the function $ P_\kappa\left(\alpha\right)$  is not a regular probability distribution for some values of $\kappa$, as it can assume 
negative values or even be singular in certain points of the phase space \cite{agarwal1981relation, schleich2011quantum, THAPLIYAL2015261, THAPLIYAL2016148, malpani2019quantum}. An exception is the case  
$\kappa=-1$, which corresponds to the Husimi ‘$Q$-function’ \cite{husimi1940some} and represents a non-negative and 
regular distribution for any quantum state $\rho$. Zeros of $Q$ function are a witness of non-classicality \cite{korsch1997zeros}. The case $\kappa=0$ corresponds to the so-called 
Wigner  ‘$W$-function' \cite{kenfack2004negativity}. Wigner function's negativity is another witness of non-classicality. The symmetrically ordered 
characteristic function would thus be $\chi\left(\gamma,0 \right)$. Finally,   $\kappa=1$ yields the so-called 
‘$P$-representation’, which was introduced independently by Glauber  and Sudarshan 
\cite{sudarshan1963equivalence}. The $P$ distribution can become negative or even singular (namely, more singular than a Dirac $\delta$) when the state $\rho$ deviates from a mixture of coherent states. For this reason, the regularity and positivity of the $P$-representation is often adopted as an indicator of ‘classicality’ of a CV state $\rho$ \cite{glauber1963quantum}. 

We now take up the states introduced above and obtain their characteristic function and thence the probability distribution function. This is done for both the input as well as the output states, obtained by sending the input state through the bosonic Gaussian channel. This enables a characterization of quantumness in the states as well as the contrast obtained due to the effect of the channel, which brings out the impact of noise.

\subsection{Thermal States}

\paragraph{A. Photon Added Thermal State\\}
\subsubsection{Input State}

We begin with the characteristic function (CF),
\begin{align}
    \chi_{}\left(\gamma,\kappa\right)={Tr\left[e^{\gamma\hat{a}^\dag}\hat{\rho}_{PATS}e^{-\gamma^{\ast} \hat{a}}\right]}e^{\frac{\kappa+1}{2}{|\gamma|^2}} \label{CFPA}.
\end{align}
Since,
\begin{align*}
    {Tr\left[e^{\gamma\hat{a}^\dag}\hat{\rho}_{PATS} e^{-\gamma^{\ast} \hat{a}}\right]}
    =\int{(\alpha^{\ast}\alpha)^{m}\exp{\left[-A|\alpha|^{2}+\gamma \alpha^\ast-\gamma^{\ast} \alpha\right]}\frac{d^2\alpha}{\pi}},
\end{align*}
making use of  Eq. \eqref{MIFS},
\begin{figure}
    \centering
    \includegraphics[width=\linewidth]{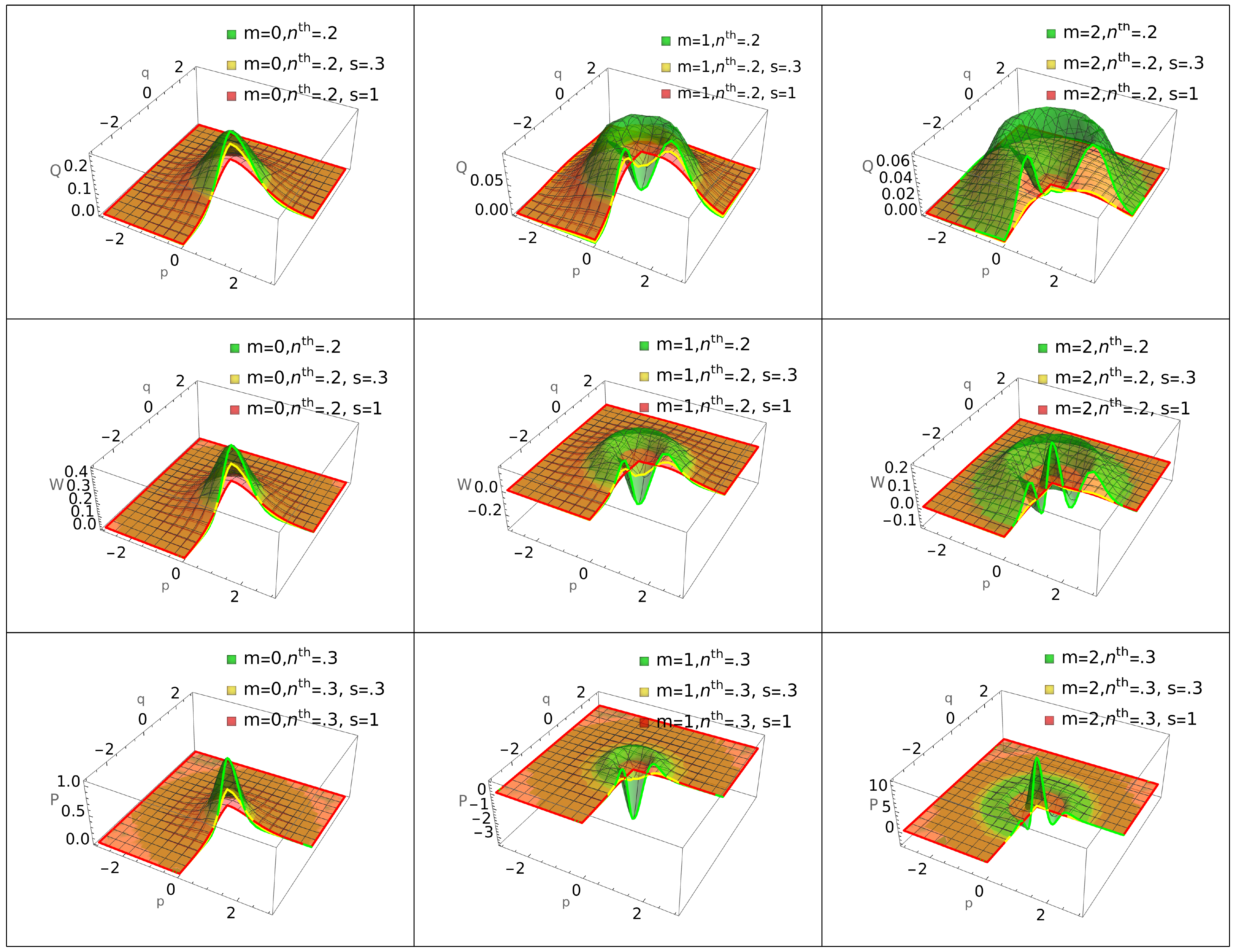}
    \caption{Quasi-Probability Distribution functions, {\it viz.}, Husimi $Q$-function (1st row), Wigner $W$-function (2nd row) and Sudarshan-Glauber $P$-function (3rd row) for thermal state (1st column), one photon added-thermal state (2nd column) and two photon-added thermal state (3rd column). Here input PATS are represented by green colour and output PATS are represented by yellow (noise parameter $s= 0.3$) and red (noise parameter $s=1$).}
    \label{PPA.eps}
\end{figure}
\begin{align*}
    {Tr\left[e^{\gamma \hat{a}^\dag}\hat{\rho}_{PATS} e^{-\gamma^{\ast} \hat{a}}\right]}=N_{m}^{-1}\sum_{l=0}^{m}{\frac{{m!}^2{({-\gamma}^\ast\gamma)}^{m-l}}{l!(({m-l)!)}^2 A^{2m-l+1}}e^{- \frac{\left|\gamma\right|^2}{A} }}.
\end{align*}
Using this in Eq. \eqref{CFPA}, the CF for the input state is,
\begin{align}
 \chi_{in}\left(\gamma,\kappa\right)&=N_{m}^{-1}\sum_{l=0}^{m}{\frac{{m!}^2{({-\gamma}^\ast\gamma)}^{m-l}}{l!(({m-l)!)}^2A^{2m-l+1}}e^{- \left(\frac{1}{A} -\frac{\kappa+1}{2}\right) \left|\gamma\right|^2}} \label{CFPATS}.
\end{align}

From Eqs. \eqref{CFPATS} and \eqref{QP},
\begin{align*}
     P_\kappa\left(\alpha\right)=\frac{N_{m}^{-1}}{\pi} \sum_{l=0}^{m}{\frac{{m!}^2{(-1)}^{m-l}}{l!(({m-l)!)}^2A^{2m-l+1}}}\int{\frac{d^{2}\gamma}{\pi}(\gamma^{\ast}\gamma)^{m-l}\exp{\left[- \left(\frac{1}{A} -\frac{\kappa+1}{2}\right) \left|\gamma\right|^2+\gamma^\ast\alpha-\gamma\alpha^\ast\right]}}.
\end{align*}
Again using Eq. \eqref{MIFS}, we get the following quasi-probability function for the input state,
\begin{align}
     P_{in}\left(\kappa,\alpha\right)=
     \frac{N_{m}^{-1}}{\pi}\sum_{l=0}^{m}{\left[\frac{{m!}^2\left(-1\right)^{m-l}}{l!\: A^{2m-l+1}}\sum_{h=0}^{m-l}{\left[\frac{(-1)^{m-l-h}|\alpha|^{2(m-l-h)}}{h! (({m-l-h)!)}^2 \left(\frac{1}{A} -\frac{\kappa+1}{2}\right)^{2m-2l-h+1}}
     \exp\left( \frac{-|\alpha|^2}{\left(\frac{1}{A} -\frac{\kappa+1}{2}\right)}\right)\right] }\right] }\label{QPPATS}.
\end{align}
\subsubsection{Output State}
{Similarly, for noisy PATS at the output}, we have,

\begin{align*}
   {Tr\left[e^{\gamma\hat{a}^\dag}\phi_{s}(\hat{\rho}_{PATS}) e^{-\gamma^{\ast} \hat{a}}\right]}=N_{m}^{-1}\sum_{l=0}^{m}\frac{{m!}^2s^l}{l!\left(As+1\right)^{2m-l+1}}\sum_{b=0}^{m}{\frac{{({-\gamma}^\ast\gamma)}^{m-l-b}}{b!\ (({m-l-b)!)}^2 {\left(\frac{A}{As+1}\right)}^{2m-l+1}}\exp{\left[-
    {\frac{As+1}{A}}\left|\gamma\right|^2\right]}}.
\end{align*}
Using the above expression for the trace in Eq. \eqref{CFPA}, we get the following CF for the output state,
\begin{align}
    \chi_{out}\left(\gamma,\kappa\right)=N_{m}^{-1}\sum_{l=0}^{m}\frac{{m!}^2s^l}{l!\left(As+1\right)^{2m-l+1}}\sum_{b=0}^{m}{\frac{{({-\gamma}^\ast\gamma)}^{m-l-b}}{b! (({m-l-b)!)}^2 {\left(\frac{A}{As+1}\right)}^{2m-l+1}}\exp{\left[-\left(
    \frac{As+1}{A}-\frac{\kappa+1}{2}\right)\left|\gamma\right|^2\right]}} \label{CFNPATS}.
\end{align}

From Eqs. \eqref{CFNPATS} and \eqref{QP},
\begin{multline*}
     P_{out}\left(\gamma,\kappa\right)=N_{m}^{-1}\sum_{l=0}^{m}\frac{{m!}^2s^l}{l!\left(As+1\right)^{2m-l+1}}\sum_{b=0}^{m}{\frac{(-1)^{m-l-h}}{b! (({m-l-b)!)}^2 {\left(\frac{A}{As+1}\right)}^{2m-l+1}}}\\
     \times \int{{({-\gamma}^\ast\gamma)}^{m-l-b}\exp{\left[-\left(
    \frac{As+1}{A}-\frac{\kappa+1}{2}\right)\left|\gamma\right|^2+\gamma \alpha^\ast-\gamma^\ast\alpha\right]}\frac{d^2\gamma}{\pi^2}}.
\end{multline*}
Using Eq. \eqref{MIFS}, the quasi-probability distribution for the output state is obtained as,
\begin{align}
    P_{out}\left(\alpha,\kappa\right) &=\frac{N_{m}^{-1}}{\pi}\sum_{l=0}^{m}\frac{{m!}^2s^l}
    {l!\left(As+1\right)^{2m-l+1}}
    \sum_{b=0}^{m-l}\frac{\left(-1\right)^{m-l-b}}{b!\: \left(\frac{A}{As+1}\right)^{2m-2l-b+1}} \nonumber\\
    &\times \sum_{h=0}^{m-l-b}\left[\frac{(-1)^{m-l-b-h}|\alpha|^{2(m-l-b-h)}}{h! (({m-l-b-h)!)}^2\left( 
    {\frac{As+1}{A}}- 
    \frac{\kappa+1}{2}\right)^{2m-2l-2b-h+1}}\right]\exp\left[\frac{-|\alpha|^2}{\left( 
    {\frac{As+1}{A}} -\frac{\kappa+1}{2}\right)}\right] \label{QPNPATS}.
\end{align}
In Figure (\ref{PPA.eps}) we have plotted quasi-probability distribution functions for several values of $n_{th}$ (characterizing temperature $T$ of the photon), $m$ (number of photon added), and noise parameter $s$.
For  $\kappa=-1,\: 0$ or  $1$ we get the Husimi $Q$, Wigner $W$ and Sudarshan-Glauber $P$ functions, respectively. The contrast between the input and output, after passing through the Gaussian channel, is brought out in each sub-figure. At $m=0$ all quasi-probability distributions are seen to be Gaussian. As photons are added to the thermal state, the quasi-probability distribution dips at the center, indicative of non-Gaussian behavior. For higher value of $m$ the $Q$-function tends to zero at the center and width also increases. The $W$- function and $P$-function have negative regions for ($m>0$), indicative of quantumness. The distributions become wider as we increase average photon number and noise.
For output PATS it can be seen that peak value decreases and the negative region shifts to the positive (for the $W$ and $P$-function) region and zeros of $Q$ also tend to shift peak values. These are signatures of decoherence, heralding the onset of classicality.

\paragraph{B. Photon Subtracted Thermal State\\}
\subsubsection{Input State}
We begin, as before, with the characteristic function,
\begin{align}
     \chi\left(\gamma,\kappa\right)={Tr\left[e^{-\gamma^{\ast} \hat{a}}\hat{\rho}_{PSTS}e^{\gamma\hat{a}^\dag}\right]}e^\frac{\kappa-1}{2}{|\gamma|^2} \label{CFPS},
\end{align}
where,
\begin{align*}
   {Tr\left[e^{-\gamma^{\ast} \hat{a}}\hat{\rho}_{PSTS}e^{\gamma\hat{a}^\dag}\right]}
    =N_{m^-}^{-1}\sum_{l=0}^{m}{\frac{{m!}^2{({-\gamma}^\ast\gamma)}^{m-l}}{l!(({m-l)!)}^2 (\frac{1}{n_{th}})^{2m-l+1}}e^{- n_{th} \left|\gamma\right|^2}}.
\end{align*}
Using this in Eq. \eqref{CFPS}, the characteristic function comes out to be,

\begin{align}
   \chi_{in}\left(\gamma,\kappa\right)= N_{m^-}^{-1}\sum_{l=0}^{m}{\frac{{m!}^2{({-\gamma}^\ast\gamma)}^{m-l}}{l!(({m-l)!)}^2 (\frac{1}{n_{th}})^{2m-l+1}}e^{- \left(n_{th}-\frac{(\kappa-1)}{2}\right) \left|\gamma\right|^2}}.\label{CFPSTS}
\end{align}
From Eqs. \eqref{CFPSTS} and \eqref{QP},
\begin{align}
     P_{in}\left(\kappa,\alpha\right)=\frac{N_{m^-}^{-1}}{\pi}\sum_{l=0}^{m}{\left[\frac{{m!}^2\left(-1\right)^{m-l}}{l!\ (\frac{1}{n_{th}})^{2m-l+1}}\sum_{h=0}^{m-l}{\left[\frac{(-1)^{m-l-h}\left|\alpha\right|^{2(m-l-h)}}{h!\ (({m-l-h)!)}^2\ \left(n_{th} -\frac{\kappa-1}{2}\right)^{2m-2l-h+1}}
     \exp\bigg( \frac{-|\alpha|^2}{\left(n_{th} -\frac{\kappa-1}{2}\right)}\bigg)\right]\ }\right]} \label{QP_PSTS}.
\end{align}
\begin{figure}
    \centering
    \includegraphics[width=\linewidth]{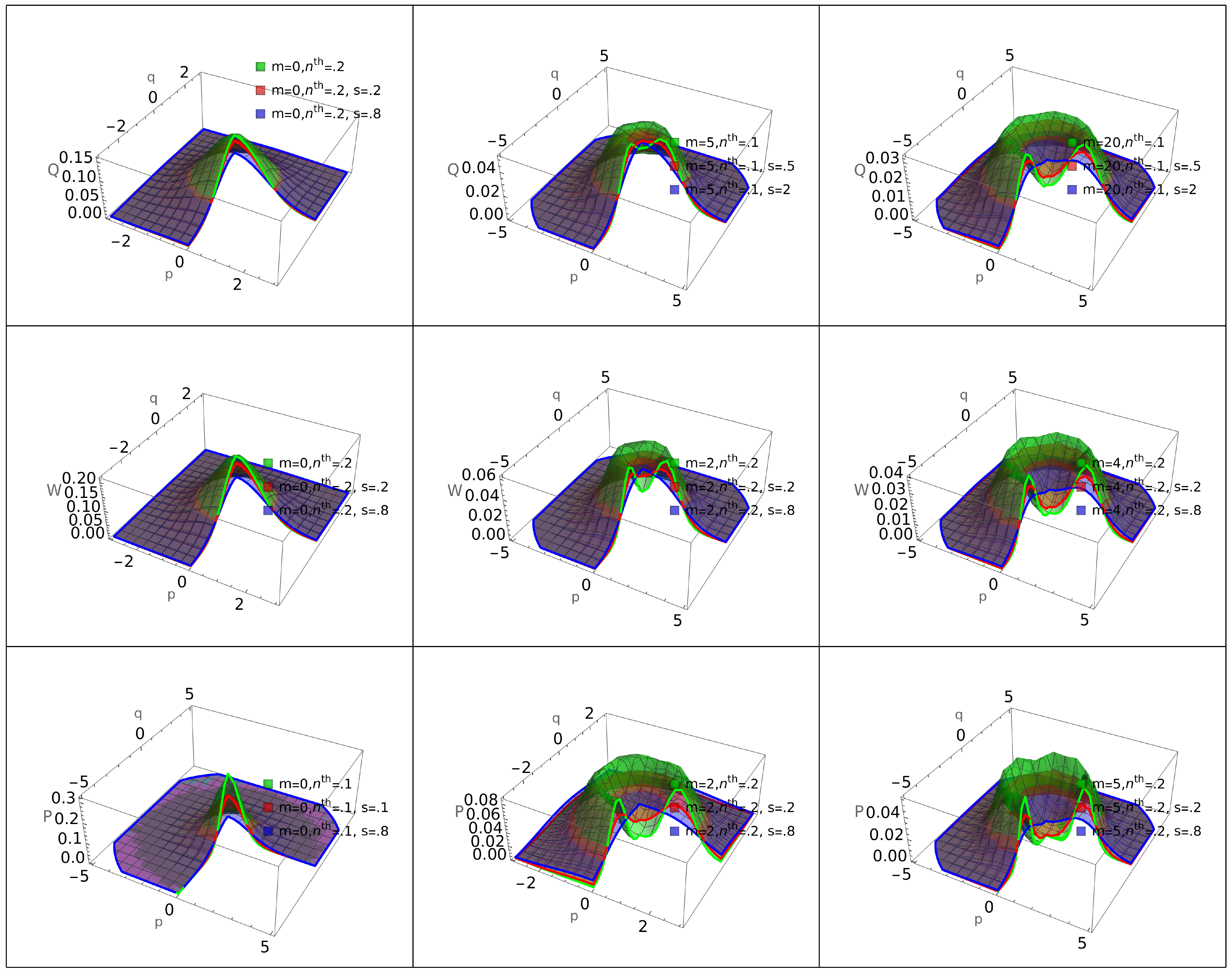}
     \caption{Quasi-Probability Distribution functions e.g Husimi $Q$-function (1st row), Wigner $W$-function (2nd row) and Sudarshan-Glauber $P$-function (3rd row) for thermal state (1st column), various photon subtracted-thermal state (2nd and 3rd columns). Input PSTS are represented by green colour and output PSTS are represented by red (noise parameter $s=0.2$) and blue (noise parameter $s=0.8$).}
    \label{PPS.eps}
\end{figure}
\subsubsection{Output State}
 For the noisy PSTS at the output, we have
 \begin{align}
    {Tr\left[e^{-\gamma^{\ast}\hat{a}} {\Phi}_s\left(\hat{\rho}_{PSTS}\right)e^{\gamma\hat{a}^\dag}\right]}
   & =N_{m^-}^{-1}
   \int\int{\frac{d^2 \alpha}{\pi}\frac{d^2 z}{\pi}{(\alpha^{\ast}\alpha)^m}\exp{\bigg[-\frac{|\alpha|^2}{n_{th}}-\frac{|z|^2}{s}+\gamma(\alpha^\ast+z^\ast)-\gamma^{\ast}(\alpha+z)\bigg]}}\nonumber\\
    &=N_{m^-}^{-1}\sum_{l=0}^{m}{\frac{{m!}^2{({-\gamma}^\ast\gamma)}^{m-l}}{l!(({m-l)!)}^2 (\frac{1}{n_{th}})^{2m-l+1}}e^{- \left(n_{th}+s\right) \left|\gamma\right|^2}}
\end{align}
Using the trace in  Eq. \eqref{CFPS}, we get the following expression for the characteristic function,
\begin{align}
    \chi_{out}\left(\gamma,\kappa\right)= N_{m^-}^{-1}\sum_{l=0}^{m}{\frac{{m!}^2{({-\gamma}^\ast\gamma)}^{m-l}}{l!(({m-l)!)}^2(\frac{1}{n_{th}})^{2m-l+1}}e^{- \left(n_{th}+s-\frac{(\kappa-1)}{2}\right) \left|\gamma\right|^2}} .\label{CFNPSTS}
\end{align}

From Eqs. \eqref{CFNPSTS} and \eqref{QP}, we can get the quasi-probability distribution function for the output state as:
\begin{align}
    P_{out}\left(\kappa,\alpha\right)=\frac{N_{m^-}^{-1}}{\pi}\sum_{n=0}^{m}{\left[\frac{{m!}^2\left(-1\right)^{m-l}}{l! (\frac{1}{n_{th}})^{2m-l+1}}\sum_{n=0}^{m-l}{\left[\frac{(-1)^{m-l-h}|\alpha|^{2(m-l-h)}}{h! (({m-l-h)!)}^2 \left(n_{th}+s -\frac{\kappa+1}{2}\right)^{2m-2l-h+1}}
     \exp\left[ \frac{-|\alpha|^2}{\left(n_{th}+s -\frac{\kappa+1}{2}\right)}\right]\right] }\right]}. 
\end{align}
In figure (\ref{PPS.eps}) we have plotted quasi-probability distribution functions for several values of $n_{th}$ (characterizing temperature $T$ of the photon), $m$ (number of photon subtracted), and noise parameter $s$.
For  $\kappa=-1$, $\kappa=0$ and  $\kappa=1$ we get Husimi $Q$, Wigner $W$ and Sudarshan-Glauber $P$-functions, respectively. At $m=0$ all quasi-probability distribution functions are Gaussian but as we subtract photons from the  thermal state we get very small dip  at the center. For higher values ($~m>40$, not shown here) of $m$ the $Q$-function can be observed to tend to zero at the center, with a corresponding increase in width. The $W$ and $P$ functions also dip at the center but do not attain negative values. The dip becomes wider as we increase the average photon number and for higher values of $m$, one can have regions which exhibit negative values.
For output PSTS it can be seen that peak values decrease and the dip (negative for higher $m$) region shifts upward, quicker than that for the PATS, exhibiting signs of decoherence. With increase in noise, the states again attain their Gaussian behavior.
Hence, increase in noise facilitates the transition from non-Gaussian to Gaussian behavior.

\paragraph{C. PAKFTS\\}
\subsubsection{Input State}
Now, from Eq. \eqref{PAKFTS}, we have,
\begin{align*}
    \text{Tr}\left\lbrack e^{\gamma{\hat{a}}^{\dag}}\rho e^{- \gamma^{*}\hat{\text{\ a}}} \right\rbrack = N_{\text{km}}^{- 1}\text{\ Tr}\left\lbrack {e^{\gamma{\hat{a}}^{\dag}}:{{\hat{a}}^{\dag m}e^{- A{\hat{a}}^{\dag}\hat{a}}\hat{a}}^{m}:e^{- \gamma^{*}\hat{\text{\ a}}}}^{\ } \right\rbrack - \ N_{\text{km}}^{- 1}\ \frac{e^{- \beta\hbar\omega k}}{k!}\text{\ Tr}\left\lbrack {e^{\gamma{\hat{a}}^{\dag}}:{{\hat{a}}^{\dag (m + k)}e}^{- {\hat{a}}^{\dag}\hat{a}}{\hat{a}}^{(m + k)}:\ e^{- \gamma^{*}\hat{\text{\ a}}}}^{\ } \right\rbrack.
\end{align*}
Using Eq. \eqref{CFPA} and the above, we get the following CF,
\begin{multline}
    \chi_{in}(\gamma,\kappa) =
    N_{\text{km}}^{- 1}\bigg[\exp\left\{  - \left(\frac{1}{A} - \frac{\kappa + 1}{2} \right)|\gamma|^{2} \right\}\sum_{l = 0}^{m}\frac{{m!}^{2}{\left( - \gamma^{*}\gamma \right)^{\ }}^{m - l}s^{l}}{l!\:((m - l)!)^{2}A^{2m - l + 1}} \\
    - \frac{e^{- \beta\hbar\omega k}}{k!}\exp\left\{- \left( 1 - \frac{\kappa + 1}{2} \right)|\gamma|^{2} \right\}\sum_{l = 0}^{m+k}\frac{{(m+k)!}^{2}{\left( - \gamma^{*}\gamma \right)^{\ }}^{m+k - l}s^{l}}{l!\:((m+k - l)!)^{2}} \bigg]. \label{CFPAKFTS}
\end{multline}

From Eqs. \eqref{CFPAKFTS}, \eqref{QP} and \eqref{MIF}, the corresponding probability distribution is obtained as,
\begin{align}
     P_{in}\left(\kappa,\alpha \right)&=\frac{N_{\text{km}}^{- 1}}{\mathbf{\pi}}\bigg[ \sum_{l = 0}^{m}{\frac{{m!}^{2}\ ( - 1)^{m - l}}{l!\:A^{2m - l + 1}}\sum_{h = 0}^{m - l}\frac{(-1)^{m - l - h} |\alpha|^{2(m - l - h)} \exp\left(- \frac{\left| \mathbf{\alpha} \right|^{2}}{\left( \frac{1}{A} - \frac{\kappa + 1}{2}\right) }\right)} {h!\:((m- l-h)!)^{2}\left( \frac{1}{A} - \frac{\kappa + 1}{2} \right)^{2m - 2l - h + 1}}}\nonumber\\
    &- \frac{e^{- \beta\hbar\omega k}}{k!}\sum_{l = 0}^{m + k}{\frac{{(m + k)!}^{2}( - 1)^{m + k - l}}{l!(s + 1)^{2(m + k) - l + 1}}\sum_{h = 0}^{m +k- l}\frac{(-1)^{m+k - l - h} |\alpha|^{2(m +k- l - h)} \exp\left( - \frac{\left| \mathbf{\alpha} \right|^{2}}{\left( 1 - \frac{\kappa + 1}{2} \right)} \right)}{h!\:((m+k - l-h)!)^{2}\left( 1 - \frac{\kappa + 1}{2} \right)^{2m + 2k - 2l - h + 1}}} \bigg].\label{QPPAKFTS}
\end{align} 
\begin{figure}
    \centering
    \includegraphics[width=\linewidth]{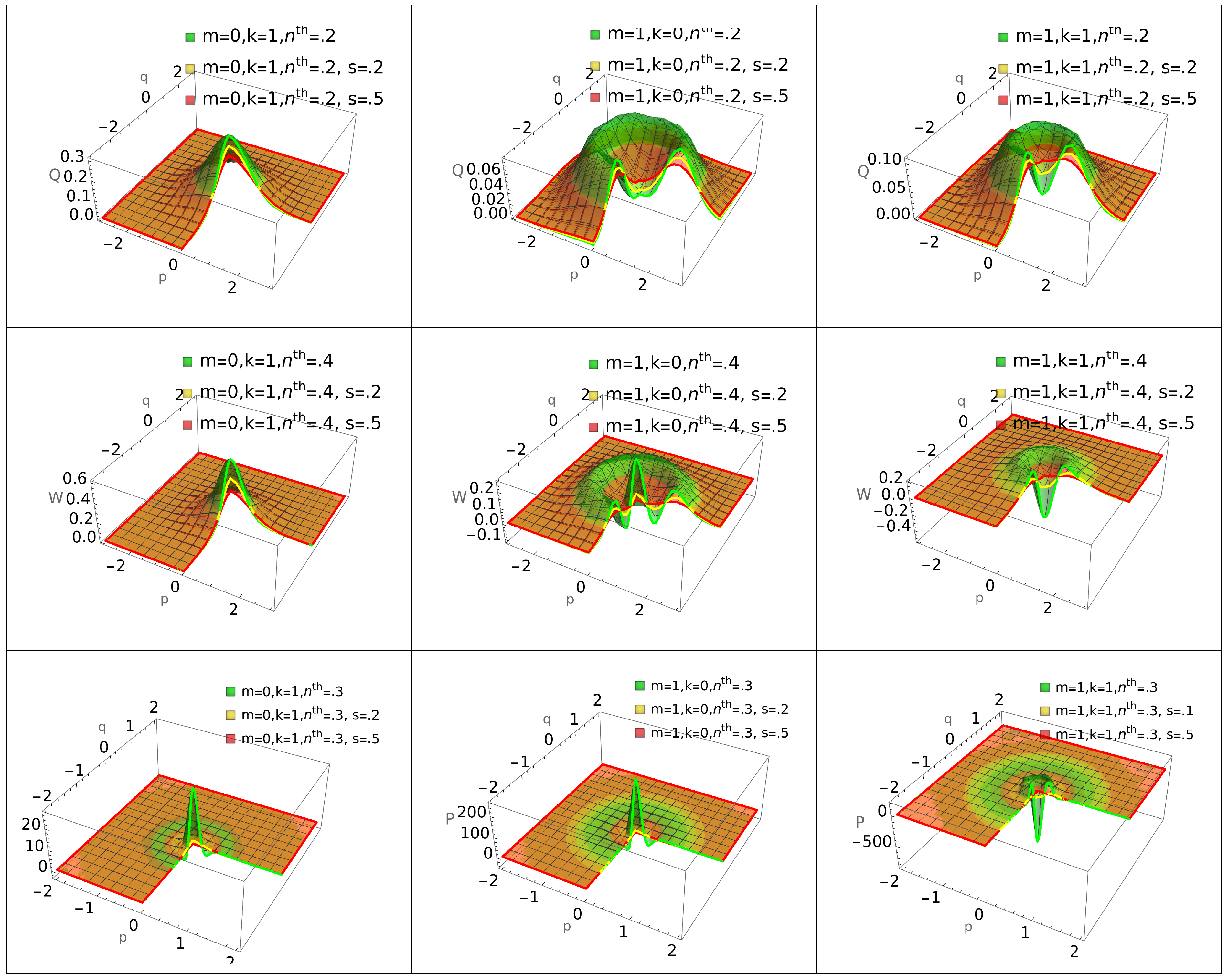}
     \caption{Quasi-Probability Distribution functions, Husimi $Q$-function (1st row), Wigner $W$-function (2nd row) and Sudarshan-Glauber $P$-function (3rd row). The  input PAKFTS are represented by green colour, while the output PAKFTS are depicted by yellow (noise parameter $s=0.2$) and red (noise parameter $s=0.5$). Here  1st column ($m=0$, $k=1$), 2nd column ($m=1$, $k=0$) and 3rd column ($m=1$, $k=l$) are PAKFTS (corresponding to input and output states).}
    \label{PPAF.eps}
\end{figure} 
\subsubsection{Output State}
Similarly, for noisy PAKFTS at the output, we have,
\begin{align*}
    \text{Tr}\left\lbrack e^{\gamma{\hat{a}}^{\dag}}{\phi_{s}(\rho_{PAKFTS})e}^{- \gamma^{*}\hat{\text{\ a}}} \right\rbrack &= N_{\text{km}}^{- 1}\bigg[\sum_{l = 0}^{m}\frac{{m!}^{2}\:\text{Tr}\left\lbrack {e^{\gamma{\hat{a}}^{\dag}}:{{\hat{a}}^{\dag m - l}\exp\left( - \frac{A{\hat{a}}^{\dag}\hat{a}}{\text{As} + 1} \right)\hat{a}^{m - l}}:e^{- \gamma^{*}\hat{\text{\ a}}}}^{\ } \right\rbrack\ s^{l}}{l!\:((m- l)!)^{2}(As + 1)^{2m - l + 1}} \\
    &- \frac{e^{- \beta k}}{k!}\sum_{l = 0}^{m + k}\frac{{(m + k)!}^{2}\:\text{Tr}\left\lbrack {e^{\gamma{\hat{a}}^{\dag}}:{{\hat{a}}^{\dag m +k- l}\ {\exp\left( - \frac{{\hat{a}}^{\dag}\hat{a}}{s + 1} \right)}\hat{a}^{m+k - l}}:e^{- \gamma^{\ast}\hat{\text{\ a}}}}^{\ } \right\rbrack s^{l}}{l!\:((m+k - l)!)^{2}(s + 1)^{2(m + k) - l + 1}} \bigg] .
\end{align*}

Using the above equation with Eq. \eqref{CFPA}, we get the following CF,
\begin{align}
    \chi_{out}(\gamma,\kappa) &= N_{\text{km}}^{- 1}\bigg[ \sum_{l = 0}^{m}{\frac{{m!}^{2}\ s^{l}}{l!\:(As + 1)^{2m - l + 1}}\sum_{h = 0}^{m - l}\frac{{\left( - \gamma^{*}\gamma \right)^{\ }}^{m - l - h}\exp\left\{ - \left( \frac{As + 1}{A} - \frac{\kappa + 1}{2} \right)|\gamma|^{2} \right\}}{h!\:((m - l-h)!)^{2}\left( \frac{As + 1}{A} \right)^{2m - 2l - h + 1}}}\nonumber\\
    &- \frac{e^{- \beta k}}{k!}\sum_{l = 0}^{m + k}{\frac{{(m + k)!}^{2}s^{l}}{l!\:(s + 1)^{2(m + k) - l + 1}}\sum_{h = 0}^{m+k - l}\frac{{\left( - \gamma^{*}\gamma \right)^{\ }}^{m + k - l - h}\exp\left\{ - \left( (s + 1) - \frac{\kappa + 1}{2} \right)|\gamma|^{2} \right\}}{h!\:((m+k - l-h)!)^{2}\:(s + 1)^{2m + 2k - 2l - h + 1}}} \bigg]. \label{CFNPAKFTS}
\end{align}
From Eqs. \eqref{CFNPAKFTS}, \eqref{QP} and \eqref{MIF},
\begin{multline*}
    {P}_{out}=\frac{N_{\text{km}}^{- 1}}{\mathbf{\pi}}\bigg[\sum_{l = 0}^{m}{\frac{{m!}^{2}\ s^{l}}{l!(As + 1)^{2m - l + 1}}\sum_{h = 0}^{m - l}\frac{I}{h!\:((m- l-h)!)^{2}\left( \frac{As + 1}{A} \right)^{2m - 2l - h + 1}}}\\
    - \frac{e^{- \beta k}}{k!}\sum_{l = 0}^{m + k}{\frac{{(m + k)!}^{2}s^{l}}{l!\:(s + 1)^{2(m + k) - l + 1}}\sum_{h = 0}^{m+k - l}{\text{\  }\frac{I^{`}}{h!\:((m + k - l - h)!)^{2}\:(s + 1)^{2m + 2k - 2l - h + 1}}}} \bigg], \label{QPNPAKFTS}
\end{multline*}
   
 where,
 \begin{equation}
   I = \sum_{b = 0}^{m - l - h}{\frac{{(m - l - h)!}^{2}{(-1)^{m-l-h-b}|\alpha|^{2(m-l-h-b)}} }{b!{(m - l - h - b)!}^{2}\left( \frac{As + 1}{A} - \frac{\kappa + 1}{2} \right)^{2m - 2l - h + 1}}\exp\left[ - \frac{\left| \mathbf{\alpha} \right|^{2}}{\left( \frac{As + 1}{A} - \frac{\kappa + 1}{2} \right)} \right]},  
 \end{equation}

and
\begin{equation}
    I^{`} = \sum_{b = 0}^{m+k - l - h}{\frac{{(m + k - l - h)!}^{2}{{(-1)^{m+k-l-h-b}|\alpha|^{2(m+k-l-h-b)}}} }{b!{(m + k - l - h - b)!}^{2}\left( s + 1 - \frac{\kappa + 1}{2} \right)^{2m + 2k - 2l - h + 1}}\exp\left[ - \frac{\left| \mathbf{\alpha} \right|^{2}}{\left( s + 1 - \frac{\kappa + 1}{2} \right)} \right]}.
\end{equation}
In the figure (\ref{PPAF.eps}) quasi-probability distribution functions are depicted for several values of $n_{th}$, $m$, and noise parameter $s$.
For  $\kappa=-1$, $\kappa=0$ and  $\kappa=1$ we get Husimi $Q$, Wigner $W$ and Sudarshan-Glauber $P$-functions, respectively. At $m=0$, $k=1$ all quasi-probability distributions are Gaussian but as we add photons to the KFTS  we get  a dip at the center for the $Q$-function and negative regions for $W$ and $P$ functions. It was observed  that for small values of $k$, the nonclassicality achieved, indicated by the negative values of the $W$ and $P$ functions, was higher as compared to the larger $k$ scenario. This suggests that hole burning around the $n = 0$ state impacts the nature of the thermal state, Gaussian in nature, more than for larger $n$, where the tail regions of the Gaussian are targetted.
For output states, as we increase noise parameter $s$, there is a decrease in the  centered dip and the states tend to their original Gaussian form. Further increase in the noise $s$ would eventually revert the states to their Gaussian form.

\subsection{Squeezed Thermal States }

\paragraph{A. PASTS\\}
\subsubsection{Input State}
From Eq. \eqref{PASTS}, we have,
\begin{align*}
{Tr\left[e^{\gamma \hat{a}^{\dag}}\hat{\rho}_{PASTS} e^{-\gamma^{\ast} \hat{a}}\right]}
&=\frac{N_{a,m}^{-1}}{\sqrt A}\int{\langle\alpha|}:\exp{\left(\gamma \hat{a}^\dag\right)}\hat{a}^{\dag m}\exp{\left[\frac{C}{2}(\hat{a}^{\dag 2}+{\hat{a}}^2)+(B-1){\hat{a}}^\dag\hat{a}\right]} \hat{a}^{m}\exp\left(-\gamma^{\ast} \hat{a}\right):{|\alpha\rangle}\frac{d^2\alpha}{\pi}\\
&=\frac{N_{a,m}^{-1}}{\sqrt A}\partial_X^{m}\int{\exp{\left[\frac{C}{2}(\alpha^{\ast 2}+\alpha^{2})+X \alpha^{\ast}\alpha+\gamma\alpha^{\ast}-\gamma^{\ast}\alpha\right]}\frac{d^2\alpha}{\pi}}.
\end{align*}
Using the Eq. \eqref{MIF},
\begin{align*}
{Tr\left[e^{\gamma \hat{a}^{\dag}}\hat{\rho}_{PASTS} e^{-\gamma^{\ast} \hat{a}}\right]}
=\frac{N_{a,m}^{-1}}{\sqrt A}\partial_X^{m}{(X^2- 
C^2)}^{-1/2}\exp\left[\frac{\frac{C}{2}({{\gamma}^{\ast}}^2+\gamma^{2}) - X
{\left|\gamma\right|^{2}}}{(X^2-C^2)}\right],
\end{align*}
where, $X=(1-B)$.
Using this in  Eq. \eqref{CFPA}, the following expression for the characteristic function is obtained,
\begin{align}
   \chi_{in}\left(\gamma,\kappa\right) =\frac{N_{a,m}^{-1}}{\sqrt A}\partial_X^{m}{(X^2- C^2)}^{-1/2} \exp\left[\frac{\frac{C}{2}({{\gamma}^{\ast}}^2+\gamma^{2}) - X {\left|\gamma\right|^{2}}}{(X^2-C^2)}+\frac{\kappa + 1}{2} {|\gamma|}^2\right] \label{CFPASTS}.
\end{align}
From Eqs. \eqref{CFPASTS} and \eqref{QP}, the quasi-probability distribution is seen to be
\begin{align}
    P_{in}\left(\alpha,\kappa\right)=\frac{N_{a,m}^{-1}}{\sqrt A}\partial_X^{m}\left[{(X^2- C^2)}^{-1/2}{(X_{*}^2- C_{*}^2)}^{-1/2}\exp\left(\frac{(C_{*}/2)({{\alpha}^{\ast}}^2+\alpha^{2})-X_{*}{|\alpha|}^2}{(X_{*}^2- C_{*}^2)}\right)\right], \label{}
\end{align}
where, $C_{*}=\frac{C}{(X^2-C^2)}$ and $X_{*}=\frac{X}{(X^2-C^2)}-\frac{\kappa+1}{2}$.
\begin{figure}
    \centering
    \includegraphics[width=\linewidth]{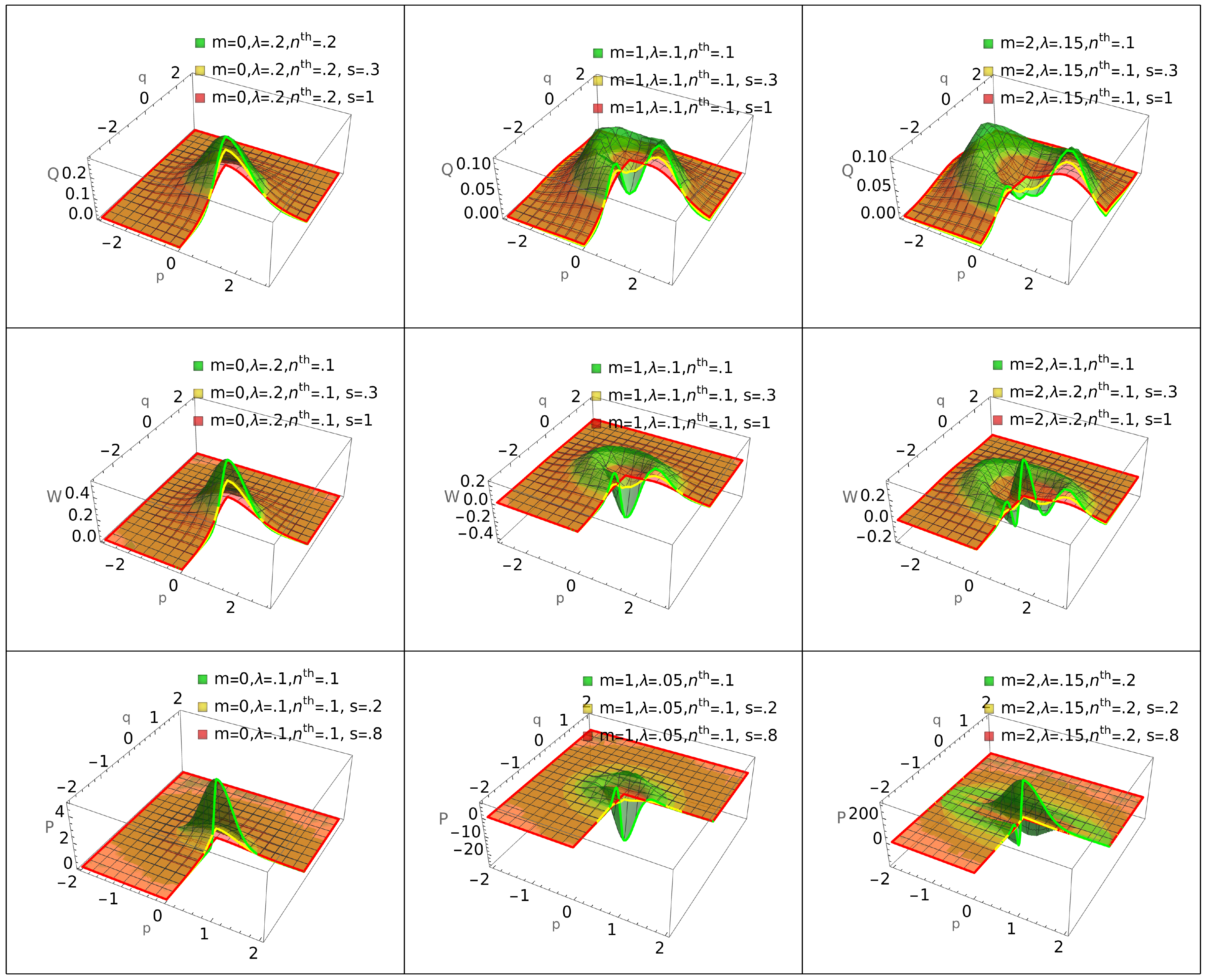}
     \caption{Quasi-Probability Distribution functions, Husimi $Q$-function (1st row), Wigner $W$-function (2nd row) and Sudarshan-Glauber $P$-function (3rd row) for  input PASTS are represented by green and output PASTS are represented by yellow and red colors, respectively. Here  the 1st column represents STS, the 2nd column one photon added-STS and the 3rd column two photon added STS, depicting both input and output PASTS  for various values of the noise parameter $s$, $n_{th}$ and the squeezing parameter $\lambda$.}
    \label{PPAS.eps}
\end{figure}
\subsubsection{Output State}
Similarly, for noisy PASTS at the output,
we get the following expression for the characteristic function,
\begin{align}
    \chi_{out}\left(\gamma,\kappa\right)=\frac{N_{a,m}^{-1}}{s \sqrt A}\partial_Y^{m}\left(Y^2-C^2\right)^{-1/2}\left(\left(Y_0+1/s\right)^2-{C_0}^2\right)^{-1/2}\exp\left[\frac{C_{1}}{2}\left({\gamma^\ast}^2+\gamma
    ^2\right)-Y_1\gamma^\ast\gamma\right]  \label{CFNPASTS1} .
\end{align}
The calculations are detailed in Appendix [B].

From Eqs. \eqref{CFNPASTS1} and \eqref{QP},
\begin{multline*}
    P_{out}\left(\alpha,\kappa\right)
    =\frac{1}{\pi}\frac{N_{a,m}^{-1}}{s \sqrt A}\partial_Y^{m}\left(Y^2-C^2\right)^{-1/2}\left(\left(Y_0+1/s\right)^2-{C_0}^2\right)^{-1/2}\\
    \times\int{\exp\left[\frac{C_{1}}{2}\left({\gamma^\ast}^2+\gamma
    ^2\right)-Y_1\gamma^\ast\gamma+\gamma^{\ast}\alpha-\gamma\alpha^{\ast}\right]}\frac{d^2\alpha}{\pi}.\\
\end{multline*}
Using Eq. \eqref{MIF}, we get following expression for quasi-probability distribution for output PASTS,
\begin{multline}
P_{out}\left(\alpha,\kappa\right)=\frac{1}{\pi}\frac{N_{a,m}^{-1}}{s \sqrt A}\partial_Y^{m}\left(Y^2-C^2\right)^{-1/2}\left(\left(Y_0+1/s\right)^2-{C_0}^2\right)^{-1/2}\left(Y_1^2-C_1^2\right)^{-1/2}\\
   \times \exp\left[\frac{\frac{C_1}{2}\left({\alpha^\ast}^2+\alpha^2\right)-Y_1\alpha^\ast\alpha}{\left({Y_1}^2-{C_1}^2\right)}\right] .
\end{multline}

Figure (\ref{PPAS.eps}) depicts quasi-probability distribution functions as a function of $n_{th}$, number of photons added $m$, squeezing parameter $\lambda$, and noise parameterized by $s$. Even though the PASTS are seen to have qualitatively similar phase space distributions as PATS, they exhibit comparatively more negative regions for the $W$ and $P$ functions for $m>0$ and for smaller values of \{$n_{th},\lambda$\}. This brings out the role of squeezing in highlighting the quantum features of the states. Furthermore, the quantum signatures in the PASTS quasi-probability distributions  are seen up to a (small) threshold value of the squeezing parameter $\lambda$. For the output states, the cutoff value of squeezing parameter (for exhibiting quantumness) decreases as we increase the noise parameter $s$.

\paragraph{B. PSSTS\\}
\subsubsection{Input State}
\begin{figure}
    \centering
    \includegraphics[width=\linewidth]{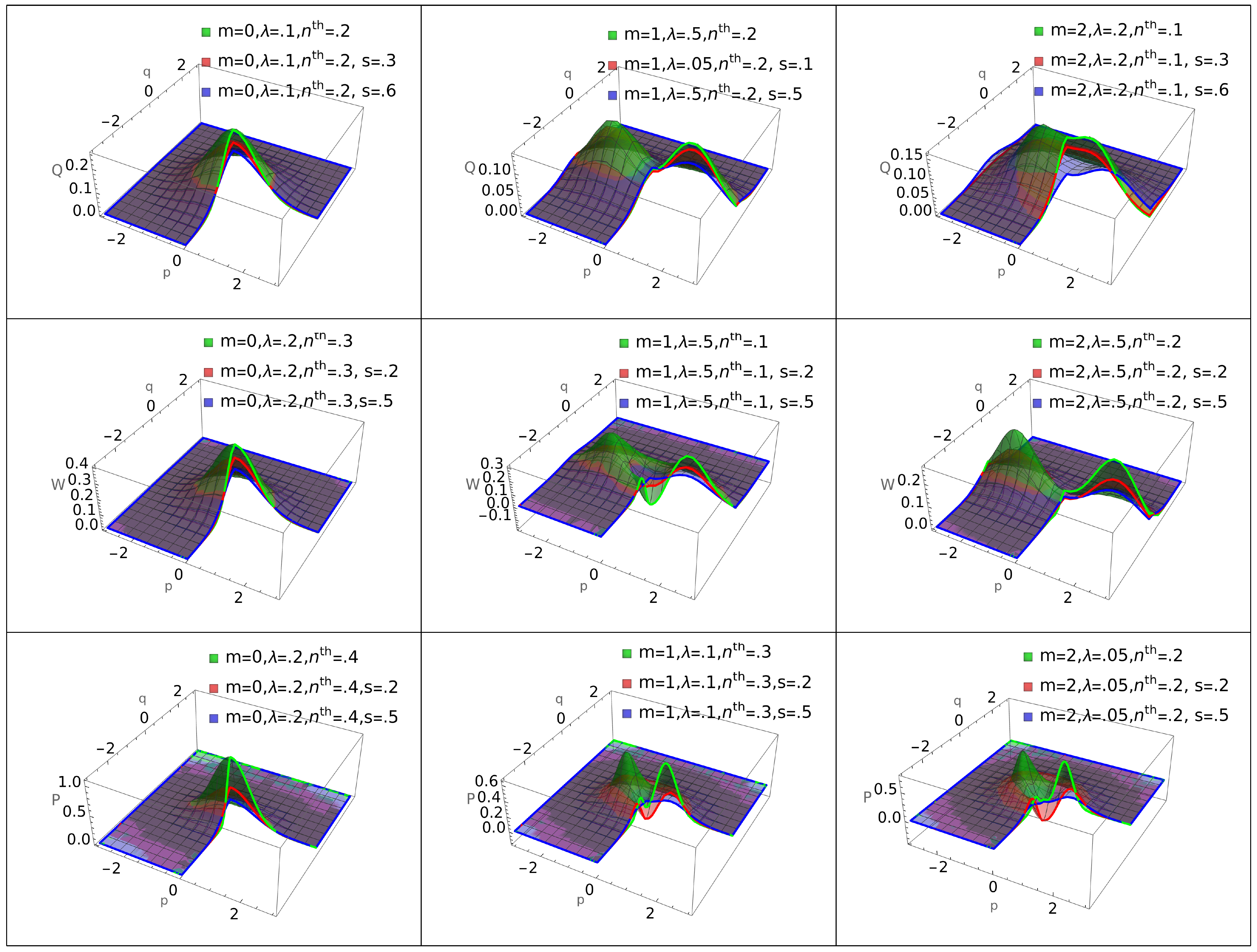}
         \caption{Quasi-Probability Distribution functions, Husimi $Q$-function (1st row), Wigner $W$-function (2nd row) and Sudarshan-Glauber $P$-function (3rd row) for  input PSSTS are represented by green color while output PSSTS are in red and blue. The  1st column represents the STS, while the 2nd and 3rd columns depict, one photon subtracted-STS and two photon subtracted-STS, respectively, for both  input and output states, for various values of parameters, such as, average number of thermal photons, noise parameter $s$ and squeezing parameter $\lambda$.}
    \label{PPSS.eps}
\end{figure}
From Eqs. \eqref{PSSTS} and \eqref{CF},
\begin{align}
    \chi_{in}\left(\gamma,\kappa\right)=\frac{N_{a,m^{-}}^{-1}}{\sqrt A}\partial_{u}^{m}\bigg[\left((1-Bu)^2-C^2u^2\right)^{-1/2}\exp{\left(A_1\gamma^{\ast 2}+A_2\gamma^{ 2}-A_3|\gamma|^2\right)}\bigg] \bigg|_{u=1} \label{CFPSSTS1},
\end{align}
which is calculated in appendix [B].\\
From Eqs. \eqref{CFPSSTS1} and \eqref{QP}, we get following expression for quasi-probability distribution for input PSSTS,
\begin{align}
    P_{in}\left(\alpha,\kappa\right)&=
    \frac{N_{a,m^{-}}^{-1}}{\pi\sqrt A}\partial_{u}^{m}\bigg[\left((1-Bu)^2-C^2u^2\right)^{-1/2}\int\frac{d^2\gamma}{\pi} \exp{\left(A_1\gamma^{\ast 2}+A_2\gamma^{ 2}-A_3|\gamma|^2+\gamma^\ast\alpha-\gamma\alpha^\ast\right)}\bigg]\bigg|_{u=1} \nonumber\\
    &=\frac{N_{a,m^{-}}^{-1}}{\pi\sqrt A}\partial_{u}^{m}\bigg[\left((1-Bu)^2-C^2u^2\right)^{-1/2}(A_3^2-4A_1A_2)^{-1/2}\exp{\left(\frac{A_1\alpha^2+A_2\alpha^{\ast 2}-A_3|\alpha|^2}{(A_3^2-4A_1A_2)}\right)}\bigg]\bigg|_{u=1}.
\end{align}

\subsubsection{Output State}
Similarly, For noisy PSSTS at the output, we have following characteristic function,
\begin{align}
    \chi_{out}\left(\gamma,\kappa\right)=\frac{N_{a,m^{-}}^{-1}}{\sqrt A}\partial_{u}^{m}\bigg[\left((1-Bu)^2-C^2u^2\right)^{-1/2}\exp{\left(-\left(N_1-\frac{(\kappa-1)}{2}\right)|\gamma|^2+N_2\gamma^2+N_3\gamma^{\ast 2}\right)}\bigg]\bigg|_{u=1}, \label{CFNPSSTS1}
\end{align}
which is calculated in Appendix [B].\\
From Eqs. \eqref{CFNPSSTS1} and \eqref{QP}, we get following expression for quasi probability distribution,
\begin{multline}
    P_{out}\left(\alpha,\kappa\right)=\frac{N_{a,m^{-}}^{-1}}{\sqrt A}\partial_{u}^{m}\left((1-Bu)^2-C^2u^2\right)^{-1/2}\left(\left(N_1-\frac{(\kappa-1)}{2}\right)^2-4N_2 N_3\right)^{-1/2}\\
    \times \exp{\left(\frac{-\left(N_1-\frac{(\kappa-1)}{2}\right)|\alpha|^2+N_2\alpha^{\ast 2}+N_3\alpha^{2}}{\left(\left(N_1-\frac{(\kappa-1)}{2}\right)^2-4N_2 N_3\right)}\right)}\bigg|_{u=1}.
\end{multline}

In figure (\ref{PPSS.eps}), we have plotted the quasi-probability distribution functions for average number of thermal photon $n_{th}$, subtracted photons $m$, squeezing parameter $\lambda$, and noise parameter $s$  for PSSTS (input and output states). It can be seen that the PSSTS $Q$-function is similar to its non-squeezed counterpart, i.e., PSTS. However, in contrast to the PSTS scenario, figure (\ref{PPS.eps}), the $W$ and $P$ functions exhibit negative regions for small values of $m$. This once again brings out the positive role of squeezing in highlighting the quantumness. Furthermore, we have seen that $W$ and $P$ functions exhibit more negative regions when odd number of photons are subtracted from the squeezed thermal state.

\section{Photon Statistics}
The intrinsic statistical properties of photons in a light source can be ascertained by the experimental and theoretical study of photon statistics. Broadly, three types of statistics can be obtained: Poissonian, super-Poissonian, and sub-Poissonian. These are arrived at by an analysis of the variance and average number of photon counts of the distribution. A semi-classical theory of light can be used to describe both Poissonian and super-Poissonian light, in which an electromagnetic wave models the light source and atoms are modeled according to quantum mechanics. In contrast, sub-Poissonian light requires the quantization of the electromagnetic field for a proper description \cite{loudon2000quantum}. Here, we will highlight three facets of photon statistics, {\it viz.,} photon number distribution (PND), second-order correlation and Mandel's $Q_{M}$ parameter.

\subsection{Photon Number Distribution}
The photon-number distribution (PND) is a key characteristic of every optical field \cite{mandel1979sub}. The PND, i.e., the probability of finding $n$ photons in a quantum state described by the density operator $\rho$ is 
\begin{equation}
P\left(n\right)=Tr[\rho|{n}\rangle\langle{n}|]=\langle{n}|\rho|{n}\rangle \label{PND}.
\end{equation}
The photon number distribution function sums up to one for all input and output states.
\subsubsection{PND for Thermal States}
\paragraph{A. PATS\\}
From Eqs. \eqref{nth} and \eqref{PND},
\begin{align}
    \langle{n}|\rho|n\rangle &=N_{m}^{-1}\langle{n}|:\hat{a}^{\dag m} e^{-A\hat{a}^\dag\hat{a}}\hat{a}^{m}:|n\rangle=N_{m}^{-1}\frac{n!}{(n-m!)}\langle{n-m}|: e^{-A\hat{a}^\dag\hat{a}}:|n-m\rangle \nonumber \\
    &=N_{m}^{-1}\frac{n!}{(n-m!)}\sum_{k=0}\langle{n-m}|: (-A)^{k}\frac{\hat{a}^{\dag k}\hat{a}^{k}}{k!}:|n-m\rangle \nonumber \\
    &= N_{m}^{-1}\frac{n!}{(n-m)!}\sum_{k=0}(-A)^{k}\frac{(n-m)!}{k!(n-m-k)!}.
\end{align}
From the binomial series,
\begin{equation}
    (1-x)^{n}=\sum_{k=0}{\frac{n!}{k!(n-k)!} (-x)^{k}} \label{BNS}.
\end{equation}
Now, from Eqs. \eqref{PND} and \eqref{BNS}, we will get following expression  of the the photon number distribution for photon added thermal state $P_{in}\left(n\right)$, used as input to the Gaussian channel, is seen to be,
\begin{equation}
    P_{in}\left(n\right)=N^{-1} \frac{n!}{\left(n-m\right)!} \left(1-A\right)^{n-m} \label{PNDPATS}.
\end{equation}

\begin{figure}
    \centering
    \includegraphics[width=\textwidth]{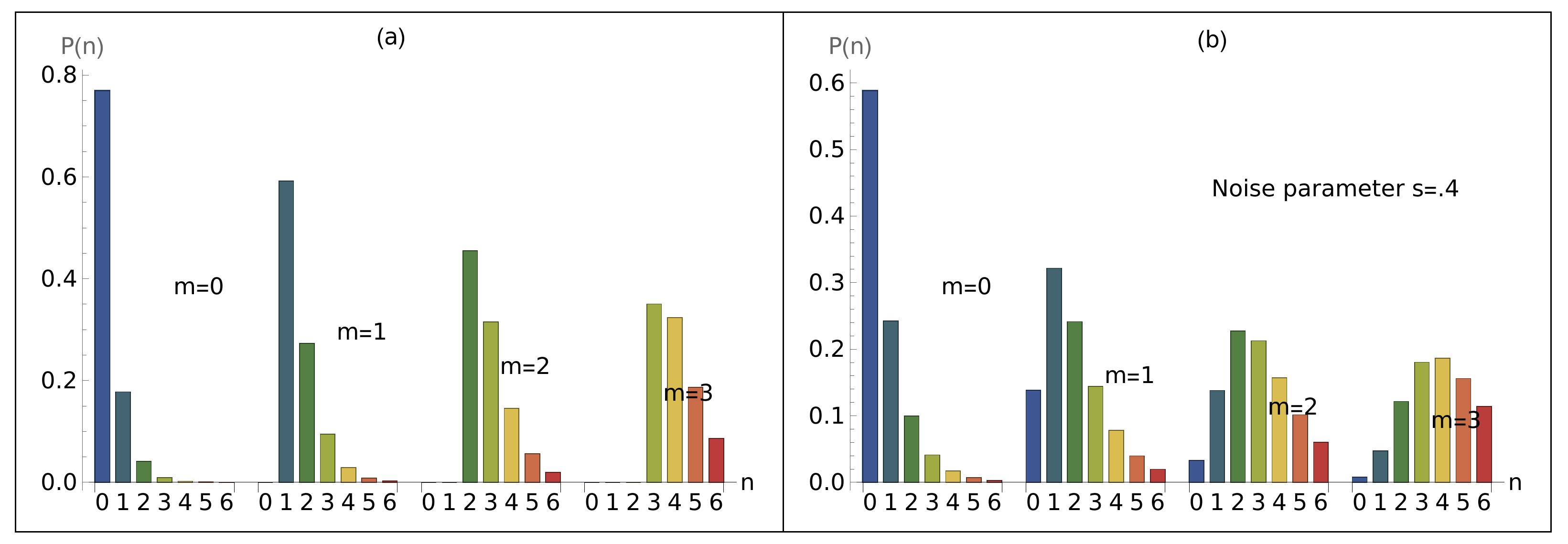}
    \caption{Photon number distributions for (a) PATS at the input with average thermal photon $n_{th}=0.2$,  and (b) noisy PATS at the output  with average thermal photon $n_{th}=0.2$  and noise parameter $s=0.4$.}
    \label{PNDPA.eps}
\end{figure}
Similarly,
we can get following $P_{out}\left(n\right)$ of photon number distribution for noisy-PATS at the output of the Gaussian channel as
\begin{equation}
    P_{out}(n)=N_{m}^{-1}\sum_{l=0}^m\frac{ m!^2 s^l}{l!\:(m-l)!^2 \left(As+1\right)^{2 m-l+1}}\frac{n!}{(n-m+l)!}\left(1-\frac{A}{A s+1}\right)^{n-m+l}. \label{PNDNPATS}
\end{equation}

\paragraph{B. PSTS\\}
From Eqs. \eqref{cth} and \eqref{PND},
\begin{align}
    P_{in}\left(n\right) & =\langle{n}|\rho_{PSTS}|{n}\rangle
    =N_{m^-}^{-1}\int{\frac{d^2 \alpha}{\pi}{(\alpha^{\ast}\alpha)^m} e^{{-\frac{|\alpha|^{2}}{n_{th}}}}\langle{n}|\alpha\rangle\langle\alpha|{n}\rangle}\nonumber \\
   & =N_{m^-}^{-1}\int{\frac{d^2 \alpha}{\pi}\frac{(\alpha^{\ast}\alpha)^{m+n}}{n!}\exp{\bigg[-\frac{n_{th}+1}{n_{th}}|\alpha|^2\bigg]}} \nonumber \\
   &=N_{m^-}^{-1}\frac{(m+n)!}{n!}\left(\frac{n_{th}}{n_{th}+1}\right)^{m+n+1}. \label{PNDPSTS}
\end{align}
\begin{figure}
    \centering
    \includegraphics[width=\linewidth]{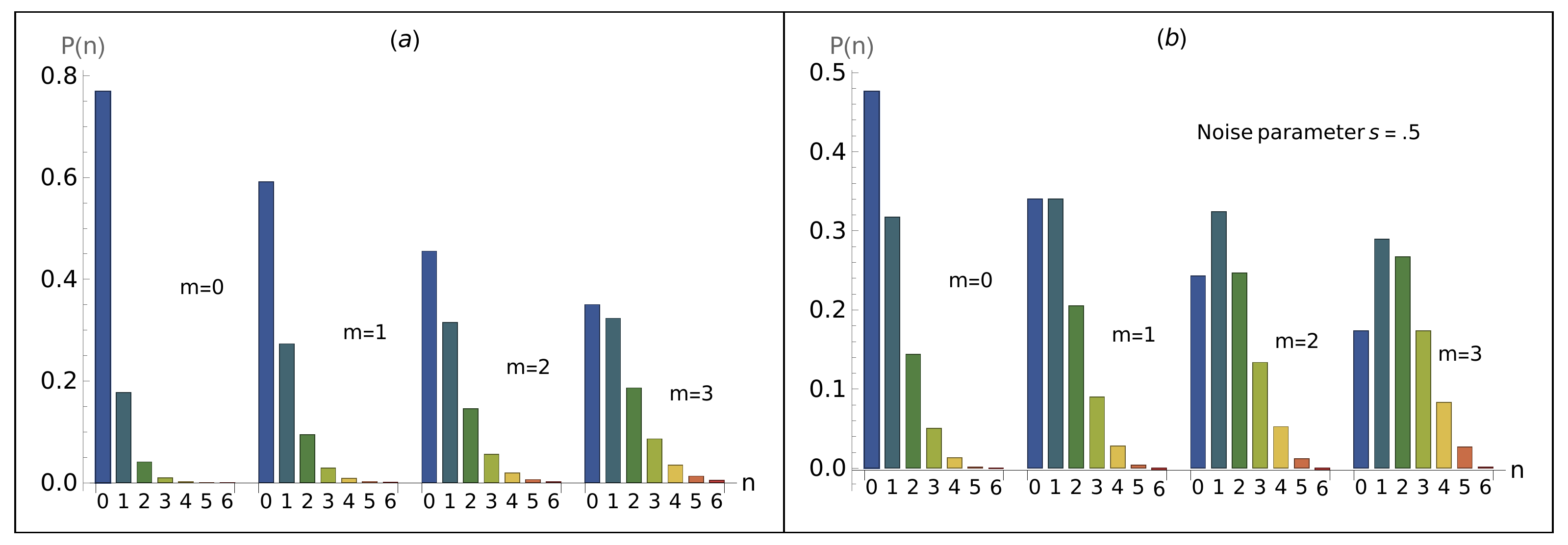}
    \caption{Photon number distributions for (a) PSTS at the input with average thermal photon $n_{th}=0.2$,  and (b) noisy PSTS at the output  with average thermal photon $n_{th}=0.2$  and noise parameter $s=0.5$.}
    \label{PNDPS.eps}
\end{figure}
Similarly, PND for noisy PSTS at the outout $P_{out}(n)$ can be shown to be,
\begin{equation}
       P_{out}\left(n\right)=N_{m^-}^{-1}\partial_{u}^{n}\left[\frac{1}{ n!\: s\left(\frac{1}{s}-u\right)}\frac{m!}{\left(
       \frac{1}{n_{th}}-u-\frac{u^2}{\frac{1}{s}-u}\right)^{m+1}}\right]\bigg|_{u=-1} \label{PNDNPSTS}.
\end{equation}

\paragraph{C. PAKFTS\\}
From Eqs. \eqref{cth}, \eqref{PND} and \eqref{PNDPATS}, we can get the following expression of photon number distribution for PAKFTS,
\begin{equation}
    P_{in}(n) = N_{\text{km}}^{- 1}\bigg[ \frac{n!}{(n - m)!}(1 - A)^{n - m} - \frac{e^{- \beta\hbar\omega k}}{k!}\frac{n!}{(n
    - m - k)!}\left| \left\langle n - m - k \middle| 0 \right\rangle \right|^{2} \bigg] \label{PNDPAKFTS}.
\end{equation}
\begin{figure}
    \centering
    \includegraphics[width=\linewidth]{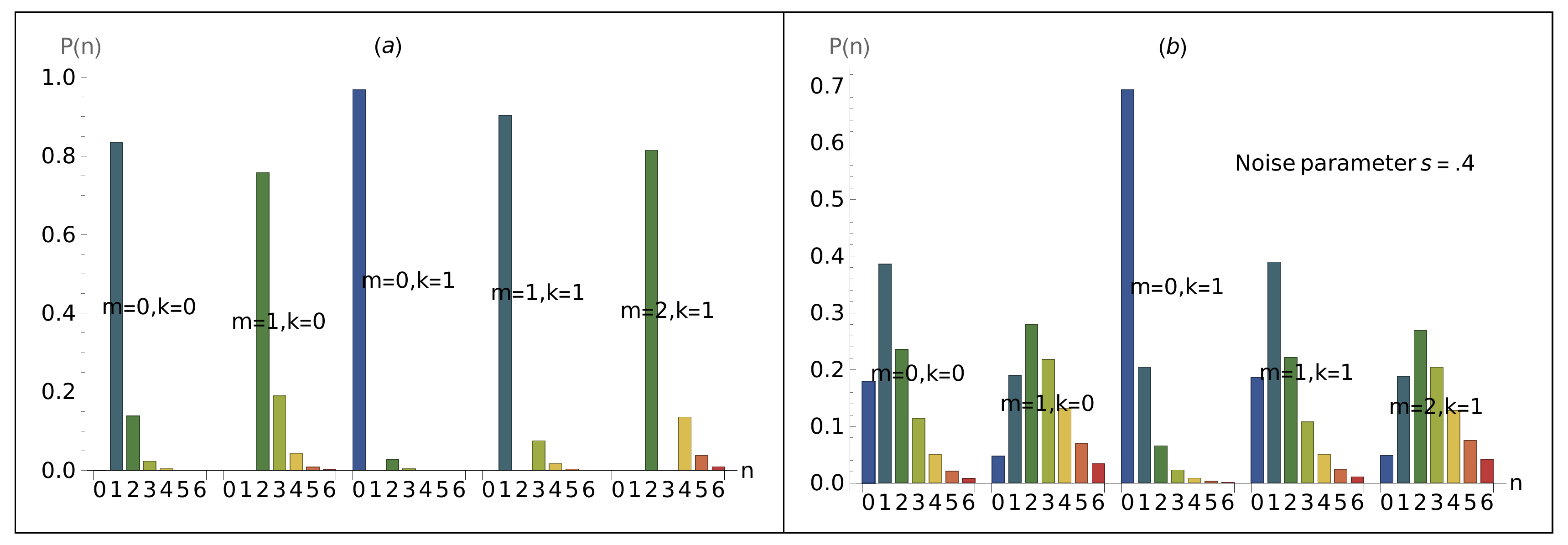}
    \caption{Photon number distributions for (a) PAKFTS at the input with average thermal photon $n_{th}=0.2$,  and (b) noisy PAKFTS at the output with average thermal photon $n_{th}=0.2$ and noise parameter $s=0.4$.}
    \label{PNDPAF.eps}
\end{figure}
Likewise, the photon number distribution for the output state of PAKFTS, comes out to be
\begin{align}
    P_{out}(n) & =
    N_{\text{km}}^{- 1}\bigg[ \sum_{l = 0}^{m}{\frac{{m!}^{2}{n!s}^{l}}{l!\:((m - l)!)^{2}(n - m - l)!\left( \text{As} + 1 \right)^{2m - l + 1}}\left( \frac{\text{As} + A - 1}{\text{As} + 1} \right)^{n - m - l}}  \nonumber \\
    & -\frac{e^{- \beta\hbar\omega k}}{k!}\sum_{l = 0}^{m + k}{\frac{{(m + k)!}^{2}n!s^{l}}{l!\:((m + k - l)!)^{2}\:(n - m - k +l)!\:(s + 1)^{2(m + k) - l + 1}}\left( \frac{s}{s + 1} \right)^{n - m - k +l}}\bigg] \label{PNDNPAKFTS}.
\end{align}

\subsubsection{PND for  Squeezed Thermal States}
\paragraph{A. Photon Added Squeezed Thermal States\\}
For photon added squeezed thermal state, the  PND, using Eq. \eqref{PND}, is
\begin{equation}
    P_{in}(n)=\left\langle\left.n\right|\right.\rho_{PASTS}\left|\left.n\right\rangle\right. \label{PNDST}
\end{equation}
To facilitate the computations, we write Eq. \eqref{PASTS} in terms of the coherent states basis as:
\begin{equation}
    \rho_{PASTS}
    =\frac{N_{a,m}^{-1}}{\sqrt{A}}\int\int\frac{d^2\alpha}{\pi}\frac{d^2\beta}{\pi}(\alpha^\ast\beta)^{ m}\exp{\left[\frac{C}{2}(\alpha^{\ast 2}+\beta^2)+B\alpha^\ast\beta-\frac{|\alpha|^2}{2}-\frac{|\beta|^2}{2}\right]}
    |\alpha\rangle\langle\beta|\label{CVPASTS}.
\end{equation}
\begin{figure}
    \centering
    \includegraphics[width=\linewidth]{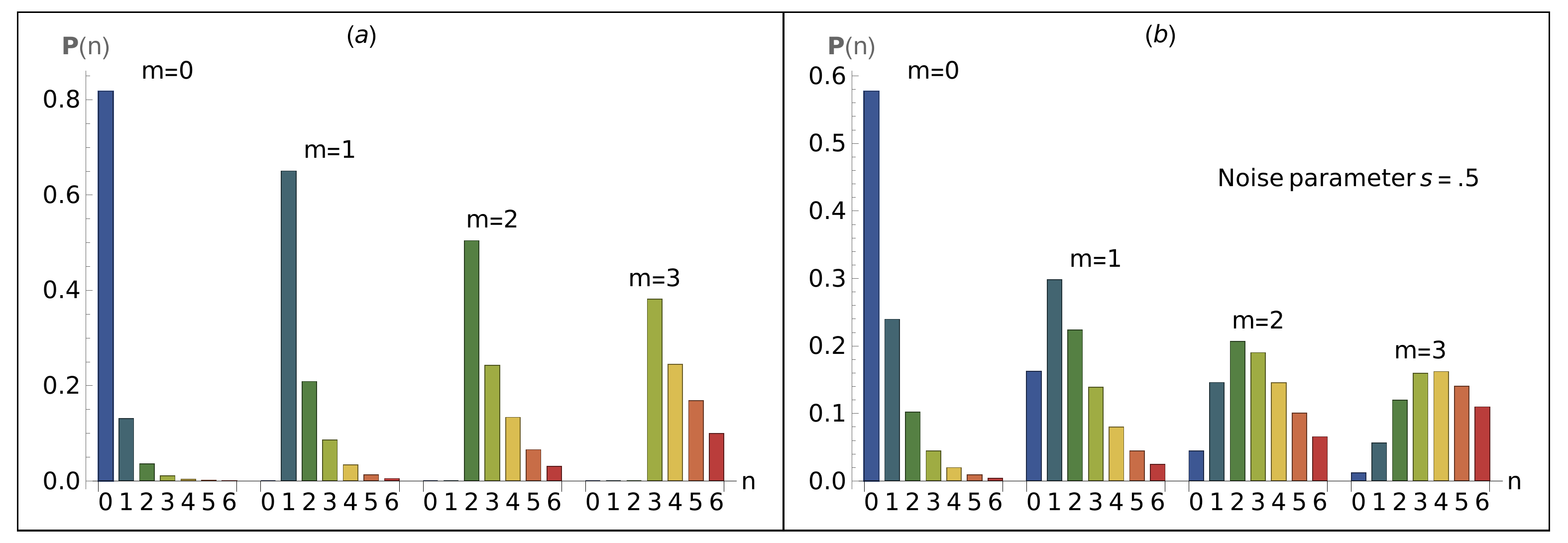}
    \caption{Photon number distributions for  (a) PASTS at the input with average thermal photon $n_{th}=0.2$,  squeezing parameter $\lambda=0.2$, and (b) noisy PASTS at the output with average thermal photon $n_{th}=0.2$, squeezing parameter $\lambda=0.2$ and noise parameter $s=0.4$.}
    \label{PNDPAS.eps}
\end{figure}
From Eqs. \eqref{PNDST} and \eqref{CVPASTS},
\begin{align*}
    \left\langle\left.n\right|\right.\rho_{PASTS}\left|\left.n\right\rangle\right.
    & =\frac{N_{a,m}^{-1}}{\sqrt{A}}\int\int\frac{d^2\alpha}{\pi}\frac{d^2\beta}{\pi}(\alpha^\ast\beta)^{ m}\exp{\left[\frac{C}{2}(\alpha^{\ast 2}+\beta^2)+B\alpha^\ast\beta-\frac{|\alpha|^2}{2}-\frac{|\beta|^2}{2}\right]}\langle n|\alpha\rangle\langle\beta|n\rangle \\
    & =\frac{N_{a,m}^{-1}}{n!\sqrt{A}}\partial_{x}^{m}\partial_{u}^{n}\int\int\frac{d^2\alpha}{\pi}\frac{d^2\beta}{\pi}\exp{\left[\frac{C}{2}(\alpha^{\ast 2}+\beta^2)+{x}\alpha^\ast\beta+u\alpha\beta^\ast-|\alpha|^2-|\beta|^2\right]}\bigg{|}_{u=1,x=B} \\
    &=\frac{N_{a,m}^{-1}}{n!\sqrt{A}}\partial_{x}^{m}\partial_{u}^{n}\int\frac{d^2\alpha}{\pi}\exp{\left[\frac{C}{2}(\alpha^{\ast 2}+u^2\alpha^2)-|\alpha|^2(1-x u)\right]}\bigg{|}_{u=1,x=B}.
\end{align*}
From the above and Eq. \eqref{PNDST}, we get the following PND for PASTS,
\begin{equation}
    P_{in}(n)=\frac{N_{a,m}^{-1}}{n!\sqrt{A}}\partial_{x}^{m}\partial_{u}^{n}\left[(1-x u)^2-C^2u^2\right]^{-1/2}\bigg{|}_{u=1,x=B} \label{PNDPASTS} .
\end{equation}

Similarly, the photon number distribution $P_{out}(n)$ for the output of PASTS can be shown to be
\begin{equation}
    P_{out}(n)=\frac{N_{a,m}^{-1}}{s\sqrt{A}}\partial_Y^{m}\partial_{u}^{n}\left[{(Y^2- C^2)}^{-1/2}\left(\left(u\left(Y_0-1+\frac{1}{s}\right)+1\right)^2-(C_{0}u)^2\right)^{-1/2}\right]\bigg{|}_{u=0} \label{PNDNPASTS}.
\end{equation}

\paragraph{B. PSSTS\\}
From Eqs. \eqref{PSSTS} and \eqref{PND},
\begin{align*}
    \left<n|\rho_{PSSTS}|n \right> & =\frac{N_{a,m^{-}}^{-1}}{\sqrt 
    A}\int\frac{d^2\alpha}{\pi}\frac{d^2\beta}{\pi}(\beta^\ast\alpha)^m\exp{\left[\frac{C}{2}(\alpha^{\ast 
    2}+{\beta}^2)+B\alpha^\ast\beta-\frac{|\alpha|^2+|\beta|^2}{2}\right]}\langle n|\alpha\rangle\langle\beta|n\rangle\\
    & =\frac{N_{a,m^{-}}^{-1}}{n!\sqrt A}\partial_v^{m+n}\int\frac{d^2\alpha}{\pi}\frac{d^2\beta}{\pi}\exp{\left[\frac{C}{2}(\alpha^{\ast 
    2}+{\beta}^2)+B\alpha^\ast\beta+v\alpha\beta^\ast-{|\alpha|^2-|\beta|^2}\right]}\bigg|_{v=0}\\
    & =\frac{N_{a,m^{-}}^{-1}}{n!\sqrt A}\partial_v^{m+n}\int\frac{d^2\alpha}{\pi}\exp{\left[\frac{C}{2}(\alpha^{\ast 
    2}+v^2\alpha^2)-{(1-B v)|\alpha|^2}\right]}\bigg|_{v=0}\\
    &=\frac{N_{a,m^{-}}^{-1}}{n!\sqrt A}\partial_v^{m+n}\left((1-B v)^2-C^2 
    v^2\right)^{-1/2}\bigg|_{v=0} .
\end{align*}
From the above and Eq. \eqref{PND}, we get the following PND for PSSTS,
\begin{equation}
       P_{in}\left(n\right)=\frac{N_{a,m^{-}}^{-1}}{n!\sqrt A}\partial_v^{m+n}\left[(1-B v)^2-C^2 
       v^2\right]^{-1/2}\bigg|_{v=0} \label{PNDPSSTS} .
\end{equation}
\begin{figure}
    \centering
    \includegraphics[width=\linewidth]{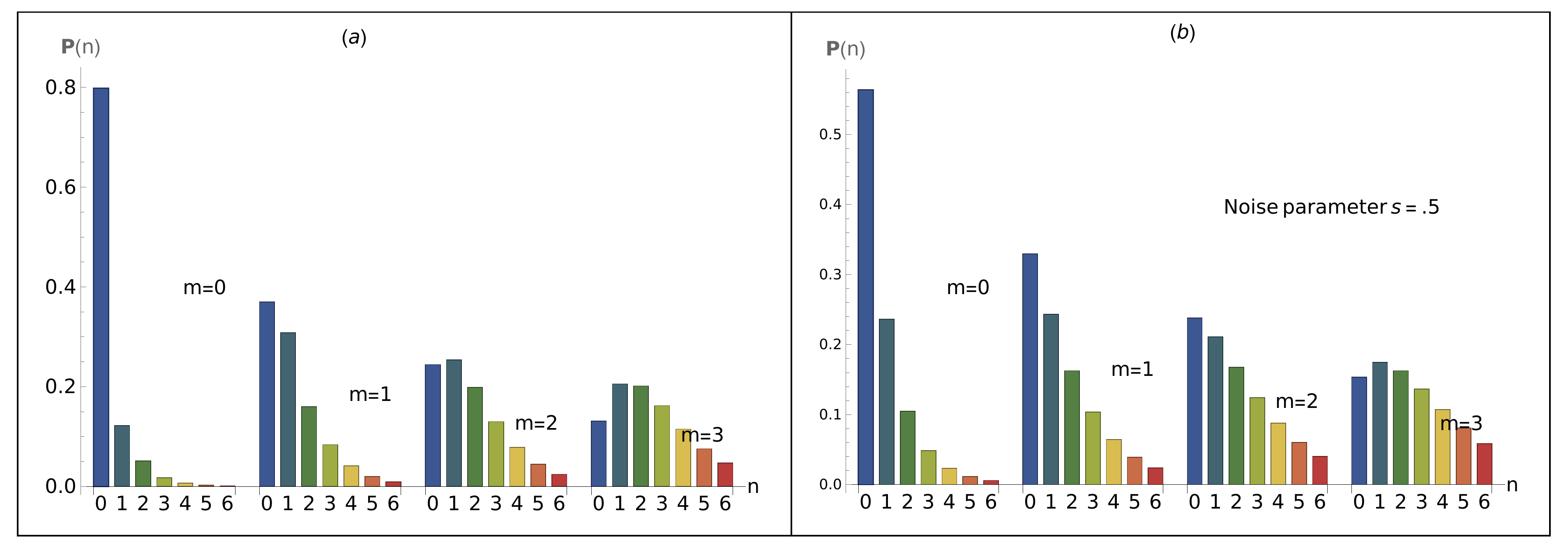}
    \caption{Photon number distributions for (a) PSSTS at the input with average thermal photon $n_{th}=0.2$, squeezing parameter $\lambda=0.3$, and (b) noisy PSSTS at the output with average thermal photon $n_{th}=0.2$, squeezing parameter $\lambda=0.3$ and noise parameter $s=0.5$.}
    \label{PNDPSS.eps}
\end{figure}
Similarly, the photon number distribution $P_{out}(n)$ for  the output of PSSTS is,
\begin{equation}
     P_{out}\left(n\right)=\frac{N_{a,m^{-}}^{-1}}{n!\sqrt A}\partial_v^n\partial_u^m
    \left[\frac{1}{s\left(1+\frac{1}{s}-v\right)}\left((1-B u)^2-C^2 u^2\right)^{-1/2}\right] \label{PNDNPSSTS},
\end{equation}
where, $u=\frac{(1-v)^2}{\left(1+\frac{1}{s}-v\right)}+v$ and $v=0$.

\
It can be seen for the thermal and squeezed thermal states that the peak for the photon number distribution is at zero photon number $n=0$. With the addition (or subtraction) of photons to (from) the thermal and squeezed thermal states, the peak shifts from zero to non-zero photons $n \neq 0$ (see figs. [7-11]). The position of the peak value of PND depends on the number of photons $m$, added (subtracted), to (from) the states. For the case of photon addition, the peak of the PND is located at $m=n$ exactly while this is not so for the photon subtraction case. Furthermore, we have  seen that in the photon addition scenario (see Figs. \ref{PNDPA.eps} and \ref{PNDPAS.eps}) there is a condition for the distribution of photons, {\it viz.} $m \le n $. Additionally, for the PAKFTS it is observed that at $n=m+k$ there is no photon distribution (see Fig. \ref{PNDPAF.eps}). It can also be seen that as we increase the noise parameter $n_{th}$ and squeezing parameter $\lambda$, a flatter and wider distribution is obtained.\\ 
At the output, we can see that the above restrictions to the photon number distribution do not apply. This is an artifact of  the randomization caused due to the interaction with the Gaussian channel. Further increase in the noise $s$ would eventually revert the states to their initial (Poissonian) form, albeit with a wider spread.

\subsection{Second Order Correlation and Mandel $Q_M$ Parameter}
To study the  statistical properties of the photon added and  subtracted states, we will now examine the second order correlation function $g^2 (s)$ as well as the Mandel $Q_{M}$-parameter.
The second order correlation is defined as,
\begin{equation}
    g^{2}\left(s\right)=\frac{\left<\hat{a}^{\dag 2}\hat{a}^2\right>} {\left<\hat{a}^{\dag}\hat{a}\right>^2}.
    \label{g2f}
\end{equation}
In the classical regime, the second order correlation function lies in the range $1\leq g^2 (s)\leq 2$ for thermal light where $s$ is the coherence time. For a  coherent state, it can be shown that $g^2 (s)=1$. Evidently $g^2 (s)<1$  is outside the allowed range of its classical counterpart and may be interpreted as an indication of the quantum regime. When $g^2 (0)<g^2 (s)$, it characterizes photon anti-bunching; its opposite is photon bunching.\\

The Mandel's $Q_{M}$-parameter is  defined as follows
\begin{equation}
     Q_{M}=\frac{\left<\hat{a}^{\dag 2}\hat{a}^2\right>-{\left<\hat{a}^\dag\hat{a}\right>}^2}{\left<\hat{a}^\dag\hat{a}\right>},
     \label{Qm}
\end{equation}
which implies the deviation of the variance of the photon number distribution of the field state under consideration  from the Poissonian distribution of the coherent state. If $Q_{M}=0$ the field is said to have Poissonian photon statistics while for $Q_{M}>0$, its super-Poissonian and  $Q_{M}<0$ implies sub-Poissonian statistics.
It is well known that the negativity of the $Q_{M}$-Parameter refers to sub-Poissonian statistics of the state. But a state can be nonclassical even though $Q_{M}$ is positive. \\

\subsubsection{For Thermal States}
\paragraph{A. PATS\\}
For $r^{t h}$ moment, 
\begin{equation}
    \left<\hat{a}^{\dag r}\hat{a}^r\right>=Tr\left[\hat{\rho}\hat{a}^{\dag r}\hat{a}^r\right]=\sum_{n=0}^{\infty}\left<n|\hat{a}^r\rho\hat{a}^{\dag r}|n \right>=N_{m}^{-1} {\sum_{n=0}^{\infty} {\frac{\left(n+r\right)!^2}{n!\left(n+r-m\right)!} \left(1-A\right)^{n+r-m}}} \label{rthmPATS} .
\end{equation}
\begin{figure}
    \centering
    \includegraphics[width=\textwidth]{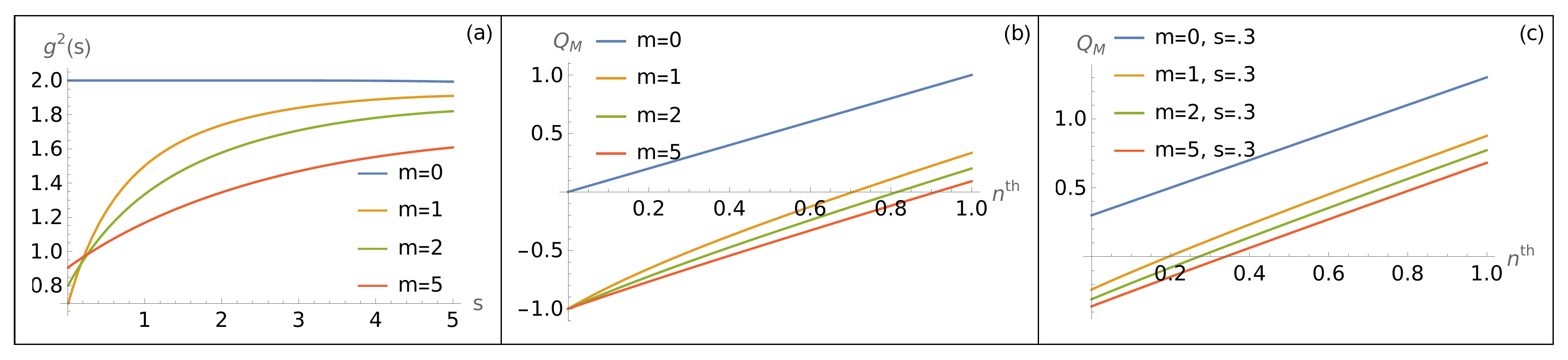}
    \caption{(a) Second ordered correlation function for output PATS with average thermal photon $n_{th}=0.3$, (b) Mandel's $Q_{M}$ function for PATS at the input, and (c) Mandel's $Q_{M}$ function for noisy  PATS at the output with noise parameter $s=0.3$.}
    \label{GQPA.eps}
\end{figure}
For the output PATS
\begin{align}
    \langle\hat{a}^{\dag r}\hat{a}^{r}\rangle=Tr\left[\phi_s(\rho)\hat{a}^{\dag r}\hat{a}^{r}\right]&=N_{m}^{-1} \sum_{l=0}^{m} \frac{{m!}^{2} s^{l}}{l!{\left(m-l\right)!)}^{2}\left(As+1\right)^{2m-l+1}}\sum_{n=0}{\frac{\left(n+r\right)!^{2}}{n!\left(n+r-m+l\right)!}}\left(1-\frac{A}{As+1}\right)^{n+r-m+l} . \label{rthmNPATS}
\end{align}

\paragraph{B. PSTS\\}
For $r^{t h}$ moment, 
\begin{equation}
   \left<\hat{a}^{\dag r}\hat{a}^r\right>
    =N_{m^-}^{-1}\int{\frac{d^2 \alpha}{\pi}{(\alpha^{\ast}\alpha)^{m+r}} e^{-\frac{\alpha^{\ast}\alpha}{n_{th}}}}=N_{m^-}^{-1} (m+r)! (n_{th})^{m+r+1}.
\end{equation}

\begin{figure}
    \centering
    \includegraphics[width=\linewidth]{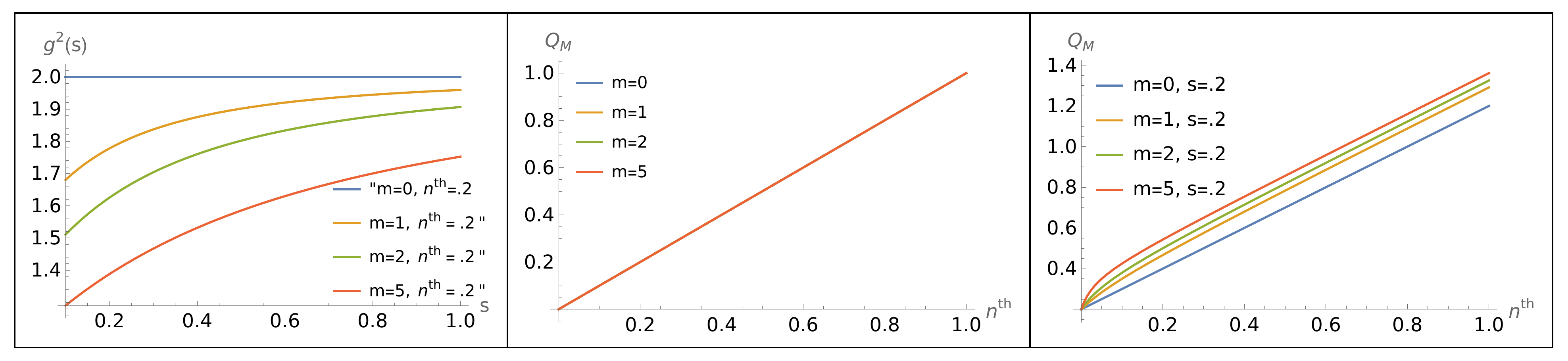}
    \caption{(a) Second ordered correlation function for output PSTS with average thermal photon $n_{th}=0.4$, (b) Mandel's $Q_{M}$ function for PSTS at the input, and (c) Mandel's $Q_{M}$ function for noisy  PSTS at the output with noise parameter $s=0.2$.}
    \label{GQPS.eps}
\end{figure}
Similarly, for the output PSTS
\begin{equation}
    \left<\hat{a}^{\dag r}\hat{a}^r\right>=N_{m^-}^{-1}\partial_{u}^{r}\left[\frac{1}{s \left(\frac{1}{s}-u\right)}\frac{m!}{\left(\frac{1}{n_{th}}-u-\frac{u^2}{\frac{1}{s}-u}\right)^{m+1}}\right]\bigg{|}_{u=0} .
\end{equation}

\paragraph{C. PAKFTS\\}
For $r^{t h}$ moment, 
\begin{equation}
    \left\langle {{\hat{a}}^{\dag r}\hat{a}}^{r} \right\rangle = N_{\text{km}}^{- 1}\sum_{n = 0}^{}\left[ \frac{{((n + r)!)}^{2}}{n!(n + r - m)!}(1 - A)^{n + r - m} - \frac{e^{- \beta\hbar\omega k}}{k!}\frac{{((n + r)!)}^{2}}{n!\ (n + r - m - k)!}\left| \left\langle n + r - m - k \middle| 0 \right\rangle \right|^{2} \right].
\end{equation}

\begin{figure}
    \centering
    \includegraphics[width=\linewidth]{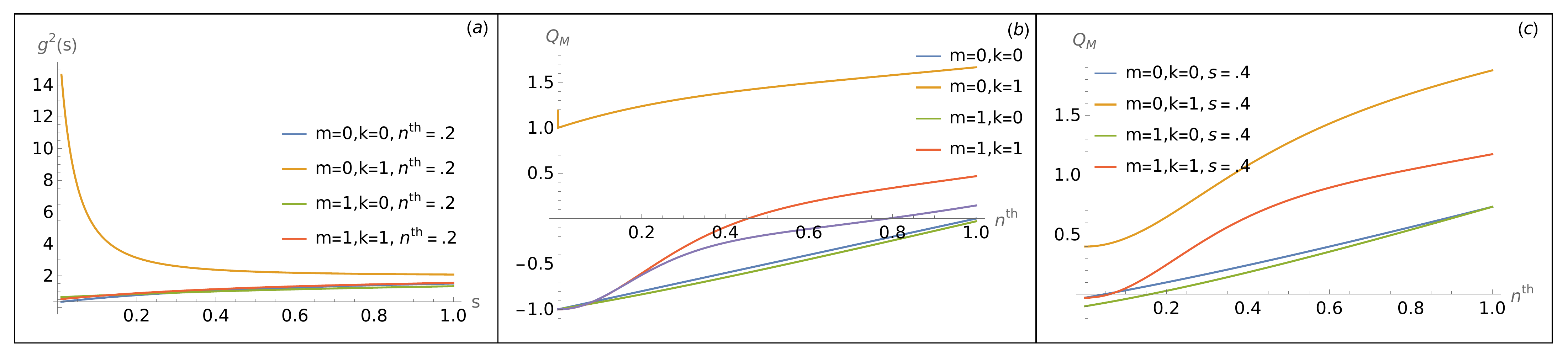}
    \caption{(a) Second ordered correlation function for output PAKFTS with average thermal photon $n_{th}=0.2$, (b) Mandel's $Q_{M}$ function for PAKFTS at the input, and (c) Mandel's $Q_{M}$ function for noisy  PAKFTS at the output with noise parameter $s=0.4$.}
    \label{GQPAF.eps}
\end{figure}
For the output PAKFTS
\begin{align}
    \left\langle {{\hat{a}}^{\dag r}\hat{a}}^{r} \right\rangle & = N_{\text{km}}^{- 1}\sum_{n = 0}^{}\bigg[ \sum_{l = 0}^{m}{\frac{{m!}^{2}{{((n + r)!)}^{2}s}^{l}}{l!\:n!\:((m - l)!)^{2}\:(n + r - m + l)!\left( \text{As} + 1 \right)^{2m - l + 1}}\left( \frac{\text{As} -A + 1}{\text{As} + 1} \right)^{n + r - m +l}} \nonumber \\
   & - \frac{e^{- \beta\hbar\omega k}}{k!}\sum_{l = 0}^{m + k}{\frac{{(m + k)!}^{2}{((n + r)!)}^{2}s^{l}}{l!\:n!\:((m + k - l)!)^{2}(n + r - m - k + l)!(s + 1)^{2(m + k) - l + 1}}\left( \frac{s}{s + 1} \right)^{n + r - m - k + l}} \bigg] .
\end{align}
\subsubsection{For Squeezed Thermal States}
\paragraph{A. PASTS\\}
For $r^{t h}$ moment, 
\begin{equation}
    \left<\hat{a}^{\dag r}\hat{a}^{r}\right>=\frac{N_{a,m}^{-1}}{\sqrt{A}}\partial_{x}^{m}\partial_{u}^{r}\left[(1-x u)^2-C^2u^2\right]^{-1/2}\bigg|_{u=1} .
\end{equation}

\begin{figure}
    \centering
    \includegraphics[width=\linewidth]{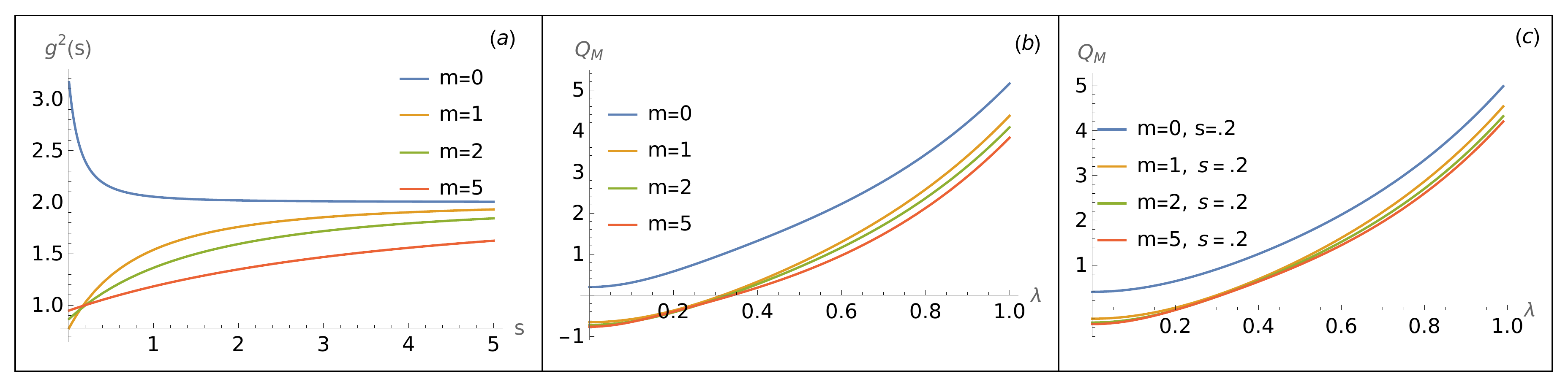}
    \caption{(a) Second ordered correlation function $g^{2}(s)$ for noisy PASTS at the output  with average thermal photon $n_{th}=0.2$ and squeezing parameter $\lambda=0.2$, (b) $Q_{M}$ for PASTS at the input with average thermal photon $n_{th}=0.2$, and (c) $Q_{M}$ for noisy PASTS at the output with average thermal photon $n_{th}=0.1$a nd noise parameter $s=0.2$.}
    \label{GQPAS.eps}
\end{figure}
The corresponding output PASTS is
\begin{equation}
    \left<\hat{a}^{\dag r}\hat{a}^{r}\right>=\frac{N_{a,m}^{-1}}{s\sqrt{A}}\partial_Y^{m}\partial_{u}^{r}\left[{(Y^2- C^2)}^{-1/2}\left(\left(u\left(Y_0-1+\frac{1}{s}\right)+1\right)^2-(C_{0}u)^2\right)^{-1/2}\right]\bigg|_{u=1} .
\end{equation} 

\paragraph{B. PSSTS\\}
For $r^{t h}$ moment, 

\begin{equation}
  \left<\hat{a}^{\dag r}\hat{a}^r\right>=tr[\hat{a}^{\dag r}\hat{a}^r\rho_{P S S T S}]=\frac{N_{a,m^{-}}^{-1}}{\sqrt A}\partial_v^{m+r}\left[(1-B v)^2-C^2 v^2\right]^{-1/2}\bigg|_{v=1} .
\end{equation}

The corresponding output PSSTS is

\begin{figure}
    \centering
    \includegraphics[width=\linewidth]{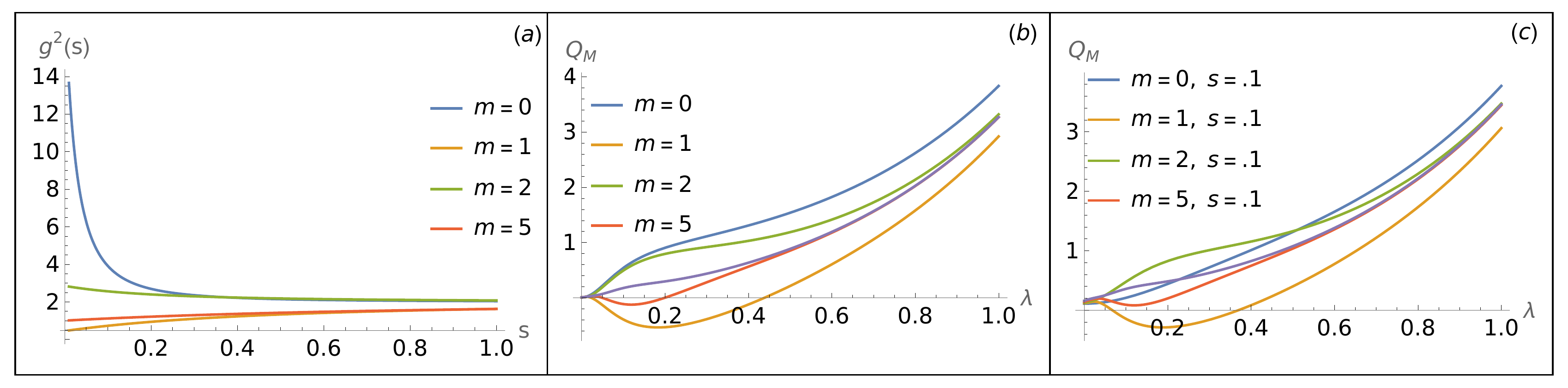}
    \caption{(a) Second ordered correlation function $g^{2}(s)$ for noisy PSSTS at the output with average thermal photon $n_{th}=0.01$, squeezing parameter $\lambda=0.2$, (b) $Q_{M}$ for PSSTS at the input with average thermal photon $n_{th}=0.01$, and (c) $Q_{M}$ for noisy PSSTS at the output with average thermal photon=0.01 and noise parameter $s=0.1$.}
    \label{GQPSS.eps}
\end{figure}
\begin{equation}
   \left<\hat{a}^{\dag r}\hat{a}^r\right>
    =\frac{N_{a,m^{-}}^{-1}}{\sqrt A}\partial_v^r\partial_u^m\left[
    \frac{1}{s\left(1+\frac{1}{s}-v\right)}\left((1-B u)^2-C^2 u^2\right)^{-1/2}\right],
\end{equation}
where, $u=\frac{(1-v)^2}{\left(1+\frac{1}{s}-v\right)}+v$ and $v=1$.\\
\\
The $r^{th}$ moment for all input and output states are calculated in Appendix C.
After putting the values of  $r^{th}$ moment in Eqs. \eqref{g2f} and \eqref{Qm} one can get the second order correlation function $g^{2}(s)$ and Mandel's $Q_{M}$-parameter, respectively.\\

From the Figs. (\ref{GQPA.eps}-\ref{GQPSS.eps}) 
we observe that with the addition of photons to the initial thermal Gaussian state, the $g^2$ function's value becomes less than one, an indicator of quantum regime. With increase in noise however, the output state's $g^2$ tends to two, corresponding to the thermal state. This reduction of $g^2$ to a value less than one is not observed for the case of photon subtraction  (see Fig. \ref{GQPS.eps}). However, for the photon subtracted squeezed thermal state (Fig. \ref{GQPSS.eps}), the reduction to less than one is observed for very small temperatures. This suggest that the quantum features of the photon added states are more robust, as compared to their photon subtracted counterparts.
\\
Negative values of $Q_{M}$, indicative of sub-Poissonian behavior, is observed for various parameters 
for input and output states, for cut-off values of $n_{th}$  and squeezing parameter $\lambda$. It's observed that the cut-off values can be increased (decreased) by adding (subtracting) photons to the thermal or the squeezed thermal states. For PSTS we see that $Q_{M}$ is only a function of $n_{th}$ and not effected by the subtraction of photons (see Fig. \ref{GQPS.eps}(b)). For PSSTS case the negative $Q_{M}$ function is observed only for odd subtracted photons (see Figs. \ref{GQPSS.eps}(b) and \ref{GQPSS.eps}(c)). A similar pattern was observed for the corresponding $W$ and  $P$ functions. However, here some negative regions were observed for even subtracted photons (Fig. \ref{PPSS.eps}). As we increase the value of the noise parameter in the output states, the negative $Q_{M}$ shifts towards positive values and cut-off values for $\{n_{th},\lambda \}$, the thermal and squeezed parameters, also decrease. For  PSTS at the input and output, negative value of Mandel's $Q_{M}$ parameter was not observed (see Fig. \ref{GQPS.eps}(b) and \ref{GQPS.eps}(c)).
\\
For thermal and squeezed thermal states, photon bunching is observed. Apart form this, some states belonging to the PAKFTS and PSSTS  exhibit bunching at $\{m=0,\:k=1\}$ and ${m=2}$, respectively (see Figs. \ref{GQPAF.eps}(a) and \ref{GQPSS.eps}(a)). Anti-bunching is observed in other photon added and subtracted states. 
The states exhibiting photon bunching  do not show negative values of $Q_{M}$, which is not so for the anti-bunching cases.


\section{Conclusion}
In this work, we have systematically studied the impact of noise, modelled by a noisy Gaussian channel, on a wide range of non-Gaussian  input states, covering both photon added, subtracted and hole burning scenarios.  The impact of noise is presented as an input-output problem. Making use of the IWOP technique allows for the derivation of analytical expressions for the various output states. The quantum nature of the states, both input and output, is developed by studying the corresponding photon statistics and quasi-probability distributions. This further allows for gauging the impact of noise (decoherence) on the non-Gaussian input states. 
It's found that photon addition has more robust quantum mechanical properties as compared to the case of photon subtraction.
It's also seen that with increase in noise, there is a tendency to lesser non-classicality as well as non-Gaussianity.
The input states considered (except PSTS) have highly non-classical properties. This is depicted by the negative region at the center of the phase space. It was seen that the threshold value of the noise parameter corresponding to the transition of quasi-probability distribution functions from partial negative ($W$ and $P$) and zero ($Q$) to completely positive definite, at the center of phase space, is dependent not only on the noise parameter, but also on the average number of thermal photons in the state and the squeezing parameter.\\
 
For photon-added (thermal and squeezed thermal states) and photon-subtracted squeezed states, by an odd number of photons, the negativity of the Mandel ${Q_{M}}$ parameter is noticeable. However, for non-classical states, Mandel's $Q_{M}$ parameter do not always indicate a negative value. In fact, for the case where an even number of photons were subtracted from the squeezed states, it was observed to be positive. It follows that the negativity of the $Q_{M}$ parameter is sufficient to distinguish the classical state from the nonclassical state.  The non-classicality of quantum states, could be also explored via measures such as those based on the volume of the negative part of the Wigner function \cite{PhysRevA.60.4034}, on the nonclassical depth \cite{PhysRevA.44.R2775}. Both the features of photon bunching and antibunching were observed in the non-Gaussian states.
For the case of photon addition and filtration, we obtained conditions indicative 
of hole burning in the input states. However, in the output these holes get filled up due to the influence of noise generated due to the passage through the Gaussian channel.\\

It was observed that photon addition was more robust than photon subtraction in withstanding the impact of noise. This could be ascribed to the non-Gaussianity (visualized by the phase space distributions) for photon added states being greater than for the subtracted photon states. In addition, for KFTS nonclassical behaviour was observed due to filtration on the thermal state that could be further enhanced by adding photon(s). PAKFTS came out to be most robust against noise that other states considered.
\\
It would be of interest to evaluate the impact of more general quantum channels on the non-Gaussian input states.

\newpage
\appendix
\appendix

\section{Appendix: Normalization Constant}
For any density, we have the following,
 \[Tr[\rho]=1.\]
\paragraph{For PATS}
\begin{equation*}
    N_m={Tr\left[:\hat{a}^{\dag m} e^{-A\hat{a}^\dag\hat{a}}\hat{a}^m:\right]}=\int{\left\langle\left.\alpha\right|\right.:\hat{a}^{\dag m} {e^{-A\hat{a}^{\dag}\hat{a}}}\hat{a}^{m}:\left|\left.\alpha\right\rangle\right.}\frac{d^{2}\alpha}{\pi}=\int{(\alpha\alpha^{\ast})^{m} e^{-A|\alpha|^2}}\frac{d^{2}\alpha}{\pi}, 
\end{equation*}
let $\alpha=x+\iota y $, $\alpha\alpha^{\ast}=x^2+y^2=r^2$ and $d^{2}\alpha=r d r d\theta$,

now we have, 
\begin{equation*}
     N_m=\int{r^{2m+1}e^{-A r^{2}}{}{d r} d\theta=m!/A^{\left(m+1\right)}} \label{NM}.
\end{equation*}
\fbox{Where we used the  Integration formula  $\int_{0}^{\infty}x^{2m+1}\exp{(-a x^{2})}{dx}=\frac{m!}{2a^{m+1}}.$}
\paragraph{For PSTS}
\begin{equation*}
     N_{m^-}=Tr\bigg[\int{\frac{d^2 \alpha}{\pi}{(\alpha^{\ast}\alpha)^m} e^{-\frac{|\alpha|^2}{n_{th}}}\left|\left.\alpha\right\rangle\right.\left\langle\left.\alpha\right|\right.}\bigg]=\int{\frac{d^2 \alpha}{\pi}{(\alpha^{\ast}\alpha)^m}e^{-\frac{|\alpha|^2}{n_{th}}}}=m!{(n_{th})}^{\left(m+1\right)}.
\end{equation*}
\paragraph{For PAKFTS}
\begin{align*}
    N_{\text{km}}^{\ }& = \text{Tr}\bigg[ :{{\hat{a}}^{\dag m}e^{- A{\hat{a}}^{\dag}\hat{a}}\hat{a}}^{m}: -\frac{e^{- \beta\hbar\omega k}}{k!}:{{\hat{a}}^{\dag (m + k)}e}^{- {\hat{a}}^{\dag}\hat{a}}{\hat{a}}^{(m + k)}: \bigg]\\
&= \int_{}^{}{\frac{d^{2}\alpha}{\pi}\langle\alpha|\ :{{\hat{a}}^{\dag m}e^{- A{\hat{a}}^{\dag}\hat{a}}\hat{a}}^{m}: - \frac{e^{- \beta\hbar\omega k}}{k!}:{{\hat{a}}^{\dag (m + k)}e}^{- {\hat{a}}^{\dag}\hat{a}}{\hat{a}}^{(m + k)}:|\alpha\rangle} \\
&= \int_{}^{}{\frac{d^{2}\alpha}{\pi}\langle\alpha|\ :{{\hat{a}}^{\dag m}e^{- A{\hat{a}}^{\dag}\hat{a}}\hat{a}}^{m}:|\alpha\rangle} - \frac{e^{- \beta\hbar\omega k}}{k!}\int_{}^{}{\frac{d^{2}\alpha}{\pi}\langle\alpha|:{{\hat{a}}^{\dag (m + k)}e}^{- {\hat{a}}^{\dag}\hat{a}}{\hat{a}}^{(m + k)}:|\alpha\rangle}.
\end{align*}
We have,
\begin{equation*}
    \int_{}^{}{\frac{d^{2}\alpha}{\pi}\langle\alpha| :{{\hat{a}}^{\dag m}e^{- A{\hat{a}}^{\dag}\hat{a}}\hat{a}}^{m}:|\alpha\rangle} = \int_{}^{}{\frac{d^{2}\alpha}{\pi}\left( \alpha\alpha^{*} \right)^{m}}e^{- A|\alpha|^{2}} = \frac{m!}{A^{m + 1}},
\end{equation*}
using the above equation we can get following expression of normalization constant For PAKFTS-
\begin{equation}
    N_{k m}=\frac{m!}{A^{m+1}}-\frac{e^{- \beta\hbar\omega k}}{k!} (m+k)! \label{Nkm}.
\end{equation}
\paragraph{For PASTS\\}
we have $tr(\rho_{PASTS})=1.$
\begin{align*}
    N_{a,m}&=\frac{1}{\sqrt A}\int\left\langle\left.\alpha\right|\right.:\hat{a}^{\dag m}\exp{\left[\frac{C}{2}(\hat{a}^{\dag 2}+{\hat{a}}^2)+(B-1){\hat{a}}^\dag\hat{a}\right]} \hat{a}^{m}:\left|\left.\alpha\right\rangle\right.\frac{d^2\alpha}{\pi}\\
    &=\frac{1}{\sqrt A}\int(\alpha^\ast\alpha)^{m}\exp{\left[\frac{C}{2}(\alpha^{\ast 2}+{\alpha}^2)+(B-1){\alpha}^{\ast}\alpha\right]} \frac{d^2\alpha}{\pi}\\
    &=\frac{1}{\sqrt A}\partial_{v}^{m}\int\exp{\left[\frac{C}{2}(\alpha^{\ast 2}+{\alpha}^2)+v{\alpha}^{\ast}\alpha\right]} \frac{d^2\alpha}{\pi}\bigg|_{v=B}.
\end{align*}
Now using the Eq. \eqref{MIF}

 \begin{equation}
    N_{a,m}=\frac{1}{\sqrt A}\partial_{v}^{m}\left[(v^2-C^2)^{-1/2}\right]\bigg|_{v=B}. \label{eq7}
\end{equation}

\paragraph{For PSSTS\\}
we have $Tr(\rho_{PSSTS})=1.$
\begin{align*}
    N_{a,m^{-}}&=Tr\bigg[\frac{1}{\sqrt A}\int\int\frac{d^2\alpha}{\pi}\frac{d^2\beta}{\pi}(\beta^\ast\alpha)^m\exp{\left[\frac{C}{2}(\alpha^{\ast 2}+{\beta}^2)+B\alpha^\ast\beta-\frac{|\alpha|^2+|\beta|^2}{2}\right]}{|\alpha\rangle\langle\beta|}\bigg]\\
   & =\frac{1}{\sqrt A}\int\int\frac{d^2\alpha}{\pi}\frac{d^2\beta}{\pi}(\beta^\ast\alpha)^m\exp{\left[\frac{C}{2}(\alpha^{\ast 2}+{\beta}^2)+B\alpha^\ast\beta-\frac{|\alpha|^2+|\beta|^2}{2}\right]}{\langle\beta|\alpha\rangle}\\
   &= \frac{1}{\sqrt A}\int\int\frac{d^2\alpha}{\pi}\frac{d^2\beta}{\pi}(\beta^\ast\alpha)^m\exp{\left[\frac{C}{2}(\alpha^{\ast 2}+{\beta}^2)+B\alpha^\ast\beta+\alpha\beta^\ast-{|\alpha|^2-|\beta|^2}\right]}\\
   &=\frac{1}{\sqrt A}\partial_{u}^{m}\int\int\frac{d^2\alpha}{\pi}\frac{d^2\beta}{\pi}\exp{\left[\frac{C}{2}(\alpha^{\ast 2}+{\beta}^2)+B\alpha^\ast\beta+u \alpha\beta^\ast-{|\alpha|^2-|\beta|^2}\right]}\bigg|_{u=1}\\
   &=\frac{1}{\sqrt A}\partial_{u}^{m}\int\frac{d^2\alpha}{\pi}\exp{\left[\frac{C}{2}(\alpha^{ \ast 2}+u^2\alpha^{ 2})-(1-Bu)|\alpha|^2\right]}\bigg|_{u=1}\\
   &=\frac{1}{\sqrt A}\partial_{u}^{m}\left[(1-Bu)^2-C^2u^2\right]^{-1/2}\bigg|_{u=1}.
\end{align*}

\section{Appendix: Characteristic Function (CF)}
\paragraph{CF For PASTS}
\begin{align*}
    {Tr\left[e^{\gamma \hat{a}^\dag}\hat{\rho} e^{-\gamma^{\ast} \hat{a}}\right]}&=\frac{N_{a,m}^{-1}}{\sqrt A}\int{\langle\alpha|}:\exp{\left(\gamma \hat{a}^\dag\right)}\hat{a}^{\dag m}\exp{\left[\frac{C}{2}(\hat{a}^{\dag 2}+{\hat{a}}^2)+(B-1){\hat{a}}^\dag\hat{a}\right]} \hat{a}^{m}\exp\left(-\gamma^{\ast} \hat{a}\right):{|\alpha\rangle}\frac{d^2\alpha}{\pi}\\&=\frac{N_{a,m}^{-1}}{\sqrt A}\int{\left(\alpha^*\alpha\right)^{m}\exp{\left[\frac{C}{2}(\alpha^{\ast 2}+\alpha^{2})+(B-1) \alpha^{\ast}\alpha+\gamma\alpha^{\ast}-\gamma^{\ast}\alpha\right]}\frac{d^2\alpha}{\pi}}\\&=\frac{N_{a,m}^{-1}}{\sqrt A}\partial_X^{m}\int{\exp{\left[\frac{C}{2}(\alpha^{\ast 2}+\alpha^{2})+X \alpha^{\ast}\alpha+\gamma\alpha^{\ast}-\gamma^{\ast}\alpha\right]}\frac{d^2\alpha}{\pi}}.
\end{align*}
Using the Eq. \eqref{MIF} 
\begin{equation*}
{Tr\left[e^{\gamma \hat{a}^\dag}\hat{\rho} e^{-\gamma^{\ast} \hat{a}}\right]}
=\frac{N_{a,m}^{-1}}{\sqrt A}\partial_X^{m}\left[{(X^2- 
C^2)}^{-1/2}\exp\bigg[\frac{\frac{C}{2}({{\gamma}^{\ast}}^2+\gamma^{2}) - X
{\left|\gamma\right|^{2}}}{(X^2-C^2)}\bigg]\right],
\end{equation*}
where $X=(1-B)$.

Thus the characteristic function for PASTS will be-
\begin{equation*}
   \chi_{in}\left(\gamma,\kappa\right) =\frac{N_{a,m}^{-1}}{\sqrt A}\partial_X^{m}\left[{(X^2- C^2)}^{-1/2} \exp\bigg[\frac{\frac{C}{2}({{\gamma}^{\ast}}^2+\gamma^{2}) - X {\left|\gamma\right|^{2}}}{(X^2-C^2)}+\frac{\kappa+1}{2}{|\gamma|}^2\bigg]\right]
\end{equation*}

Similarly, for noisy PASTS at the output,
\begin{align*}
    Tr\left[e^{\gamma{\hat{a}}^\dag}\rho e^{-\gamma^\ast\hat{a}}\right]
    &=\frac{N_{a,m}^{-1}}{s \sqrt 
    A}\partial_Y^m\left[\int\frac{d^2\alpha}{\pi }\left\langle\alpha\middle| 
    e^{\gamma{\hat{a}}^\dag}:e^{-\frac{{\hat{a}}^\dag\hat{a}}{s}}\left(Y^2-C^2\right)^{-1/2} 
    \exp\left[\frac{\frac{C}{2}\left({\hat{a}}^{{\dag 2}}+{\hat{a}}^2\right)-Y{\hat{a}}^\dag\hat{a}}{s^2\left(Y^2-C^2\right)}\right]:e^{-\gamma^\ast\hat{a}}\middle|\alpha\right\rangle \right]\\
    &=\frac{N_{a,m}^{-1}}{s \sqrt A}\partial_Y^m\bigg[\left(Y^2-C^2\right)^{-1/2}\int\frac{d^2z}{\pi }\exp\left[-\left(Y_{0}+1/s\right)\alpha^\ast\alpha+\left(C_{0}/2\right)\left({\alpha^\ast}^2+\alpha^2\right)+\gamma\alpha^\ast-\gamma^\ast\alpha\right]\bigg],
\end{align*} 

where, $Y_{0}=\frac{Y}{\left(Y^2-C^2\right)}$ and $ C_{0}=\frac{C}{\left(Y^2-C^2\right)}.$ \\

Using the Eq. \eqref{MIF}

\begin{multline*}
   Tr\left[e^{\gamma{\hat{a}}^\dag}\rho e^{-\gamma^\ast\hat{a}}\right]
   =\frac{N_{a,m}^{-1}}{s \sqrt A}\partial_Y^m\bigg[\left(Y^2-C^2\right)^{-1/2}\left(\left(Y_{0}+1/s\right)^2-{C_{0}}^2\right)^{-1/2}\exp\bigg[\frac{\frac{C_{0}}{2}\left({\gamma^\ast}^2+\gamma^2\right)-\left(Y_{0}+1/s\right)\gamma^\ast\gamma}{\left(\left(Y_{0}+1/s\right)^2-{C_{0}}^2\right)}\bigg]\bigg].
\end{multline*}

After putting the value of trace in  Eq. \eqref{CFPA}, we get the following expression for characteristic function for the output state of PASTS -

\begin{multline*}
    \chi_{out}\left(\gamma.\kappa\right)
    =\frac{N_{a,m}^{-1}}{s \sqrt A}\partial_Y^{m}\left(Y^2-C^2\right)^{-1/2}\left(\left(Y_{0}+1/s\right)^2-{C_{0}}^2\right)^{-1/2}\\
    \times \exp\left\{\frac{\frac{C_{0}}{2}\left({\gamma^\ast}^2+\gamma^2\right)}{\left(\left(Y_{0}+1/s\right)^2-{C_{0}}^2\right)}-\left(\frac{\left(Y_{0}+1/s\right)}{\left(\left(Y_{0}+1/s\right)^2-{C_{0}}^2\right)} -\frac{\beta+1}{2}\right)|\gamma|^2\right\},
\end{multline*}

let we denote $Y_{1}=\frac{\left(Y_{0}+1/s\right)}{\left(\left(Y_{0}+1/s\right)^2-{C_{0}}^2\right)}\ -\frac{\kappa+1}{2}$, $C_{1}=\frac{C_{0}}{\left(\left(Y_{0}+1/s\right)^2-{C_{0}}^2\right)}.$

\begin{equation}
    \chi_{out}\left(\gamma,\kappa\right)=\frac{N_{a,m}^{-1}}{s \sqrt A}\partial_Y^{m}\bigg[\left(Y^2-C^2\right)^{-1/2}\left(\left(Y_0+1/s\right)^2-{C_0}^2\right)^{-1/2}exp\bigg[\frac{C_{1}}{2}\left({\gamma^\ast}^2+\gamma^2\right)-Y_1\gamma^\ast\gamma\bigg] \bigg] .\label{CFNPASTS} 
\end{equation}

\paragraph{CF For PSSTS}
\begin{align*}
    &{Tr\left[e^{-{{\gamma}^{\ast}}\hat{a}}\hat{\rho}e^{\gamma\hat{a}^{\dag}}\right]}\\
    &=Tr\bigg[\frac{N_{a,m^{-}}^{-1}}{\sqrt{A}}\int\int\frac{d^{2}\alpha}{\pi}\frac{d^{2}\beta}{\pi}(\beta^{\ast}\alpha)^{m} \exp{\left[\frac{C}{2}(\alpha^{\ast 2}+\beta^{2})+B \alpha^{\ast}\beta-\frac{|\alpha|^{2}+|\beta|^{2}}{2}\right]} e^{-\gamma^{\ast} \hat{a}}|\alpha\rangle\langle\beta|e^{\gamma\hat{a}^{\dag}}\bigg]\\
     &=\frac{N_{a,m^{-}}^{-1}}{\sqrt A}\int\int\frac{d^2\alpha}{\pi}\frac{d^2\beta}{\pi}(\beta^\ast\alpha)^m\exp{\left[\frac{C}{2}(\alpha^{\ast 2}+{\beta}^2)+B\alpha^\ast\beta-\gamma^\ast\alpha+\gamma\beta^\ast-\frac{|\alpha|^2+|\beta|^2}{2}\right]}\langle\beta|\alpha\rangle \\
     &=\frac{N_{a,m^{-}}^{-1}}{\sqrt A}\int\int\frac{d^2\alpha}{\pi}\frac{d^2\beta}{\pi}(\beta^\ast\alpha)^m\exp{\left[\frac{C}{2}(\alpha^{\ast 2}+{\beta}^2)+B\alpha^\ast\beta+\beta^\ast\alpha-\gamma^\ast\alpha+\gamma\beta^\ast-{|\alpha|^2-|\beta|^2}\right]}\\
     &=\frac{N_{a,m^{-}}^{-1}}{\sqrt A}\partial_{u}^{m}\int\int\frac{d^2\alpha}{\pi}\frac{d^2\beta}{\pi}\exp{\left[\frac{C}{2}(\alpha^{\ast 2}+{\beta}^2)+B\alpha^\ast\beta+u \beta^\ast\alpha-\gamma^\ast\alpha+\gamma\beta^\ast-{|\alpha|^2-|\beta|^2}\right]}\bigg|_{u=1}\\
     &=\frac{N_{a,m^{-}}^{-1}}{\sqrt A}\partial_{u}^{m}\int\frac{d^2\alpha}{\pi}\exp{\left[\frac{C}{2}(\alpha^{\ast 2})-\gamma^\ast\alpha-{|\alpha|^2}\right]}\exp{\big[B\alpha^\ast(u \alpha+\gamma)+\frac{C}{2}(u\alpha+\gamma)^{2}\big]}\bigg|_{u=1}\\
    &=\frac{N_{a,m^{-}}^{-1}}{\sqrt A}\partial_{u}^{m}\int\frac{d^2\alpha}{\pi}\exp{\bigg[\frac{C}{2}(\alpha^{\ast 2}+u^2\alpha^2)-(1-Bu)|\alpha|^2-\alpha(\gamma^\ast-\gamma C u)+B\gamma\alpha^\ast+\frac{C}{2}\gamma^2\bigg]}\bigg|_{u=1}\\
     &=\frac{N_{a,m^{-}}^{-1}}{\sqrt A}\partial_{u}^{m}\bigg[\left((1-Bu)^2-C^2u^2\right)^{-1/2}\exp{\bigg[\frac{-(1-Bu)(\gamma^\ast-\gamma C u)B \gamma+\frac{C}{2}((\gamma^\ast-\gamma C u)^2+u^2 B^2\gamma^2)}{(\left(1-Bu)^2-C^2u^2\right)}+\frac{C}{2}\gamma^2\bigg]}\bigg]\bigg|_{u=1}\\
     &=\frac{N_{a,m^{-}}^{-1}}{\sqrt A}\partial_{u}^{m}\bigg[\left((1-Bu)^2-C^2u^2\right)^{-1/2}\\
     &\times\exp{\bigg[\frac{\frac{C}{2}\gamma^{\ast 2}+\gamma^{2}(\frac{C(u^2 B^2+C^2 u^2)}{2}+C u B(1-Bu))-|\gamma|^2(B(1-Bu)+C^2 u)}{(\left(1-Bu)^2-C^2u^2\right)}+\frac{C}{2}\gamma^2\bigg]}\bigg]\bigg|_{u=1}.
\end{align*}
After putting the value of trace in  Eq. \eqref{CFPS}, we get the following expression for characteristic function,
\begin{align}
    \chi_{PSSTS}\left(\gamma,\kappa\right)&=\frac{N_{a,m^{-}}^{-1}}{\sqrt A}\partial_{u}^{m}\big[\left((1-Bu)^2-C^2u^2\right)^{-1/2}\exp{\left(A_1\gamma^{\ast 2}+A_2\gamma^{ 2}-A_3|\gamma|^2\right)}\big]\bigg|_{u=1},\label{CFPSSTS}
\end{align}
where 
$A_1=\frac{\left(C/2\right)}{(\left(1-Bu)^2-C^2u^2\right)}$, $A_2=\frac{(\frac{C(u^2 B^2+C^2 u^2)}{2}+B C u(1-Bu))}{(\left(1-Bu)^2-C^2u^2\right)}+\frac{C}{2}$, $A_3=\frac{(B(1-Bu)+C^2 u)}{\left(1-Bu)^2-C^2u^2\right)}-\frac{\kappa-1}{2}.\\$
\\
Similarly, for noisy PSSTS at the output
 \begin{align*}
    & {Tr\left[e^{-\gamma^{\ast} \hat{a}}\phi_{s}(\hat{\rho})e^{\gamma\hat{a}^\dag}\right]}\\
     &=Tr\bigg[\frac{N_{a,m^{-}}^{-1}}{\sqrt A}\int\frac{d^2z}{s\pi}\frac{d^2\alpha}{\pi}\frac{d^2\beta}{\pi}
    (\beta^\ast\alpha)^m \exp{\left(-\gamma^{\ast} \hat{a}\right)}\left|\right.{\alpha+z}\left\rangle\right.\left\langle\right.{\beta+z}\left|\right.\exp{\left(\gamma\hat{a}^\dag\right)}\\
    &\times\exp{\left[\frac{C}{2}(\alpha^{\ast 2}+{\beta}^2)+B\alpha^\ast\beta+\frac{1}{2}\{z(\alpha^\ast-\beta^\ast)-z^\ast(\alpha-\beta)\}-\frac{|\alpha|^2+|\beta|^2}{2}-\frac{|z|^2}{s}\right]}
    \bigg]\\
    &=\frac{N_{a,m^{-}}^{-1}}{\sqrt A}\int\frac{d^2z}{s\pi}\frac{d^2\alpha}{\pi}\frac{d^2\beta}{\pi}
    (\beta^\ast\alpha)^m \left\langle\right.{\beta+z}\left|\right.{\alpha+z}\left\rangle\right.\\
    &\times\exp{\left[\frac{C}{2}(\alpha^{\ast 2}+{\beta}^2)+B\alpha^\ast\beta+\frac{1}{2}\{(z(\alpha^\ast-\beta^\ast)-z^\ast(\alpha-\beta))\}+\gamma(z^\ast+\beta^\ast)-\gamma^\ast(z+\alpha)-\frac{|\alpha|^2+|\beta|^2}{2}-\frac{|z|^2}{s}\right]},
 \end{align*}
 
 one can have,
 
\begin{align*}
    \left\langle\right.{\beta+z}\left|\right.{\alpha+z}\left\rangle\right.&=\exp{\bigg[-\frac{|\alpha+z|^2}{2}-\frac{|\beta+z|^2}{2}+(\beta^{\ast}+z^\ast)(\alpha+z)\bigg]}\\
    &=\exp{\bigg[-\frac{|\alpha|^2+|z|^2+\alpha z^\ast+z \alpha^\ast}{2}-\frac{|\beta|^2+|z|^2+\beta z^\ast+z \beta^\ast}{2}+(\beta^{\ast}\alpha+\beta^{\ast}z+z^\ast\alpha+|z|^2)\bigg]}\\
    &=\exp{\bigg[-\frac{|\alpha|^2}{2}-\frac{|\beta|^2}{2}-\frac{1}{2}\{z(\alpha^\ast-\beta^\ast)-z^\ast(\alpha-\beta)\}+\beta^\ast\alpha\bigg]}.
\end{align*}  
\begin{align*}
     &{Tr\left[e^{-\gamma^{\ast} \hat{a}}\phi_{s}(\hat{\rho})e^{\gamma\hat{a}^\dag}\right]}\\
    &=\frac{N_{a,m^{-}}^{-1}}{\sqrt A}\int\frac{d^2z}{s\pi}\frac{d^2\alpha}{\pi}\frac{d^2\beta}{\pi}
    (\beta^\ast\alpha)^m\exp{\left[\frac{C}{2}(\alpha^{\ast 2}+{\beta}^2)+B\alpha^\ast\beta+\beta^\ast\alpha-z\gamma^\ast+z^\ast\gamma+\gamma\beta^\ast-\gamma^\ast\alpha-|\alpha|^2-|\beta|^2-\frac{|z|^2}{s}\right]}\\
    &=\frac{N_{a,m^{-}}^{-1}}{\sqrt A}\int\frac{d^2\alpha}{\pi}\frac{d^2\beta}{\pi}
    (\beta^\ast\alpha)^m
     \exp{\left[\frac{C}{2}(\alpha^{\ast 2}+{\beta}^2)+B\alpha^\ast\beta+\beta^\ast\alpha+\gamma\beta^\ast-\gamma^\ast\alpha-|\alpha|^2-|\beta|^2-{|\gamma|^2}{s}\right]}\\
      &=\frac{N_{a,m^{-}}^{-1}}{\sqrt A}\int\int\frac{d^2\alpha}{\pi}\frac{d^2\beta}{\pi}(\beta^\ast\alpha)^m\exp{\left[\frac{C}{2}(\alpha^{\ast 2}+{\beta}^2)+B\alpha^\ast\beta+\beta^\ast\alpha-\gamma^\ast\alpha+\gamma\beta^\ast-{|\alpha|^2-|\beta|^2}\right]}\\
     &=\frac{N_{a,m^{-}}^{-1}}{\sqrt A}\partial_{u}^{m}\int\int\frac{d^2\alpha}{\pi}\frac{d^2\beta}{\pi}\exp{\left[\frac{C}{2}(\alpha^{\ast 2}+{\beta}^2)+B\alpha^\ast\beta+u \beta^\ast\alpha-\gamma^\ast\alpha+\gamma\beta^\ast-{|\alpha|^2-|\beta|^2}-{|\gamma|^2}{s}\right]}\bigg|_{u=1}\\
     &=\frac{N_{a,m^{-}}^{-1}}{\sqrt A}\partial_{u}^{m}\int\frac{d^2\alpha}{\pi}\exp{\left[\frac{C}{2}(\alpha^{\ast 2})-\gamma^\ast\alpha-{|\alpha|^2}\right]}\exp{\bigg[B\alpha^\ast(u \alpha+\gamma)+\frac{C}{2}(u\alpha+\gamma)^{2}-{|\gamma|^2}{s}\bigg]}\bigg|_{u=1}\\
     &=\frac{N_{a,m^{-}}^{-1}}{\sqrt A}\partial_{u}^{m}\bigg[\int\frac{d^2\alpha}{\pi}\exp{\bigg[\frac{C}{2}(\alpha^{\ast 2}+u^2\alpha^2)-(1-Bu)|\alpha|^2-\alpha(\gamma^\ast-\gamma C u)+B\gamma\alpha^\ast+\frac{C}{2}\gamma^2-{|\gamma|^2}{s}\bigg]}\bigg]\bigg|_{u=1}\\
     &=\frac{N_{a,m^{-}}^{-1}}{\sqrt A}\partial_{u}^{m}\left((1-Bu)^2-C^2u^2\right)^{-\frac{1}{2}}\exp{\bigg[\frac{-(1-Bu)(\gamma^\ast-\gamma C u)B \gamma+\frac{C}{2}((\gamma^\ast-\gamma C u)^2+u^2 B^2\gamma^2)}{(\left(1-Bu)^2-C^2u^2\right)}+\frac{C}{2}\gamma^2-{|\gamma|^2}{s}\bigg]}\bigg|_{u=1}\\
     &=\frac{N_{a,m^{-}}^{-1}}{\sqrt A}\partial_{u}^{m}\bigg[\left((1-Bu)^2-C^2u^2\right)^{-1/2}\\
     &\times\exp{\bigg[\frac{\frac{C}{2}\gamma^{\ast 2}+\gamma^{2}(\frac{C(u^2 B^2+C^2 u^2)}{2}+C u B(1-Bu))-|\gamma|^2(B(1-Bu)+C^2 u)}{(\left(1-Bu)^2-C^2u^2\right)}+\frac{C}{2}\gamma^2-{|\gamma|^2}{s}\bigg]}\bigg]\bigg|_{u=1}\\
   &=\frac{N_{a,m^{-}}^{-1}}{\sqrt A}\partial_{u}^{m}\bigg[\left((1-Bu)^2-C^2u^2\right)^{-1/2}\exp{\left(N_1\gamma^{\ast 2}+N_2\gamma^{ 2}-N_3|\gamma|^2\right)}\bigg]\bigg|_{u=1},
\end{align*}
where 
$N_1=\frac{\left({C/2}\right)}{\left((1-Bu)^2-C^2u^2\right)},N_2=\frac{(\frac{C(u^2 B^2+C^2 u^2)}{2}+B C u(1-Bu))}{\left((1-Bu)^2-C^2u^2\right)}+\frac{C}{2},N_3=\frac{(B(1-Bu)+C^2 u)}{\left((1-Bu)^2-C^2u^2\right)}+s$.\\

After putting the value of trace in  Eq. \eqref{CFPS}. we get following expression for characteristic function

\begin{equation}
    \chi\left(\gamma,\kappa\right)=\frac{N_{a,m^{-}}^{-1}}{\sqrt A}\partial_{u}^{m}\bigg[\left((1-Bu)^2-C^2u^2\right)^{-1/2}\exp{\bigg[-\left(N_1-\frac{(\kappa-1)}{2}\right)|\gamma|^2+N_2\gamma^2+N_3\gamma^{\ast 2}\bigg]}\bigg]\bigg|_{u=1}. \label{CFNPSSTS}
\end{equation}

\section{Appendix: $r^{th}$ Moment For Input And Output States}
For the derivation of $r^{th}$ moment for input and output states, we consider the following 
\begin{align}
    \left<\hat{a}^{\dag r}\hat{a}^r\right>=Tr\left[\hat{\rho}\hat{a}^{\dag r}\hat{a}^r\right]\label{rthm}.
\end{align}
\paragraph{For PATS\\}
$r^{t h}$-moment of PATS can be find using  above Eq. \eqref{rthm},
\begin{align*}
    \langle{n}|\hat{a}^r\rho\hat{a}^{\dag r}|n\rangle &=N_{m}^{-1}\langle{n}|\hat{a}^r:\hat{a}^{\dag m} e^{-A\hat{a}^\dag\hat{a}}\hat{a}^{m}:\hat{a}^{\dag r}|n\rangle=N_{m}^{-1}\frac{(n+r)!}{n!}\langle{n+r}|:\hat{a}^{\dag m} e^{-A\hat{a}^\dag\hat{a}}\hat{a}^{ m}:|n+r\rangle\\
    &=N_{m}^{-1}\frac{(n+r)!}{n!}\sum_{l=0}{\frac{(n+r)!}{(n+r-m)!}}\langle{n+r-m}|: (-A)^{l}\frac{\hat{a}^{\dag k}\hat{a}^{l}}{l!}:|n+r-m\rangle\\
    &=N_{m}^{-1}\frac{(n+r)!}{n!}\sum_{l=0}{\frac{(n+r)!}{(n+r-m)!}}\frac{(-A)^{l}(n+r-m)}{l!\:(n+r-m-l)}\\
    &=N_{m}^{-1}\frac{(n+r)!}{n!}{\frac{(n+r)!}{(n+r-m)!}}(1-A)^{n+r-m}.
\end{align*}
Thus one can see,
\begin{align}
   \langle\hat{a}^{\dag r}\hat{a}^r\rangle=Tr\left[\hat{\rho}\hat{a}^{\dag r}\hat{a}^r\right]=\sum_{n=0}^{\infty}\langle{n}|\hat{a}^r\rho\hat{a}^{\dag r}|n \rangle=N_{m}^{-1} {\sum_{n=0}^{\infty} {\frac{\left(n+r\right)!^2}{n!\left(n+r-m\right)!} \left(1-A\right)^{n+r-m}}}. 
\end{align}

Similarly, for noisy PATS at the output,
\begin{align*}  
\langle\hat{a}^{\dag r}\hat{a}^{r}\rangle &=Tr\left[\hat{{\Phi}_s\left(\rho_{PATS}\right)}\hat{a}^{\dag r}\hat{a}^r\right]=\sum_{n=0}\langle{n}|\hat{a}^r{\Phi}_s\left(\rho_{PATS}\right)\hat{a}^{\dag r}|n \rangle\\
     &=N_{m}^{-1}:\sum_{l=0}^{m}\frac{{m!}^2 (n+r)! \left\langle\left.{n+r}\right|\right.\left({\hat{a}}^\dag\right)^{m-l}\exp\left(-\frac{A{\hat{a}}^{\dag}\hat{a}}{As+1}\right)\left(\hat{a}\right)^{m-l}\left|\left.{n+r}\right\rangle\right.s^l}{n!\:l!\:{\left(m-l\right)!)}^2\left(As+1\right)^{2m-l+1}}: ,
\end{align*}
we get the following (making use of the Eq. \eqref{BNS}),
\begin{align}
    \langle\hat{a}^{\dag r}\hat{a}^{r}\rangle=Tr\left[\phi_s(\rho)\hat{a}^{\dag r}\hat{a}^{r}\right]&=N_{m}^{-1} \sum_{l=0}^{m} \frac{{m!}^{2} s^{l}}{l!{\left(m-l\right)!)}^{2}\left(As+1\right)^{2m-l+1}}\sum_{n=0}{\frac{\left(n+r\right)!^{2}}{n!\left(n+r-m+l\right)!}}\left(1-\frac{A}{As+1}\right)^{n+r-m+l} .
\end{align}
\paragraph{For PSTS\\}
For $r^{t h}$ moment
\begin{align}
\langle\hat{a}^{\dag r}\hat{a}^r\rangle&=Tr[\hat{a}^{\dag r}\hat{a}^r\rho_{P S T S}]=N_{m^-}^{-1}Tr\bigg[\int{\frac{d^2 \alpha}{\pi}{(\alpha^{\ast}\alpha)^m}e^{-\frac{|\alpha|^2}{n_{th}}}\hat{a}^{\dag r}\hat{a}^r\left|\left.\alpha\right\rangle\right.\left\langle\left.\alpha\right|\right.\bigg]}\nonumber\\
&=N_{m^-}^{-1}\int{\frac{d^2 \alpha}{\pi}{(\alpha^{\ast}\alpha)^{m+r}}e^{-\frac{|\alpha|^2}{n_{th}}}}=N_{m^-}^{-1} (m+r)! (n_{th})^{m+r+1}.
\end{align}

Similarly for noisy PSTS at the output,

for $r^{t h}$ moment
\begin{align}
\left<\hat{a}^{\dag r}\hat{a}^r\right>&=Tr[\hat{a}^{\dag r}\hat{a}^r\phi_{s}(\rho)]\nonumber\\
&=N_{m^-}^{-1}Tr\bigg[\int\int{\frac{d^2 \alpha}{\pi}\frac{d^2 z}{\pi s}{(\alpha^{\ast}\alpha)^m}\exp{\bigg\{-\frac{|\alpha|^2}{n_{th}}-\frac{|z|^2}{s}\bigg\}}\hat{a}^r\left|\left.\alpha+z\right\rangle\right.\left\langle\left.\alpha+z\right|\right.\hat{a}^{\dag r}}\bigg]\nonumber\\
    &=N_{m^-}^{-1}\int\int{\frac{d^2 \alpha}{\pi}\frac{d^2 z}{\pi s}{(\alpha^{\ast}\alpha)^m}{|\alpha+z|^{2r}}\exp{\bigg[-\frac{|\alpha|^2}{n_{th}}-\frac{|z|^2}{s}\bigg]}\left\langle\left.\alpha+z\right|\left.\alpha+z\right\rangle\right.}\nonumber\\
    &=N_{m^-}^{-1}\partial_{u}^{r}\int\int{\frac{d^2 \alpha}{\pi}\frac{d^2 z}{\pi s}{(\alpha^{\ast}\alpha)^m}\exp{\bigg[-\frac{|\alpha|^2}{n_{th}}-\frac{|z|^2}{s}+u|\alpha+z|^2\bigg]}}\bigg|_{u=0}\nonumber\\
    &=N_{m^-}^{-1}\partial_{u}^{r}\int\int{\frac{d^2 \alpha}{\pi}\frac{d^2 z}{\pi s}{(\alpha^{\ast}\alpha)^m}\exp{\bigg[-(1/n_{th}-u)|\alpha|^2-\left(\frac{1}{s}-u\right)|z|^2+u(z\alpha^{\ast}+z^\ast\alpha)\bigg]}}\bigg|_{u=0}\nonumber\\
    &=N_{m^-}^{-1}\partial_{u}^{r}\left[\frac{1}{\left(\frac{1}{s}-u\right) s}\int{\frac{d^2 \alpha}{\pi}{(\alpha^{\ast}\alpha)^m}\exp{\bigg[-\left(\frac{1}{n_{th}}-u-\frac{u^2}{\frac{1}{s}-u}\right)|\alpha|^2\bigg]}}\right]\bigg|_{u=0}\nonumber\\
    &=N_{m^-}^{-1}\partial_{u}^{r}\left[\frac{1}{\left(\frac{1}{s}-u\right) s}\frac{m!}{\left(\frac{1}{n_{th}}-u-\frac{u^2}{\frac{1}{s}-u}\right)^{m+1}}\right]\bigg|_{u=0}.
\end{align}

\paragraph{For PAKFTS\\}
We have density operator for PAKFTS
\begin{align*}
    \rho_{\text{PAKFT}} = N_{\text{km}}^{- 1}\left[ :{{\hat{a}}^{\dag m}e^{- A{\hat{a}}^{\dag}\hat{a}}\hat{a}}^{m}: - \frac{e^{- \beta\hbar\omega k}}{k!}:{{\hat{a}}^{\dag (m + k)}e}^{- {\hat{a}}^{\dag}\hat{a}}{\hat{a}}^{(m + k)}: \right],
\end{align*}
using  the above and Eq. \eqref{rthm}, $r^{t h}$ moment-
\begin{align}
    \left\langle {{\hat{a}}^{\dag r}\hat{a}}^{r} \right\rangle &= N_{\text{km}}^{- 1}\ \text{Tr}\left\lbrack {{\hat{a}}^{r}:{{\hat{a}}^{\dag m}e^{- A{\hat{a}}^{\dag}\hat{a}}\hat{a}}^{m}:\hat{a}}^{\dag r} \right\rbrack - \ N_{\text{km}}^{- 1}\ \frac{e^{- \beta\hbar\omega k}}{k!}\ \text{Tr}\left\lbrack {{\hat{a}}^{r}:{{\hat{a}}^{\dag (m + k)}e}^{- {\hat{a}}^{\dag}\hat{a}}{\hat{a}}^{(m + k)}:\hat{a}}^{\dag r} \right\rbrack \\
    &= N_{\text{km}}^{- 1}\sum_{n = 0}^{}\left[ \frac{{((n + r)!)}^{2}}{n!(n + r - m)!}(1 - A)^{n + r - m} - \frac{e^{- \beta\hbar\omega k}}{k!}\frac{{((n + r)!)}^{2}}{n!\ (n + r - m - k)!}\left| \left\langle n + r - m - k \middle| 0 \right\rangle \right|^{2} \right].
\end{align}

Similarly, for noisy PAKFTS the output,

\begin{align}
    \left\langle {{\hat{a}}^{\dag r}\hat{a}}^{r} \right\rangle & = N_{\text{km}}^{- 1}\sum_{n = 0}^{}\bigg[ \sum_{l = 0}^{m}{\frac{{m!}^{2}{{((n + r)!)}^{2}s}^{l}}{l!\:n!\:((m - l)!)^{2}\:(n + r - m + l)!\left( \text{As} + 1 \right)^{2m - l + 1}}\left( \frac{\text{As} -A + 1}{\text{As} + 1} \right)^{n + r - m +l}} \nonumber \\
   & - \frac{e^{- \beta\hbar\omega k}}{k!}\sum_{l = 0}^{m + k}{\frac{{(m + k)!}^{2}{((n + r)!)}^{2}s^{l}}{l!\:n!\:((m + k - l)!)^{2}(n + r - m - k + l)!(s + 1)^{2(m + k) - l + 1}}\left( \frac{s}{s + 1} \right)^{n + r - m - k + l}} \bigg] .
\end{align}
\paragraph{For PASTS\\}
For $r^{t h}$ moment,
\begin{align}
    \left<\hat{a}^{\dag r}\hat{a}^{r}\right>&=Tr{\left[\hat{a}^{\dag r}\hat{a}^{r}\rho_{PASTS}\right]}\nonumber\\
    &=\frac{N_{a,m}^{-1}}{\sqrt{A}}Tr\bigg[\int\int\frac{d^2\alpha}{\pi}\frac{d^2\beta}{\pi}(\alpha^\ast\beta)^{ m}\exp{\left[\frac{C}{2}(\alpha^{\ast 2}+\beta^2)+B\alpha^\ast\beta-\frac{|\alpha|^2}{2}-\frac{|\beta|^2}{2}\right]}\hat{a}^{\dag r}\hat{a}^r|\alpha\rangle\langle\beta|\bigg] \nonumber\\
    &=\frac{N_{a,m}^{-1}}{\sqrt{A}}\int\int\frac{d^2\alpha}{\pi}\frac{d^2\beta}{\pi}(\alpha^\ast\beta)^{ m}\exp{\left[\frac{C}{2}(\alpha^{\ast 2}+\beta^2)+B\alpha^\ast\beta-\frac{|\alpha|^2}{2}-\frac{|\beta|^2}{2}\right]}\left\langle\left.\beta\right|\right.\hat{a}^{\dag r}\hat{a}^r\left|\left.\alpha\right\rangle\right.\nonumber\\
    &=\frac{N_{a,m}^{-1}}{\sqrt{A}}\int\int\frac{d^2\alpha}{\pi}\frac{d^2\beta}{\pi}(\alpha^\ast\beta)^{m}(\alpha\beta^\ast)^{r}\exp{\left[\frac{C}{2}(\alpha^{\ast 2}+\beta^2)+B\alpha^\ast\beta+\alpha\beta^\ast-|\alpha|^2-|\beta|^2\right]}\nonumber\\
    &=\frac{N_{a,m}^{-1}}{\sqrt{A}}\partial_{x}^{m}\partial_{u}^{r}\int\int\frac{d^2\alpha}{\pi}\frac{d^2\beta}{\pi}\exp{\left[\frac{C}{2}(\alpha^{\ast 2}+\beta^2)+x\alpha^\ast\beta+u\alpha\beta^\ast-|\alpha|^2-|\beta|^2\right]}\nonumber\\
    &=\frac{N_{a,m}^{-1}}{\sqrt{A}}\partial_{x}^{m}\partial_{u}^{r}\int\frac{d^2\alpha}{\pi}\exp{\left[\frac{C}{2}\alpha^{\ast 2}-|\alpha|^2+( x u |\alpha^2|+\frac{C}{2}u^2\alpha^2)\right]}\nonumber\\
    &=\frac{N_{a,m}^{-1}}{\sqrt{A}}\partial_{x}^{m}\partial_{u}^{r}\int\frac{d^2\alpha}{\pi}\exp{\left[\frac{C}{2}(\alpha^{\ast 2}+u^2\alpha^2)-|\alpha|^2(1-x u)\right]}\nonumber\\
   &=\frac{N_{a,m}^{-1}}{\sqrt{A}}\partial_{x}^{m}\partial_{u}^{r}\left[(1-x u)^2-C^2u^2\right]^{-1/2} ,
\end{align}
{where $u=1$ and $x=B$}.

Similarly, for noisy PASTS at the output,
\begin{align}
&\left<\hat{a}^{\dag r}\hat{a}^{r}\right>=tr{\left[\hat{a}^{\dag k}\hat{a}^{r}\phi_s(\rho)\right]}\nonumber\\
    &=\frac{N_{a,m}^{-1}}{s\sqrt{A}}Tr\bigg[\int\int\frac{d^2\alpha}{\pi}\frac{d^2\beta}{\pi}\partial_Y^{m}
    \left[{(Y^2- C^2)}^{-1/2} \exp\left[\frac{C_0}{2}({{\alpha}^{\ast 2}}+\beta^2) - (Y_0-1+1/s) {\alpha^\ast\beta}-\frac{|\alpha|^2+|\beta|^2}{2}\right]\right]{\hat{a}^{\dag r}\hat{a}^{r}}|\alpha\rangle\langle\beta|\bigg]\nonumber\\
    &=\frac{N_{a,m}^{-1}}{s\sqrt{A}}\int\int\frac{d^2\alpha}{\pi}\frac{d^2\beta}{\pi}\partial_Y^{m}
    \left[{(Y^2- C^2)}^{-1/2} \exp\left[\frac{C_0}{2}({{\alpha}^{\ast 2}}+\beta^2) - (Y_0-1+1/s) {\alpha^\ast\beta}-\frac{|\alpha|^2+|\beta|^2}{2}\right]\right]\left\langle\left.\beta\right|\right.\hat{a}^{\dag r}\hat{a}^r\left|\left.\alpha\right\rangle\right.\nonumber\\
    &=\frac{N_{a,m}^{-1}}{s\sqrt{A}}
    \int\int\frac{d^2\alpha}{\pi}\frac{d^2\beta}{\pi}\partial_Y^{m}\left[{(Y^2- C^2)}^{-1/2}(\alpha \beta^\ast)^r \exp\left[\frac{C_0}{2}({{\alpha}^{\ast 2}}+\beta^2) - (Y_0-1+\frac{1}{s}) {\alpha^\ast\beta}+\alpha\beta^\ast-{|\alpha|^2-|\beta|^2}\right]\right]\nonumber\\
    &=\frac{N_{a,m}^{-1}}{s\sqrt{A}}\int\int\frac{d^2\alpha}{\pi}\frac{d^2\beta}{\pi}\partial_Y^{m}\partial_{u}^{r}\left[{(Y^2- C^2)}^{-1/2} \exp\left[\frac{C_0}{2}({{\alpha}^{\ast 2}}+\beta^2) - (Y_0-1+\frac{1}{s}) {\alpha^\ast\beta}+u{\alpha\beta^\ast}-{|\alpha|^2-|\beta|^2}\right]\right]\nonumber\\
    &=\frac{N_{a,m}^{-1}}{s\sqrt{A}}\int\frac{d^2\alpha}{\pi}\partial_Y^{m}\partial_{u}^{r}\left[{(Y^2- C^2)}^{-1/2} \exp\left[\frac{C_0}{2}({{\alpha}^{\ast 2}})+\frac{C_0}{2}u^2{{\alpha}^{\ast 2}}-{\left(u\left(Y_0-1+\frac{1}{s}\right)+1\right)|\alpha|^2}\right]\right]\nonumber\\
    &=\frac{N_{a,m}^{-1}}{s\sqrt{A}}\partial_Y^{m}\partial_{u}^{r}\left[{(Y^2- C^2)}^{-1/2}\left(\left(u\left(Y_0-1+\frac{1}{s}\right)+1\right)^2-(C_{0}u)^2\right)^{-1/2}\right],
\end{align}
{where $u=1$}.

\paragraph{For PSSTS\\}
For $r^{th}$ moment,
\begin{align}
  \left<\hat{a}^{\dag r}\hat{a}^r\right> &=Tr[\hat{a}^{\dag k}\hat{a}^k\rho_{P S S T S}]\nonumber\\
  &=Tr\bigg[\frac{N_{a,m^{-}}^{-1}}{\sqrt A}\int\frac{d^2\alpha}{\pi}\frac{d^2\beta}{\pi}(\beta^\ast\alpha)^m\exp{\left[\frac{C}{2}(\alpha^{\ast 2}+{\beta}^2)+B\alpha^\ast\beta-\frac{|\alpha|^2+|\beta|^2}{2}\right]}\hat{a}^{\dag r}\hat{a}^r|\alpha\rangle\langle\beta|\bigg]\nonumber\\
  &=\frac{N_{a,m^{-}}^{-1}}{\sqrt A}\int\frac{d^2\alpha}{\pi}\frac{d^2\beta}{\pi}(\beta^\ast\alpha)^m\exp{\left[\frac{C}{2}(\alpha^{\ast 2}+{\beta}^2)+B\alpha^\ast\beta-\frac{|\alpha|^2+|\beta|^2}{2}\right]}\langle\beta|\hat{a}^{\dag r}\hat{a}^r|\alpha\rangle\nonumber\\
  &=\frac{N_{a,m^{-}}^{-1}}{\sqrt A}\int\frac{d^2\alpha}{\pi}\frac{d^2\beta}{\pi}(\beta^\ast\alpha)^{m+r}\exp{\left[\frac{C}{2}(\alpha^{\ast 2}+{\beta}^2)+B\alpha^\ast\beta-\frac{|\alpha|^2+|\beta|^2}{2}\right]}\langle\beta|\alpha\rangle\nonumber\\
    &=\frac{N_{a,m^{-}}^{-1}}{\sqrt A}\int\frac{d^2\alpha}{\pi}\frac{d^2\beta}{\pi}(\beta^\ast\alpha)^{m+r}\exp{\left[\frac{C}{2}(\alpha^{\ast 2}+{\beta}^2)+B\alpha^\ast\beta+\alpha\beta^\ast-{|\alpha|^2-|\beta|^2}\right]}\nonumber\\
    &=\frac{N_{a,m^{-}}^{-1}}{\sqrt A}\partial_v^{m+r}\int\frac{d^2\alpha}{\pi}\frac{d^2\beta}{\pi}\exp{\left[\frac{C}{2}(\alpha^{\ast 2}+{\beta}^2)+B\alpha^\ast\beta+v\alpha\beta^\ast-{|\alpha|^2-|\beta|^2}\right]}\bigg|_{v=1}\nonumber\\
    &=\frac{N_{a,m^{-}}^{-1}}{\sqrt A}\partial_v^{m+r}\int\frac{d^2\alpha}{\pi}\exp{\left[\frac{C}{2}(\alpha^{\ast 2}+v^2\alpha^2)-{(1-B v)|\alpha|^2}\right]}\nonumber\\
    &=\frac{N_{a,m^{-}}^{-1}}{\sqrt A}\partial_v^{m+r}\left((1-B v)^2-C^2 v^2\right)^{-1/2}\bigg|_{v=1}.
\end{align}

Similarly, for noisy PSSTS at the output,
for $r^{th}$ moment,
\begin{align*}
   &\left<\hat{a}^{\dag r}\hat{a}^r\right>=Tr[\hat{a}^{\dag k}\hat{a}^k\phi_s(\rho_{P S T S})]\\
   &=Tr\bigg[ \frac{N_{a,m^{-}}^{-1}}{\sqrt A}\int\frac{d^2 z}{s\pi}\frac{d^2\alpha}{\pi}\frac{d^2\beta}{\pi}(\beta^\ast\alpha)^m\\
   &\times\exp{\left[\frac{C}{2}(\alpha^{\ast 2}+{\beta}^2)+B\alpha^\ast\beta+\frac{1}{2}{z(\alpha^\ast-\beta^\ast)-z^\ast(\alpha-\beta)}-\frac{|\alpha|^2+|\beta|^2}{2}-\frac{|Z|^2}{s}\right]}\hat{a}^{\dag r}\hat{a}^r|{\alpha+z}\rangle\langle{\beta+z}|\bigg]\\
    &=\frac{N_{a,m^{-}}^{-1}}{\sqrt A}\int \frac{d^2 z}{\pi s}\frac{d^2\alpha}{\pi}\frac{d^2\beta}{\pi}
    (\beta^\ast\alpha)^m((\alpha+z)(\beta^\ast+z^\ast))^r\\
    &\times\exp{\left[\frac{C}{2}(\alpha^{\ast 2}+{\beta}^2)+B\alpha^\ast\beta+\frac{1}{2}\{z(\alpha^\ast-\beta^\ast)-z^\ast(\alpha-\beta)\}-\frac{|\alpha|^2+|\beta|^2}{2}-\frac{|Z|^2}{s}\right]}\langle{\beta+z}|{\alpha+z}\rangle,
\end{align*}

using above, one can see,
\begin{align*}
   &\left<\hat{a}^{\dag r}\hat{a}^r\right>=\frac{N_{a,m^{-}}^{-1}}{\sqrt A}\int\frac{d^2 z}{s\pi}\frac{d^2\alpha}{\pi}\frac{d^2\beta}{\pi}
    (\beta^\ast\alpha)^m((\alpha+z)(\beta^\ast+z^\ast))^r\\
    &\times\exp{\left[\frac{C}{2}(\alpha^{\ast 2}+{\beta}^2)+B\alpha^\ast\beta-\beta^\ast z-z^\ast\alpha-{|\alpha|^2+|\beta|^2}-(1+\frac{1}{s})|z|^2\right]}\\
    &=\frac{N_{a,m^{-}}^{-1}}{\sqrt A}\partial_v^r\int\frac{d^2z}{s\pi}\frac{d^2\alpha}{\pi}\frac{d^2\beta}{\pi}
    (\beta^\ast\alpha)^m\\
    &\times\exp{\left[\frac{C}{2}(\alpha^{\ast 2}+{\beta}^2)+B\alpha^\ast\beta-\beta^\ast z-z^\ast\alpha+v(\alpha+z)(\beta^\ast+z^\ast)-{|\alpha|^2+|\beta|^2}-(1+\frac{1}{s})|z|^2\right]}\\
    &=\frac{N_{a,m^{-}}^{-1}}{\sqrt A}\partial_v^r\int\frac{d^2z}{s\pi}\frac{d^2\alpha}{\pi}\frac{d^2\beta}{\pi}
    (\beta^\ast\alpha)^m\\
    &\times\exp{\left[\frac{C}{2}(\alpha^{\ast 2}+{\beta}^2)+B\alpha^\ast\beta+v\alpha\beta^\ast-{|\alpha|^2+|\beta|^2}-(1+\frac{1}{s}-v)|z|^2-z^\ast\alpha(1-v)-\beta^\ast z(1-v)\right]}\\
    &=\frac{N_{a,m^{-}}^{-1}}{\sqrt A}\partial_v^r\int\frac{d^2\alpha}{\pi}\frac{d^2\beta}{\pi}
    \frac{(\beta^\ast\alpha)^m}{s\left(1+\frac{1}{s}-v\right)}\\
    &\times\exp{\left[\frac{C}{2}(\alpha^{\ast 2}+{\beta}^2)+B\alpha^\ast\beta+v\alpha\beta^\ast-{|\alpha|^2+|\beta|^2}+\frac{(1-v)^2}{\left(1+\frac{1}{s}-v\right)}\alpha\beta^\ast\right]}\\
    &=\frac{N_{a,m^{-}}^{-1}}{\sqrt A}\partial_v^r\partial_u^m\int\frac{d^2\alpha}{\pi}\frac{d^2\beta}{\pi}
    \frac{1}{s\left(1+\frac{1}{s}-v\right)}
    \exp{\left[\frac{C}{2}(\alpha^{\ast 2}+{\beta}^2)+B\alpha^\ast\beta-{|\alpha|^2+|\beta|^2}+u\alpha\beta^\ast\right]}\\
    &=\frac{N_{a,m^{-}}^{-1}}{\sqrt A}\partial_v^r\partial_u^m\int\frac{d^2\alpha}{\pi}
    \frac{1}{s\left(1+\frac{1}{s}-v\right)}\exp{\left(\frac{C}{2}(\alpha^{\ast 2}+u^2\alpha^2)-(1-B u)|\alpha|^2\right)}\\
    &=\frac{N_{a,m^{-}}^{-1}}{\sqrt A}\partial_v^r\partial_u^m
    \frac{1}{s\left(1+\frac{1}{s}-v\right)}\left((1-B u)^2-C^2 u^2\right)^{-1/2},
\end{align*}
where $u=\frac{(1-v)^2}{\left(1+\frac{1}{s}-v\right)}+v$ and $v=1$.

\newpage

\bibliographystyle{ieeetr}
\bibliography{source}

\begin{thebibliography}{10}

\bibitem{nielsen_chuang_2010}
M.~A. Nielsen and I.~L. Chuang, {\em Quantum Computation and Quantum
  Information: 10th Anniversary Edition}.
\newblock Cambridge University Press, 2010.

\bibitem{wilde_2013}
M.~M. Wilde, {\em Quantum Information Theory}.
\newblock Cambridge University Press, 2013.

\bibitem{gruska1999quantum}
J.~Gruska, {\em Quantum computing}, vol.~2005.
\newblock McGraw-Hill London, 1999.

\bibitem{bennett1992quantum}
C.~H. Bennett, G.~Brassard, and A.~K. Ekert, ``Quantum cryptography,'' {\em
  Scientific American}, vol.~\textbf{267}, no.~4, pp.~50--57, 1992.

\bibitem{buluta2009quantum}
I.~Buluta and F.~Nori, ``Quantum simulators,'' {\em Science},
  vol.~\textbf{326}, no.~5949, pp.~108--111, 2009.

\bibitem{dell2006multiphoton}
F.~Dell’Anno, S.~De~Siena, and F.~Illuminati, ``Multiphoton quantum optics
  and quantum state engineering,'' {\em Physics reports}, vol.~\textbf{428},
  no.~2-3, pp.~53--168, 2006.

\bibitem{makhlin2001quantum}
Y.~Makhlin, G.~Sch{\"o}n, and A.~Shnirman, ``Quantum-state engineering with
  josephson-junction devices,'' {\em Reviews of modern physics},
  vol.~\textbf{73}, no.~2, p.~357, 2001.

\bibitem{subhashish2019open}
S.~Banerjee, {\em \textit{Open Quantum System: Dynamics of Nonclassical
  Evolution}}.
\newblock Springer Singapore, 2019.

\bibitem{louisell1973quantum}
W.~H. Louisell, {\em Quantum statistical properties of radiation}.
\newblock John Wiley and Sons, Inc., New York, 1973.

\bibitem{paris2003purity}
M.~G. Paris, F.~Illuminati, A.~Serafini, and S.~De~Siena, ``Purity of gaussian
  states: Measurement schemes and time evolution in noisy channels,'' {\em
  Physical Review A}, vol.~\textbf{68}, no.~1, p.~012314, 2003.

\bibitem{lvovsky2020production}
A.~Lvovsky, P.~Grangier, A.~Ourjoumtsev, V.~Parigi, M.~Sasaki, and
  R.~Tualle-Brouri, ``Production and applications of non-gaussian quantum
  states of light,'' {\em arXiv preprint arXiv:2006.16985}, 2020.

\bibitem{RevModPhys.84.621}
C.~Weedbrook, S.~Pirandola, R.~Garc\'{\i}a-Patr\'on, N.~J. Cerf, T.~C. Ralph,
  J.~H. Shapiro, and S.~Lloyd, ``Gaussian quantum information,'' {\em Rev. Mod.
  Phys.}, vol.~\textbf{84}, pp.~621--669, May 2012.

\bibitem{o2009photonic}
J.~L. O'brien, A.~Furusawa, and J.~Vu{\v{c}}kovi{\'c}, ``Photonic quantum
  technologies,'' {\em Nature Photonics}, vol.~\textbf{3}, no.~12,
  pp.~687--695, 2009.

\bibitem{adesso2014continuous}
G.~Adesso, S.~Ragy, and A.~R. Lee, ``Continuous variable quantum information:
  Gaussian states and beyond,'' {\em Open Systems \& Information Dynamics},
  vol.~\textbf{21}, no.~01n02, p.~1440001, 2014.

\bibitem{ourjoumtsev2007increasing}
A.~Ourjoumtsev, A.~Dantan, R.~Tualle-Brouri, and P.~Grangier, ``Increasing
  entanglement between gaussian states by coherent photon subtraction,'' {\em
  Physical review letters}, vol.~\textbf{98}, no.~3, p.~030502, 2007.

\bibitem{laurat2005entanglement}
J.~Laurat, G.~Keller, J.~A. Oliveira-Huguenin, C.~Fabre, T.~Coudreau,
  A.~Serafini, G.~Adesso, and F.~Illuminati, ``Entanglement of two-mode
  gaussian states: characterization and experimental production and
  manipulation,'' {\em Journal of Optics B: Quantum and Semiclassical Optics},
  vol.~\textbf{7}, no.~12, p.~S577, 2005.

\bibitem{PhysRevLett.88.097904}
S.~D. Bartlett, B.~C. Sanders, S.~L. Braunstein, and K.~Nemoto, ``Efficient
  classical simulation of continuous variable quantum information processes,''
  {\em Phys. Rev. Lett.}, vol.~\textbf{88}, p.~097904, Feb 2002.

\bibitem{PhysRevLett.82.1784}
S.~Lloyd and S.~L. Braunstein, ``Quantum computation over continuous
  variables,'' {\em Phys. Rev. Lett.}, vol.~\textbf{82}, pp.~1784--1787, Feb
  1999.

\bibitem{PhysRevLett.97.110501}
N.~C. Menicucci, P.~van Loock, M.~Gu, C.~Weedbrook, T.~C. Ralph, and M.~A.
  Nielsen, ``Universal quantum computation with continuous-variable cluster
  states,'' {\em Phys. Rev. Lett.}, vol.~\textbf{97}, p.~110501, Sep 2006.

\bibitem{PhysRevA.79.062318}
M.~Gu, C.~Weedbrook, N.~C. Menicucci, T.~C. Ralph, and P.~van Loock, ``Quantum
  computing with continuous-variable clusters,'' {\em Phys. Rev. A},
  vol.~\textbf{79}, p.~062318, Jun 2009.

\bibitem{PhysRevLett.112.120504}
N.~C. Menicucci, ``Fault-tolerant measurement-based quantum computing with
  continuous-variable cluster states,'' {\em Phys. Rev. Lett.},
  vol.~\textbf{112}, p.~120504, Mar 2014.

\bibitem{PhysRevA.95.052352}
F.~Arzani, N.~Treps, and G.~Ferrini, ``Polynomial approximation of non-gaussian
  unitaries by counting one photon at a time,'' {\em Phys. Rev. A},
  vol.~\textbf{95}, p.~052352, May 2017.

\bibitem{PhysRevA.64.012310}
D.~Gottesman, A.~Kitaev, and J.~Preskill, ``Encoding a qubit in an
  oscillator,'' {\em Phys. Rev. A}, vol.~\textbf{64}, p.~012310, Jun 2001.

\bibitem{PhysRevLett.123.200502}
B.~Q. Baragiola, G.~Pantaleoni, R.~N. Alexander, A.~Karanjai, and N.~C.
  Menicucci, ``All-gaussian universality and fault tolerance with the
  gottesman-kitaev-preskill code,'' {\em Phys. Rev. Lett.}, vol.~\textbf{123},
  p.~200502, Nov 2019.

\bibitem{Kim_2008}
M.~S. Kim, ``Recent developments in photon-level operations on travelling light
  fields,'' {\em Journal of Physics B: Atomic, Molecular and Optical Physics},
  vol.~\textbf{41}, p.~133001, jun 2008.

\bibitem{Xu_2014}
Y.~H.-C. Xu~Xue-Xiang and W.~Yan, ``Comparison between photon
  annihilation-then-creation and photon creation-then-annihilation thermal
  states: Non-classical and non-gaussian properties,'' {\em Chinese Physics B},
  vol.~\textbf{23}, p.~070301, may 2014.

\bibitem{Zhou}
J.~Zhou, H.-y. Fan, and J.~Song, ``Photon-subtracted two-mode squeezed thermal
  state and its photon-number distribution,'' {\em International Journal of
  Theoretical Physics}, vol.~51, pp.~1591--1599, May 2012.

\bibitem{Yuan_Hong_Chun_2010}
Y.~Hong-Chun, X.~Xue-Xiang, and F.~Hong-Yi, ``Generalized photon-added coherent
  state and its quantum statistical properties,'' {\em Chinese Physics B},
  vol.~\textbf{19}, p.~104205, oct 2010.

\bibitem{PhysRevA.82.053812}
S.-Y. Lee and H.~Nha, ``Quantum state engineering by a coherent superposition
  of photon subtraction and addition,'' {\em Phys. Rev. A}, vol.~\textbf{82},
  p.~053812, Nov 2010.

\bibitem{PhysRevA.84.012302}
S.-Y. Lee, S.-W. Ji, H.-J. Kim, and H.~Nha, ``Enhancing quantum entanglement
  for continuous variables by a coherent superposition of photon subtraction
  and addition,'' {\em Phys. Rev. A}, vol.~\textbf{84}, p.~012302, Jul 2011.

\bibitem{PhysRevA.82.043828}
M.~Scalora, M.~A. Vincenti, D.~de~Ceglia, V.~Roppo, M.~Centini, N.~Akozbek, and
  M.~J. Bloemer, ``Second- and third-harmonic generation in metal-based
  structures,'' {\em Phys. Rev. A}, vol.~\textbf{82}, p.~043828, Oct 2010.

\bibitem{Zhou_2012}
J.~Zhou, J.~Song, H.~Yuan, and B.~Zhang, ``The statistical properties of a new
  type of photon-subtracted squeezed coherent state,'' {\em Chinese Physics
  Letters}, vol.~\textbf{29}, no.~5, p.~050301, 2012.

\bibitem{Lu_2014}
L.~Dao-Ming and F.~Hong-Yi, ``Photon number cumulant expansion and generating
  function for photon added- and subtracted-two-mode squeezed states,'' {\em
  Chinese Physics B}, vol.~\textbf{23}, p.~020302, dec 2013.

\bibitem{WANG2015108}
S.~Wang, L.-L. Hou, and X.-F. Xu, ``Higher nonclassical properties and
  entanglement of photon-added two-mode squeezed coherent states,'' {\em Optics
  Communications}, vol.~\textbf{335}, pp.~108--115, 2015.

\bibitem{PhysRevLett.71.1816}
K.~Vogel, V.~M. Akulin, and W.~P. Schleich, ``Quantum state engineering of the
  radiation field,'' {\em Phys. Rev. Lett.}, vol.~\textbf{71}, pp.~1816--1819,
  Sep 1993.

\bibitem{Sperling_2015}
J.~Sperling, W.~Vogel, and G.~S. Agarwal, ``Balanced homodyne detection with
  on-off detector systems: Observable nonclassicality criteria,'' {\em
  Europhysics Letters}, vol.~\textbf{109}, p.~34001, feb 2015.

\bibitem{miranowicz2004dissipation}
A.~Miranowicz and W.~Leo{\'n}ski, ``Dissipation in systems of linear and
  nonlinear quantum scissors,'' {\em Journal of Optics B: Quantum and
  Semiclassical Optics}, vol.~\textbf{6}, no.~3, p.~S43, 2004.

\bibitem{marchiolli2004engineering}
M.~A. Marchiolli and W.~D. Jos{\'e}, ``Engineering superpositions of displaced
  number states of a trapped ion,'' {\em Physica A: Statistical Mechanics and
  its Applications}, vol.~\textbf{337}, no.~1-2, pp.~89--108, 2004.

\bibitem{pathak2013elements}
A.~Pathak, {\em Elements of quantum computation and quantum communication}.
\newblock CRC Press Boca Raton, 2013.

\bibitem{agarwal2012quantum}
G.~S. Agarwal, {\em \textit{Quantum optics}}.
\newblock Cambridge University Press, 2012.

\bibitem{agarwal1991nonclassical}
G.~Agarwal and K.~Tara, ``Nonclassical properties of states generated by the
  excitations on a coherent state,'' {\em Physical Review A}, vol.~\textbf{43},
  no.~1, p.~492, 1991.

\bibitem{lee2010quantum}
S.-Y. Lee and H.~Nha, ``Quantum state engineering by a coherent superposition
  of photon subtraction and addition,'' {\em Physical Review A},
  vol.~\textbf{82}, no.~5, p.~053812, 2010.

\bibitem{PhysRevA.43.492}
G.~S. Agarwal and K.~Tara, ``Nonclassical properties of states generated by the
  excitations on a coherent state,'' {\em Phys. Rev. A}, vol.~\textbf{43},
  pp.~492--497, Jan 1991.

\bibitem{Yang:09}
Y.~Yang and F.-L. Li, ``Nonclassicality of photon-subtracted and
  photon-added-then-subtracted gaussian states,'' {\em J. Opt. Soc. Am. B},
  vol.~\textbf{26}, pp.~830--835, Apr 2009.

\bibitem{escher2004controlled}
B.~Escher, A.~Avelar, T.~da~Rocha~Filho, and B.~Baseia, ``Controlled hole
  burning in the fock space via conditional measurements on beam splitters,''
  {\em Physical Review A}, vol.~\textbf{70}, no.~2, p.~025801, 2004.

\bibitem{zavatta2004quantum}
A.~Zavatta, S.~Viciani, and M.~Bellini, ``Quantum-to-classical transition with
  single-photon-added coherent states of light,'' {\em science},
  vol.~\textbf{306}, no.~5696, pp.~660--662, 2004.

\bibitem{podoshvedov2014extraction}
S.~A. Podoshvedov, ``Extraction of displaced number states,'' {\em JOSA B},
  vol.~\textbf{31}, no.~10, pp.~2491--2503, 2014.

\bibitem{Priya}
P.~Malpani, N.~Alam, K.~Thapliyal, A.~Pathak, V.~Narayanan, and S.~Banerjee,
  ``Lower- and higher-order nonclassical properties of photon added and
  subtracted displaced fock states,'' {\em Annalen der Physik},
  vol.~\textbf{531}, no.~2, p.~1800318, 2019.

\bibitem{MALPANI2020124964}
P.~Malpani, K.~Thapliyal, N.~Alam, A.~Pathak, V.~Narayanan, and S.~Banerjee,
  ``Impact of photon addition and subtraction on nonclassical and phase
  properties of a displaced fock state,'' {\em Optics Communications},
  vol.~\textbf{459}, p.~124964, 2020.

\bibitem{Malpani2020}
P.~Malpani, N.~Alam, K.~Thapliyal, A.~Pathak, V.~Narayanan, and S.~Banerjee,
  ``Manipulating nonclassicality via quantum state engineering processes:
  Vacuum filtration and single photon addition,'' {\em Annalen der Physik},
  vol.~\textbf{532}, no.~1, p.~1900337, 2020.

\bibitem{PhysRevA.100.023813}
K.~Debnath, A.~H. Kiilerich, A.~Benseny, and K.~M\o{}lmer, ``Coherent spectral
  hole burning and qubit isolation by stimulated raman adiabatic passage,''
  {\em Phys. Rev. A}, vol.~\textbf{100}, p.~023813, Aug 2019.

\bibitem{xu2019dynamical}
X.-X. Xu and H.-C. Yuan, ``Dynamical evolution of photon-added thermal state in
  thermal reservoir,'' {\em Chinese Physics B}, vol.~\textbf{28}, no.~11,
  p.~110301, 2019.

\bibitem{hu2012nonclassicality}
L.-Y. Hu and Z.-M. Zhang, ``Nonclassicality and decoherence of photon-added
  squeezed thermal state in thermal environment,'' {\em JOSA B},
  vol.~\textbf{29}, no.~4, pp.~529--537, 2012.

\bibitem{hu2010photon}
L.-y. Hu, X.-x. Xu, Z.-s. Wang, and X.-f. Xu, ``Photon-subtracted squeezed
  thermal state: nonclassicality and decoherence,'' {\em Physical Review A},
  vol.~\textbf{82}, no.~4, p.~043842, 2010.

\bibitem{Hong-IWOP}
H.~yi~Fan, ``Operator ordering in quantum optics theory and the development of
  dirac’s symbolic method,'' {\em Journal of Optics B: Quantum and
  Semiclassical Optics}, vol.~\textbf{5}, p.~R147, jun 2003.

\bibitem{dirac1958section}
P.~A.~M. Dirac, {\em The principles of quantum mechanics}.
\newblock No.~27, Oxford university press, 1981.

\bibitem{scully_zubairy_1997}
M.~O. Scully and M.~S. Zubairy, {\em \textit{Quantum Optics}}.
\newblock Cambridge University Press, 1997.

\bibitem{puri2001mathematical}
R.~R. Puri {\em et~al.}, {\em Mathematical methods of quantum optics},
  vol.~\textbf{79}.
\newblock Springer Berlin, Heidelberg, 2001.

\bibitem{Gerry3chep}
C.~Gerry and P.~Knight, {\em Coherent states}, p.~43–73.
\newblock Cambridge University Press, 2004.

\bibitem{THAPLIYAL2015261}
K.~Thapliyal, S.~Banerjee, A.~Pathak, S.~Omkar, and V.~Ravishankar,
  ``Quasiprobability distributions in open quantum systems: Spin-qubit
  systems,'' {\em Annals of Physics}, vol.~\textbf{362}, pp.~261--286, 2015.

\bibitem{PhysRevD.35.1831}
F.~Hong-Yi, H.~R. Zaidi, and J.~R. Klauder, ``New approach for calculating the
  normally ordered form of squeeze operators,'' {\em Phys. Rev. D},
  vol.~\textbf{35}, pp.~1831--1834, Mar 1987.

\bibitem{Dodonov_2002}
V.~V. Dodonov, ```nonclassical' states in quantum optics: a `squeezed' review
  of the first 75 years,'' {\em Journal of Optics B: Quantum and Semiclassical
  Optics}, vol.~\textbf{4}, p.~R1, jan 2002.

\bibitem{Hu:12}
L.-Y. Hu and Z.-M. Zhang, ``Nonclassicality and decoherence of photon-added
  squeezed thermal state in thermal environment,'' {\em J. Opt. Soc. Am. B},
  vol.~\textbf{29}, pp.~529--537, Apr 2012.

\bibitem{PhysRevA.50.3295}
M.~J.~W. Hall, ``Gaussian noise and quantum-optical communication,'' {\em Phys.
  Rev. A}, vol.~\textbf{50}, pp.~3295--3303, Oct 1994.

\bibitem{zhang2021quantifying}
Y.~Zhang and S.~Luo, ``Quantifying decoherence of gaussian noise channels,''
  {\em Journal of Statistical Physics}, vol.~\textbf{183}, no.~2, pp.~1--18,
  2021.

\bibitem{cahill1969ordered}
K.~E. Cahill and R.~J. Glauber, ``Ordered expansions in boson amplitude
  operators,'' {\em Physical Review}, vol.~\textbf{177}, no.~5, p.~1857, 1969.

\bibitem{agarwal1981relation}
G.~S. Agarwal, ``Relation between atomic coherent-state representation, state
  multipoles, and generalized phase-space distributions,'' {\em Physical Review
  A}, vol.~\textbf{24}, no.~6, p.~2889, 1981.

\bibitem{schleich2011quantum}
W.~P. Schleich, {\em Quantum optics in phase space}.
\newblock John Wiley \& Sons, 2011.

\bibitem{THAPLIYAL2016148}
K.~Thapliyal, S.~Banerjee, and A.~Pathak, ``Tomograms for open quantum systems:
  In(finite) dimensional optical and spin systems,'' {\em Annals of Physics},
  vol.~\textbf{366}, pp.~148--167, 2016.

\bibitem{malpani2019quantum}
P.~Malpani, K.~Thapliyal, N.~Alam, A.~Pathak, V.~Narayanan, and S.~Banerjee,
  ``Quantum phase properties of photon added and subtracted displaced fock
  states,'' {\em Annalen der Physik}, vol.~\textbf{531}, no.~11, p.~1900141,
  2019.

\bibitem{husimi1940some}
K.~Husimi, ``Some formal properties of the density matrix,'' {\em Proceedings
  of the Physico-Mathematical Society of Japan. 3rd Series}, vol.~\textbf{22},
  no.~4, pp.~264--314, 1940.

\bibitem{korsch1997zeros}
H.~Korsch, C.~M{\"u}ller, and H.~Wiescher, ``On the zeros of the husimi
  distribution,'' {\em Journal of Physics A: Mathematical and General},
  vol.~\textbf{30}, no.~20, p.~L677, 1997.

\bibitem{kenfack2004negativity}
A.~Kenfack and K.~{\.Z}yczkowski, ``Negativity of the wigner function as an
  indicator of non-classicality,'' {\em Journal of Optics B: Quantum and
  Semiclassical Optics}, vol.~\textbf{6}, no.~10, p.~396, 2004.

\bibitem{sudarshan1963equivalence}
E.~Sudarshan, ``Equivalence of semiclassical and quantum mechanical
  descriptions of statistical light beams,'' {\em Physical Review Letters},
  vol.~\textbf{10}, no.~7, p.~277, 1963.

\bibitem{glauber1963quantum}
R.~J. Glauber, ``The quantum theory of optical coherence,'' {\em Physical
  Review}, vol.~\textbf{130}, no.~6, p.~2529, 1963.

\bibitem{loudon2000quantum}
R.~Loudon, {\em The quantum theory of light}.
\newblock OUP Oxford, 2000.

\bibitem{mandel1979sub}
L.~Mandel, ``Sub-poissonian photon statistics in resonance fluorescence,'' {\em
  Optics letters}, vol.~\textbf{4}, no.~7, pp.~205--207, 1979.

\bibitem{PhysRevA.60.4034}
M.~G. Benedict and A.~Czirj\'ak, ``Wigner functions, squeezing properties, and
  slow decoherence of a mesoscopic superposition of two-level atoms,'' {\em
  Phys. Rev. A}, vol.~\textbf{60}, pp.~4034--4044, Nov 1999.

\bibitem{PhysRevA.44.R2775}
C.~T. Lee, ``Measure of the nonclassicality of nonclassical states,'' {\em
  Phys. Rev. A}, vol.~\textbf{44}, pp.~R2775--R2778, Sep 1991.

\end{thebibliography}

\end{document}